\documentclass[%
reprint,
showpacs,
longbibliography, 
nofootinbib,
amsmath,amssymb,
aps,
twocolumn,
prx,
]{revtex4-1}

\pdfoutput=1
\usepackage[utf8]{inputenc}
\usepackage[english]{babel}
\usepackage[T1]{fontenc}
\usepackage{amsmath}
\usepackage{hyperref}

\usepackage{tikz}
\usepackage{lipsum}

\usepackage{dsfont}
\usepackage[normalem]{ulem}
\usepackage{mathtools}
\usepackage[makeroom]{cancel}
\usepackage{bbm}
\usepackage[english]{babel}

\usepackage{graphicx}
\usepackage{amsthm}
\usepackage{soul}

\newcommand{\tr}{\mathrm{Tr}} 
\newcommand{\Tr}{\mathrm{Tr}} 

\newcommand{\id}{\mathbbm{1}}

\newcommand{\ket}[1]{\left.\left|{#1}\right.\right\rangle}

\newcommand{\bra}[1]{\left.\left\langle{#1}\right.\right|}

\newcommand{\braket}[2]{\left\langle #1 \middle| #2 \right\rangle}

\newcommand{\deq}[1]{\stackrel{#1}{=}}
\newcommand{\bigo}[1]{\mathcal{O}\left (#1\right)}

\newcommand{\average}[2]{\left \langle #1 \right \rangle_{#2}}
\newcommand{\rom}[1]{\uppercase\expandafter{\romannumeral #1\relax}}
\newcommand{\norbra}[1]{\left( #1\right)}
\newcommand{\sqrbra}[1]{\left[ #1\right]}
\newcommand{\curbra}[1]{\left\{ #1\right\}}
\newcommand{\lind}{\mathcal{L}}
\newcommand{\J}{\mathbb{J}}

\usepackage{amssymb}             
\newcommand{\dt}{{\rm d}t\,}
\newcommand{\de}{{\rm d}}
\newcommand{\operatorS}{\mathbb{S}}

\definecolor{ms}{rgb}{0,0,0}
\newcommand{\ms}[1]{\color[rgb]{0,0,0}{#1}}

\newcommand{\G}{\mathcal{G}}  

\begin{document}
	
	\title{Quantum work statistics close to equilibrium 
	}
	\author{Matteo Scandi}
    \affiliation{ICFO-Institut de Ciencies Fotoniques, The Barcelona Institute of
	Science and Technology, Castelldefels (Barcelona), 08860, Spain}
	\author{Harry J. D. Miller}
    \affiliation{Department of Physics and Astronomy, University of Exeter, Stocker Road, Exeter EX4 4QL, UK.}
    \affiliation{\ms{Department of Physics and Astronomy, The University of Manchester, Manchester M13 9PL, UK}}
    \author{Janet Anders}
    \affiliation{Department of Physics and Astronomy, University of Exeter, Stocker Road, Exeter EX4 4QL, UK.}
    \affiliation{\ms{Institute of Physics and Astronomy, University of Potsdam, 14476 Potsdam, Germany.}}
    \author{Mart\'i Perarnau-Llobet}
    \affiliation{Max-Planck-Institut f\"ur Quantenoptik, D-85748 Garching, Germany}
	\date{\today}
	

	\begin{abstract}
	We study the statistics of work, dissipation,  and entropy production of a quantum quasi-isothermal process,   where the system remains close to  thermal equilibrium along the transformation. We derive a general analytic expression for the work distribution and the cumulant generating function. 
 	All work cumulants split into a classical (non-coherent) and quantum (coherent) term, implying that close to equilibrium there are two independent channels of dissipation at all levels of the statistics.  For non-coherent or commuting protocols, only the first two cumulants survive, leading to a Gaussian distribution with its first two moments related through the classical fluctuation-dissipation relation. On the other hand, quantum coherence leads to positive skewness and excess kurtosis in the distribution, and we demonstrate that these non Gaussian effects are a manifestation of asymmetry in relation to the resource theory of thermodynamics. Furthermore, we also show that the non-coherent and coherent contributions to dissipation satisfy independently the Evans-Searles fluctuation theorem, which sets strong bounds on the fluctuations in dissipation, with negative values exponentially suppressed. 
	Our findings are illustrated in a driven two-level system and an Ising chain, where quantum signatures of the work distribution in the macroscopic limit are discussed. 
	\end{abstract}
	
	\maketitle

	\section{Introduction }
	
	The statistics of work, heat, and dissipation play a central role in the study of the non-equilibrium thermodynamics of small systems, both classical \cite{Jarzynski2011,Seifert_2012} and quantum \cite{Esposito2009,Campisi2011c,Goold_2016}. They are known to satisfy  fluctuation theorems \cite{Jarzynski1997d,Crooks1999,Jarzynski2011,Hnggi2015,Funo2018}, which impose some general restrictions on the form of such distributions. Yet,  a complete  characterisation of such thermodynamic   distributions is highly non-trivial, depending heavily on the details  of the specific thermodynamic protocol and the nature (either classical or quantum) of the system under consideration. This complexity contrasts with equilibrium thermodynamics, where it is known that an isothermal reversible transformation outputs a deterministic amount of work, given by the difference of free energy evaluated at the endpoints of the protocol. This observation motivates the  study of \emph{quasi}-isothermal processes, where the transformation is slow enough for the system to always remain close to thermal equilibrium. In this regime, one might hope to find some universal and simple behaviour for the {\ms probability distribution of the work,  which depends only on the equilibrium expectation value of some functional of the driving speed, as it can be expected from linear response theory. Giving such a characterisation} for quantum systems is  the aim of the article. 

	
	
The \emph{quasi}-isothermal, slow driving, {\ms or adiabiatic linear response} regime has been thoroughly studied for classical systems~\cite{nultonQuasistaticProcessesStep1985b,speckDistributionWorkIsothermal2004,Crooks2007,hoppenauWorkDistributionQuasistatic2013a,Kwon2013,Mandal2016a}. 
Notably, the work distribution is known to become Gaussian, at least for typical work values \cite{speckDistributionWorkIsothermal2004, hoppenauWorkDistributionQuasistatic2013a}. Using {\ms the} Jarzynski equality, this result implies the Fluctuation-Dissipation-Relation (FDR) \cite{Jarzynski1997d}:
	\begin{align}\label{eq:FDRclassical}
		\average{w_{\rm diss}}{}  = \frac{1}{2} \beta\, \sigma^2_{\rm diss}.
	\end{align}
	Here,  $\sigma_{{\ms {\rm diss}}}^2 \equiv \langle w^2 \rangle - \langle w \rangle^2 $  is the variance of the work distribution $p(w)$, $\langle w_{\rm diss}\rangle {\ms := \langle w \rangle - \Delta F}$ the average dissipated work along the protocol, and $\beta=1/k_B T$ with $T$ the temperature of the environment.	This relation{\ms, together with the Gaussianity, implies} that the work probability distribution for slowly driven classical systems is completely characterised by {\ms the  average dissipation alone}. 
	
	When moving to quantum systems this picture appears to be incomplete whenever the driving produces coherences between different energy levels: in fact, it has been shown in~\cite{millerWorkFluctuationsSlow2019} that a correction to Eq.~\eqref{eq:FDRclassical} is needed in order to account for the fluctuations arising from additional quantum uncertainty in the system. This result also implies that the probability distribution will deviate from a Gaussian whenever coherences are produced,  {\ms meaning that one needs more information than the average $\langle w_{\rm diss}\rangle$ to characterise $p(w)$ even in the slow driving regime.}
	
	Starting from this observation, reviewed in section~\ref{sec:QFDR}, a complete study of the {\ms production of } irreversibility {\ms in} quantum systems close to equilibrium is presented. We consider a process in which the Hamiltonian of the system is modified by a sequence of $N\gg1$ discrete quenches; after each quench, the system is allowed enough time to thermalise. This kind of protocols, which has been extensively used to describe quasi-isothermal processes \cite{nultonQuasistaticProcessesStep1985b,Crooks2007,Anders2013,Gallego2014,Bumer2019}, is particularly appealing both from the interpretational and the analytical point of view: in fact, providing a way to clearly define work and heat production (being the change of internal energy during the quench or the thermalisation procedure, respectively),  we can avoid any reference to the actual equilibration mechanism. 
	Indeed, discrete protocols were introduced in classical thermodynamics as a way to isolate the features of a slowly driven continuous protocol from the details of the relaxation dynamics~\cite{nultonQuasistaticProcessesStep1985b}.  In section~\ref{sec:contProtocol}, we show that this intuition carries {\ms over} to the quantum regime, and we explain how to generalise our method to more general dynamics. Therefore, both for its pedagogical value  and  the fact that almost no generality is lost, we focus our attention on discrete protocols.

	 In this context, we are able to identify a number of universal properties of the quantum work statistics close to equilibrium. First {\ms we are able to show that the probability distribution of the dissipation during a protocol equals the one for its time reversed. This fact, together with Crooks relation, leads to the Evans-Searles fluctuation theorem \cite{Evans2002}, which implies  that negative values of the dissipation are exponentially damped  (Eq.~\eqref{eq:modifiedCrooks}). Moreover, we  prove that the distribution is Gaussian if and only if no coherence is created during the process (Eq.~(\ref{eq:CGFpolynomial}, \ref{eq:CGFcommsplit})). We also show that the non Gaussian character arising from quantum effects produces a positive skewness and excess kurtosis, witnessing a tendency of the system for extreme deviations above the average dissipation. These results are reviewed in section~\ref{sec:CGF}, where we also numerically study the validity of the slow driving approximation.}
	 
	 
	 The main technical tool we use  is the quasistatic expansion of  the cumulant generating function (CGF), see Eqs.~\eqref{eq:CGF1} and \eqref{eq:CGFslowdriving}.  
	 The connection between the CGF and the Fourier transform of $p(w)$ (Eq.~\eqref{eq:CGFFourier}) yields a systematic procedure to reconstruct the probability distribution, which we show in section~\ref{sec:propDensity}. In particular, in section~\ref{sec:CGFcohe} we consider a protocol in which the eigenbasis of the system Hamiltonian is changed, while keeping the spectrum fixed: we obtain that the probability distribution concentrates on a discrete set even in the quasistatic limit, which is a purely quantum effect reflecting the additional freedom provided by the possibility of creating coherence between energy levels. In section~\ref{sec:CGFcentral} we review how the central limit theorem relates to our results, and we study the thermodynamic limit of an Ising chain, showing that despite the Gaussianity of the distribution, the breakdown of the FDR still {\ms witnesses} the underlying quantum character of the process.


	Finally, in section~\ref{sec:asymmetry} we identify the origin of the entropy production with the degradation of athermality and asymmetry resources \cite{mohammady2019energetic}. In particular, we show how in the quasistatic regime the two channels of entropy production decouple at all levels of statistics (Eq.~(\ref{eq:covysplitting}, \ref{eq:splitCGF})). This result also has implications for the resource theory of thermodynamics \cite{Lostaglio2019}. We prove that the family of second laws  of~\cite{brandaoSecondLawsQuantum2015b}, accounting for the non equilibrium resources in the energy spectrum, collapse into a single law in this regime, which only contributes with a Gaussian term to the probability distribution. On the other hand, we show that the additional constraints on the coherence discussed in~\cite{lostaglioDescriptionQuantumCoherence2015,lostaglioQuantumCoherenceTimeTranslation2015} explicitly account for the non Gaussian effects in the entropy production. These results constitute an additional step in the direction of binding together the quantum work statistics with the resource theoretical description of thermodynamics~\cite{guarnieriQuantumWorkStatistics2019}.
	
	Due to the number of results and the technicality of the derivations, many of the discussions and proofs are deferred to the appendices, which are integral part of the work. We always assume bounded spectra and non degeneracy in the energy levels. Dropping both assumptions would not qualitatively change the results, at the price of complicating the exposition.  {\ms It should also be noted that, unless stated otherwise, all the equalities starting from section~\ref{sec:QFDR}  are to be intended to be valid up to higher order corrections in perturbation theory.}  Lastly, a guide to the notations used is given in Appendix~\ref{app:notations}.

\section{Framework}\label{sec:Framework}

Consider a thermodynamic protocol where a system is driven while being in contact with a thermal bath at inverse temperature~$\beta$. In classical thermodynamics the amount  of irreversibility produced during the process can be quantified in terms of the Clausius inequality $ \Sigma := \Delta S - \beta \Delta Q\geq0$, where $\Delta Q$ is the heat {\ms absorbed from} the environment, $\Delta S$ the change of the system's entropy.  
This motivates the introduction of the entropy production $\Sigma$. Equivalently, one can also define the  dissipated work $w_{\rm diss} := w - \Delta F$ to be the difference between the work needed to complete the process with respect to the average minimum value given by the free energy change $\Delta F$. The duality between the two formulations is a consequence of the first law of thermodynamics:
		\begin{align}
			\Delta U &= w + Q \nonumber\\
			&=\Delta F+w_{\rm diss} + \beta^{-1} \Delta S-\beta^{-1}\Sigma \nonumber\\
			&= \Delta U +w_{\rm diss} -\beta^{-1}  \Sigma,\label{eq:equivalence}
		\end{align}
		which connects the two quantities via the relation ${\Sigma = \beta w_{\rm diss} }$ (see e.g. \cite{mohammady2019energetic} for a recent discussion for quantum systems). This identification should be kept in mind in the rest of the text, as we will often pass from a description in terms of work dissipation rate to one in terms of entropy production rate.
		
		{\ms We consider processes in which the
	Hamiltonian of the system is transformed} by a series of instantaneous quenches between two fixed endpoints $H_A$ and $H_B$, {\ms after each of which the system is allowed enough time to relax} to thermal equilibrium. Specifically, we consider protocols performed of $N$ steps, which {\ms each} consists of \cite{Anders2013}: 
\begin{enumerate}
\item \underline{Quench on the Hamiltonian}: a very fast process in which the Hamiltonian of the system changes as $H_{i}\rightarrow H_{i+1}$, while the state remains unaffected;  
\item \underline{Equilibration procedure}: in which the Hamiltonian is kept fixed, while the system is allowed enough time to perfectly thermalise (${\rho_i \rightarrow \rho_{i+1} \equiv \pi(H_{i+1})}$). 
\end{enumerate}
Since the initial and final point of the process are fixed increasing the number of steps corresponds to an increasingly slower process.

\textcolor{ms}{\subsection{Work statistics} }

In quantum thermodynamics the work depends on the measurement scheme chosen \cite{Talkner2016,baumer2018fluctuating,Debarba2019,Strasberg2019}. We will use here the standard two projective measurement scheme (TPM), which consists in measuring the energy at the beginning and at the end of each quench, and identifying the difference between the two with work \cite{Talkner2007c}. This is justified by the fact that the system is isolated during the quench, so that any change of the internal energy arises from the work performed on the system. From this definition the probability $p^{(i)}(w)$ of a work $w$ in the $i$-th quench is given by:
\begin{align}\label{eq:TPMA1mt}
p^{(i)}(w) &= \sum_{E^{(l)}_{i+1} -E^{(k)}_i = w} \bra{E^{(k)}_i}\pi_i\ket{E^{(k)}_i} \left |\braket{E^{(k)}_i}{ E^{(l)}_{i+1}}\right|^2 ,
\end{align}
where we  denoted the eigenvalues and eigenvectors of the $i$-th Hamiltonian by $E_i$ and $\ket{E_i}$, respectively. 

In order to obtain the full probability distribution we can use the fact that the work at each step is an independent random variable, as the thermalisation processes erase any memory of the previous states.  Then, the full work distribution $p(w)$ can be obtained by convoluting the work distributions at  each step. {\ms This }is in general an untreatable task. For this reason, it is more convenient to consider the cumulant generating function (CGF),
	\begin{align}\label{eq:CGF1}
		K^{-\beta w}(\lambda)&:= \log\int_{-\infty}^{\infty} \de w\, p(w) e^{-\beta\lambda w} ,
	\end{align}
	which is additive under independent random processes{\ms, polynomial of degree two for a Gaussian distribution, and non polynomial otherwise}. The probability distribution $p(w)$ can then be obtained by an appropriate  inverse Fourier transform of~\eqref{eq:CGF1}, whereas the cumulants of the work can be directly computed by differentiation of the CGF, i.e.:
	\begin{align}\label{eq:cumulants}
	\kappa_{w}^{(n)} := (-\beta)^{-n} \frac{\de^n}{\de\lambda^n}\,K^{-\beta w}(\lambda)\Big|_{\lambda=0}.
	\end{align} 

	 In the case at hand the CGF is given by (Appendix~\ref{app:TPM}):
	 	\begin{align}\label{eq:CGF1II}
		K^{-\beta w}(\lambda)=\sum_{i=1}^{N-1} \log \bigg(\Tr\sqrbra{e^{-\beta\lambda H_{i+1}}e^{\beta\lambda H_{i}}\pi_{i}}\bigg).
	\end{align}
It is insightful to rewrite Eq.~\eqref{eq:CGF1II} as~\cite{weiRelationsDissipatedWork2017}:
	\begin{align}\label{eq:CGFsplit}
		K^{-\beta w}(\lambda) &= \sum_{i=1}^{N-1}  \log \bigg(\Tr\sqrbra{\frac{e^{-\beta\lambda H_{i+1}}}{{\mathcal{Z}_{i+1}}^\lambda}\frac{e^{\beta\lambda H_{i}}}{{\mathcal{Z}_{i}}^{-\lambda}}\pi_{i}}\frac{{\mathcal{Z}^\lambda_{i+1}}}{{\mathcal{Z}^\lambda_{i}}}\bigg) \nonumber\\
		&=  -\beta \lambda \Delta F + \sum_{i=1}^{N-1}  (\lambda-1)S_\lambda(\pi_{i+1} || \pi_{i}),
	\end{align} 
	where we {\ms have} isolated the contribution coming from the {\ms increase of} free energy of equilibrium $F(H) = -\beta^{-1} \log \mathcal{Z}(H)$, and we made use of the definition of $\lambda$-Renyi divergence ${S_\lambda(\varrho|| \sigma) = \frac{1}{\lambda -1} \log \Tr\sqrbra{\varrho^\lambda \sigma^{1-\lambda}}}$. In this way we {\ms have} split the CGF in a deterministic part, independent on the particular {\ms driving}, and which only shifts the average work by a constant, and a contribution which explicitly depends on the protocol, accounting for the dissipation arising during the process.  {\ms For this reason we focus our study on} the dissipative CGF, defined as: 
	\begin{align}\label{eq:CGFdissdef}
		K^{\text{diss}}(\lambda) := K^{-\beta (w-\Delta F)}(\lambda) = \sum_{i=1}^{N-1}  (\lambda-1)S_\lambda(\pi_{i+1} || \pi_{i}).
	\end{align}
	This expression {\ms also} highlights the relation between the cumulants of dissipated work and the second laws of thermodynamics~\cite{brandaoSecondLawsQuantum2015b}, as pointed out in~\cite{guarnieriQuantumWorkStatistics2019}.
	
	\textcolor{ms}{
	In the limit $N\rightarrow \infty$, the system is always at thermal equilibrium and we have $p(w)= \delta(w - \Delta F)$, so that the probability distribution becomes independent of the specific protocol implemented. This behaviour can be verified by taking the limit of Eq.~\eqref{eq:CGFsplit}, noticing that the sum in the equation goes to zero as $\bigo{1/N}$.
}

\textcolor{ms}{The slow driving regime is then attained by considering finite but large $N$. This  corresponds to the static linear response regime,  which has been already used to characterise the average dissipation~$\langle w_{\rm diss}\rangle$ in the quantum regime \cite{Campisi2012geometric,sivakThermodynamicMetricsOptimal2012,Bonana2014,acconcia2015shortcuts,Ludovico2016}. Our goal is to go beyond these  findings by  characterising  {\it all} cumulants  of the work distribution in linear response.}
	
{\ms Before continuing, it is worth pointing out a number of remarks: (i) while, strictly speaking, the TPM scheme is invasive whenever $[H_i,H_{i+1}]\neq 0$  as the second measurement dephases  $\pi_i$ in the $H_{i+1}$ basis, this has no implications for the work statistics of the whole process as $\pi_{i}$ would be anyway dephased by the thermalisation process in which~$\pi_{i} \rightarrow \pi_{i+1}$ (note that each step is independent of the previous one);  (ii) importantly,} the previous observation  implies that the same $p(w)$ would be obtained by other schemes to estimate work such as weak or continuous measurements \cite{Allahverdyan2014,Solinas2015,Hofer2017}, so that our results do not depend on the particular measurement scheme used to measure work, {\ms as it has been verified in~\cite{millerWorkFluctuationsSlow2019} for the work fluctuations;} (iii) as it was pointed out in the introduction, while we have characterised slow processes by a discrete model, our results can be extended to more general continuous dynamics (e.g., described by time-dependent master equations) as it is sketched in Sec.~\ref{sec:contProtocol}; in this case the derivations and results become more cumbersome so we prefer to keep the more pedagogical and physically insightful discrete model along most of the manuscript. 

\medskip
\medskip
\medskip
\medskip

 \section{Quantum fluctuation dissipation relation}\label{sec:QFDR}
 
 \begin{figure*}[t]
 	\centering
 	\includegraphics[width=0.32\linewidth]{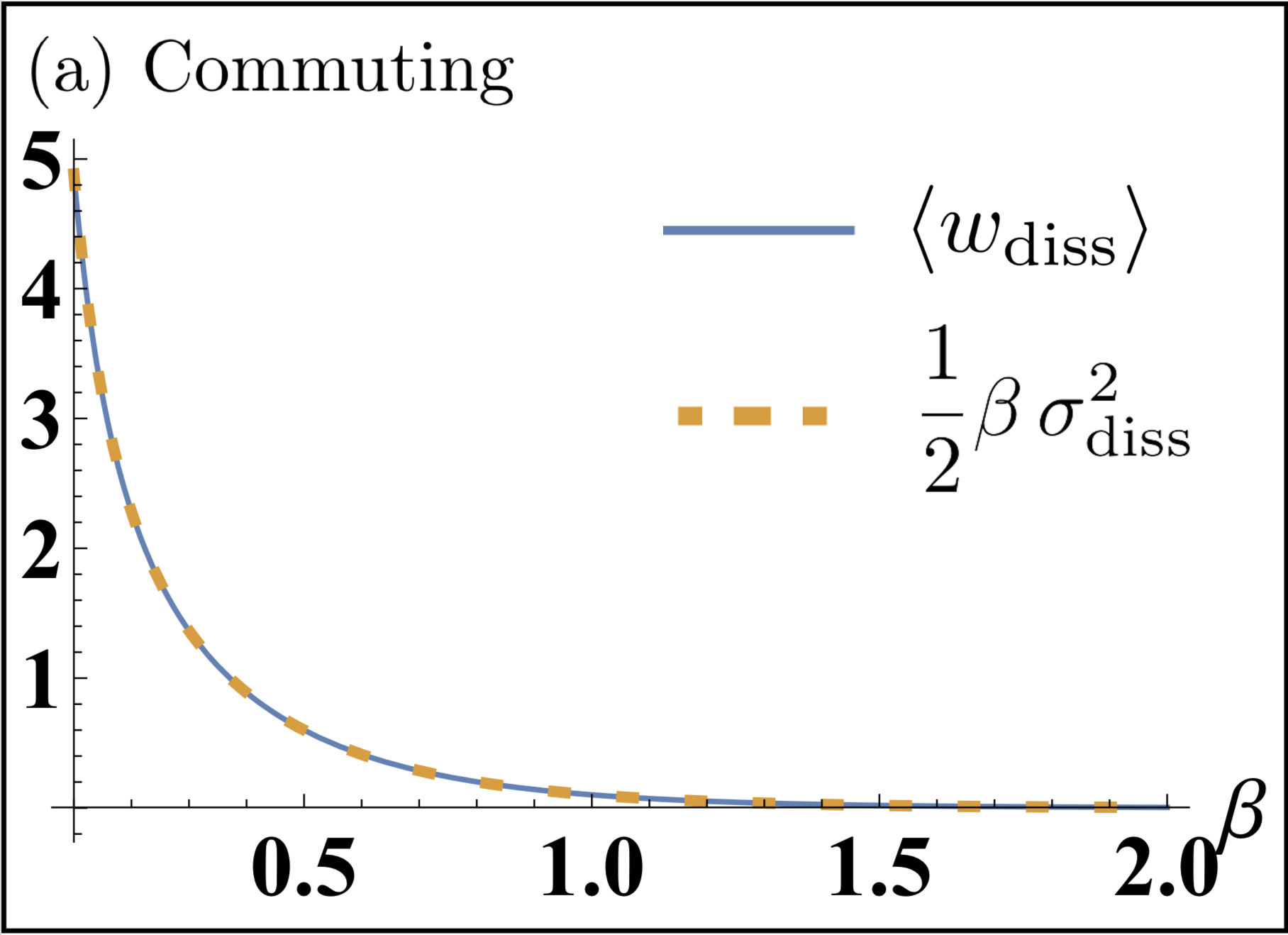}\,
 	\includegraphics[width=0.32\linewidth]{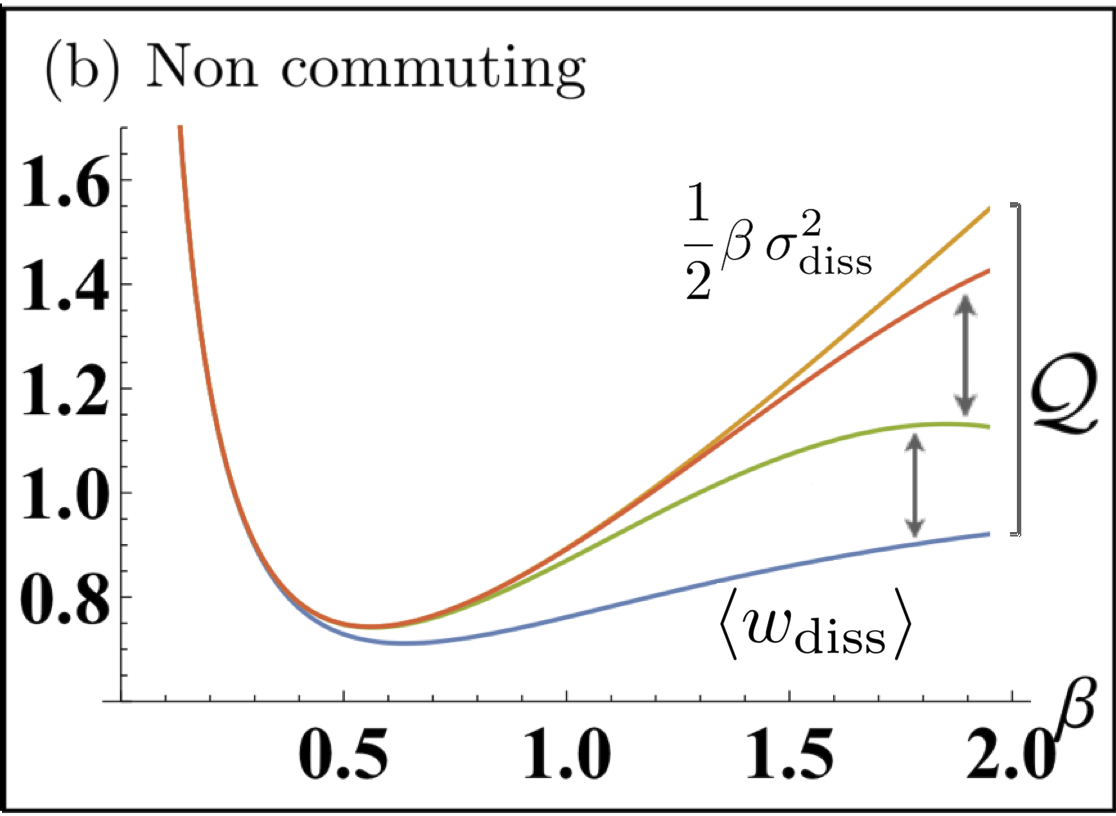}\,
 	\includegraphics[width=0.32\linewidth]{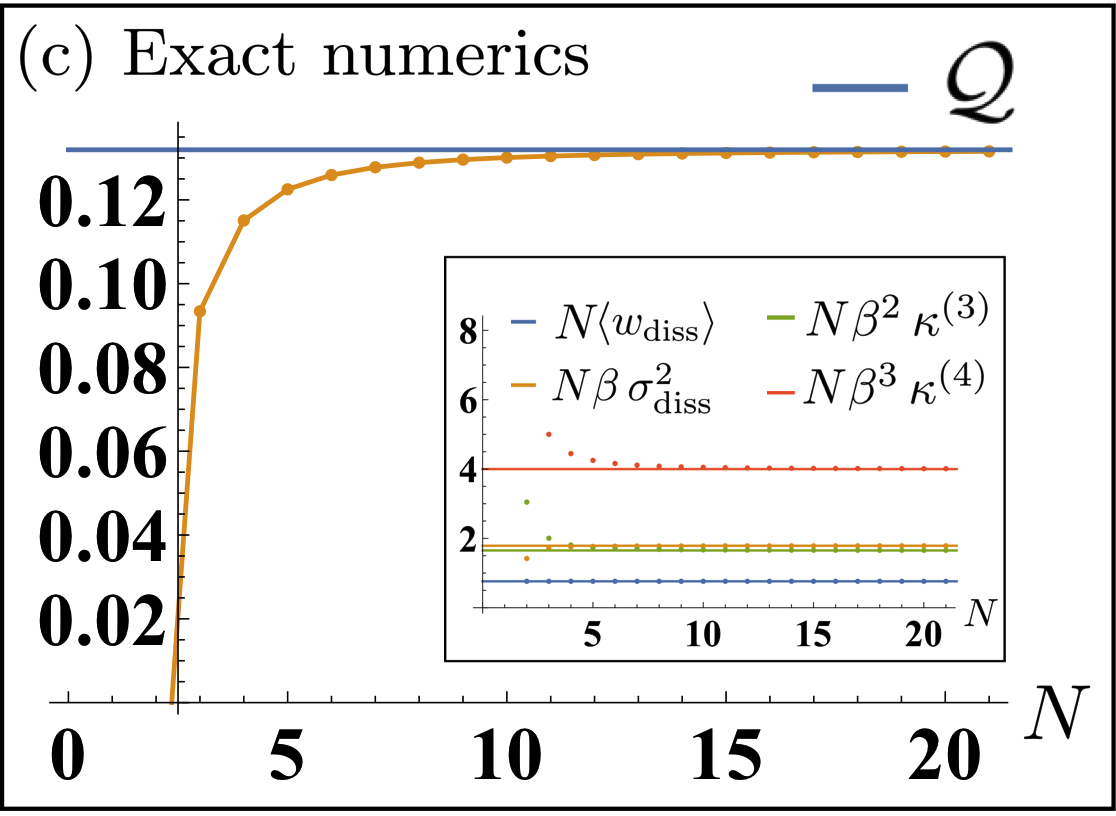}
 	\caption{Dissipated work and work fluctuations as a function of the inverse temperature $\beta$ for a qubit undergoing  two protocols: (a) $H(t) = (1+t) \,\hat\sigma^z$, and (b) $H(t) = t\, \hat\sigma^x + {(1-t)} \,\hat\sigma^z$, for $t\in[0,1]$. In figure (b) the two additional curves are obtained by adding $\sum^{4,6}_{n=3}\frac{(-\beta)^{n-1}}{n!}\kappa^{(n)}_{ w}$ to the dissipated work, showing how the sum asymptotically converges to the quantum contribution $\mathcal{Q}$. In figure (c) we compare the exact value of $\mathcal{Q}$ (dots) with the quasistatic limit presented here (line); in the inset we use the same convention to study the convergence of higher cumulants. All the values have been multiplied by N, and $\beta$ is set to one. The protocol considered is the same as in (b). }
 	\label{fig:fig1}
 \end{figure*}
 
We can pass to review how the FDR in Eq.~\eqref{eq:FDRclassical} changes when passing to quantum systems. These considerations are the natural extension to discrete processes of the work in~\cite{millerWorkFluctuationsSlow2019}, and constitute the starting point for the study of the probability distribution in the next sections.
 
 {\ms We focus on} the first two cumulants of the distribution: $\average{w_{\rm diss}}{}= \kappa_{w}^{(1)}-\Delta F$ and $\sigma^2_{\rm diss}= \kappa_{w}^{(2)}$.	From  Eq.~\eqref{eq:CGF1II} and Eq.~\eqref{eq:cumulants}, we obtain {\ms the exact expressions} (Appendix~\ref{app:FDR}):
 \begin{align}
 &\beta\,\average{w_{\rm diss}}{}   = \sum_{i=1}^{N-1}  \, S (\pi_{i} || \pi_{i+1}),\label{eq:averageDiss}\\
 &\beta^2\, \sigma^2_{\rm diss} =  \sum_{i=1}^{N-1}  \, V (\pi_{i} || \pi_{i+1}),\label{eq:averageFluct}
 \end{align}
 where $S (\varrho || \sigma)$ is the usual relative entropy, and $V(\varrho || \sigma)$ is the relative entropy variance, defined as $V(\varrho || \sigma) := \Tr\sqrbra{\varrho (\log \varrho - \log \sigma)^2}-\Tr\sqrbra{\varrho (\log \varrho - \log \sigma)}^2$. In the limit in which $N\gg 1$, both $S(.||. )$ and $V(.||.)$ go to zero as $\bigo{1/N^2}$ in the leading order. Therefore, in this regime one can expand the average dissipation and the work fluctuations as (Appendix~\ref{app:FDR}):
 
 \begin{align}\label{eq:avWork}
 &\average{w_{\rm diss}}{}  =  \frac{\beta}{2 N}\int_\gamma  \Tr\sqrbra{\dot H_t\, \J_{\pi_t} [\Delta_{\pi_t}\dot H_t]},\\
 &\frac{1}{2} \beta\, \sigma^2_{\rm diss}  = \frac{\beta}{2 N} \int_\gamma \Tr\sqrbra{\dot H_t \, \operatorS_{\pi_t} [\Delta_{\pi_t}\dot H_t]},\label{eq:fluctuations}
 \end{align}
 where we moved to the continuous description $H_t$ with $t\in (0,1)$, i.e., we approximated the discrete trajectory ($H_j $ with $N\gg 1$) by a continuous one {\ms denoted by $\gamma$} (see Appendix~\ref{app:notations} for details {\ms on the notation, and also \cite{Campisi2012geometric,Bonana2014,acconcia2015shortcuts,Ludovico2016,scandiThermodynamicLengthOpen2019} for similar expansions of $\langle w_{\rm diss}\rangle$ using  linear response theory)}. The two superoperators in Eq.~\eqref{eq:avWork} and Eq.~\eqref{eq:fluctuations} are defined by:
 \begin{align}
 &\mathbb{J}_\varrho(A):=\int^1_0 \de x \, \varrho^x \, A  \, \varrho^{1-x}\, ,\label{eq:operatorJ}\\
 &\operatorS_\varrho(A):=\frac{1}{2}\lbrace\varrho,  A\rbrace ,\label{eq:operatorS}
 \end{align}
 and $\Delta_\varrho A=A-\Tr\sqrbra{A\varrho}$ is a projector on the space of traceless operators. It should be noticed that one has ${\operatorS_\varrho \geq \J_\varrho >0}$, with equality if and only if ${\sqrbra{A, \varrho} = 0}$, in which case $\J_\varrho (A) \equiv\operatorS_\varrho (A) \equiv \varrho A$~\cite{marshallInequalitiesTraceFunction1985}. Examining {Eq.~(\ref{eq:avWork}) and (\ref{eq:fluctuations})}, this  condition means that whenever no coherence in the energy basis is created during the protocol (${[H_t, \dot H_t ] = 0}$ at all times), we have the standard work fluctuation-dissipation relation: 
 \begin{align}\label{eq:commFDR}
 \average{w_{\rm diss}}{}  = \frac{1}{2} \beta\, \sigma^2_{\rm diss} \hspace{1cm}{\rm (commuting)},
 \end{align}
 in complete analogy with the classical case of Eq.~\eqref{eq:FDRclassical}. 
 
 As anticipated in the introduction, though, in the general case we obtain a modified FDR of the form 
 \begin{align}\label{eq:noncommFDR}
 \average{w_{\rm diss}}{}  = \frac{1}{2} \beta\, \sigma^2_{\rm diss}-\mathcal{Q}, \hspace{0.5cm}{\rm (non ~commuting)},
 \end{align}
 where a non-negative quantum correction is present. This takes the form
 \begin{align}
 \mathcal{Q} = \frac{\beta}{2N} \int_\gamma \int_0^1 {\rm d}y \, I^y(\pi_t, \dot H_t), \label{eq:Qdef}
 \end{align}
 where we have introduced the Wigner-Yanase-Dyson skew information:
 \begin{align}\label{eq:skewinfoDef}
 I^{y} (\varrho, L) :&= -\frac{1}{2}\Tr\sqrbra{[\varrho^{y}, L][\varrho^{1-y}, L]},
 \end{align}
 which is a quantifier of quantum uncertainty {\ms of} the observable $L$, as measured in the state $\varrho$~\cite{wignerINFORMATIONCONTENTSDISTRIBUTIONS1963}. In particular, the skew information is positive $I^{y} (\varrho, L)\geq 0$ (it vanishes \textit{iff} $[L,\rho]=0$), decreases under classical mixing, and reduces to the usual variance if $\rho$ is pure. 
 

{\ms We stress that the quantum correction $\mathcal{Q}$ is of order~$\bigo{1/N}$, whereas any other possible  violation  of the FDR for classical systems due to the breakdown of the slow driving assumption will be of order~$\bigo{1/N^2}$.} In this sense, the breakdown of Eq.~\eqref{eq:FDRclassical} \textcolor{ms}{at first order in the driving speed} is a witness of the presence of purely quantum effects in the protocol, in particular, the creation of coherence between different energy levels. This point is made more precise in  section~\ref{sec:asymmetry}, where we connect the presence of the correction Eq.~\eqref{eq:Qdef} to the creation and degradation of asymmetry during each step of the process.
 
 As explained in \cite{millerWorkFluctuationsSlow2019}, the quantum FDR Eq.~\eqref{eq:commFDR} can be interpreted as follows: in a slow process, the work fluctuations $\sigma^2_{\rm diss}$ can be expressed as the sum of a thermal contribution (given by $2 \langle w_{\rm diss}\rangle/\beta$) and a quantum one coming from the presence of quantum coherence (given by $2 \mathcal{Q}/\beta$). As an illustration, in Fig.~\ref{fig:fig1} we present the dissipated work and the work fluctuations for a commuting and a non commuting protocol. For lower temperature the two quantities qualitatively differ in the presence of coherence, while in the {\ms limit of} high temperatures one always regains the classical picture as the thermal fluctuations dominate. 

 Interestingly, Eq.~\eqref{eq:noncommFDR} also reflects the emergence of non Gaussian behaviour in the work distribution. Indeed, if we take the logarithm of Jarzynski equality we can isolate the first two cumulants and obtain {\ms the exact relation}~\cite{Jarzynski1997d}
 \begin{align}
 0&\deq{\rm (Jarzynski)} \log \average{e^{-\beta(w-\Delta F)}}{} \nonumber\\
 &= \sum^\infty_{n=1}\frac{(-\beta)^{n}}{n!}\kappa^{(n)}_{ w} \nonumber\\
 &= -\beta \average{w_{\rm diss}}{}   +\frac{1}{2} \beta^2\, \sigma^2_{\rm diss} + \sum^\infty_{n=3}\frac{(-\beta)^{n}}{n!}\kappa^{(n)}_{ w}.\label{eq:jarzynskiderivation}
 \end{align}
 From Eq.~\eqref{eq:noncommFDR}, we can identify the sum in the last equation as $-\beta\mathcal{Q}$, that is~\cite{millerWorkFluctuationsSlow2019}:
 \begin{align}\label{eq:qzero}
 \mathcal{Q} = \sum^\infty_{n=3}\frac{(-\beta)^{n-1}}{n!}\kappa^{(n)}_{ w}.
 \end{align}
 From the properties of the skew information it can be deduced that $\mathcal{Q}\neq 0$ as soon as coherence is generated at any point along a protocol. This fact, together with Eq.~\eqref{eq:qzero}, implies that higher-order cumulants  must be non-zero. We thus conclude that coherence implies a non Gaussian work distribution \textcolor{ms}{even in the slow driving regime, in contrast to the Gaussian distribution found in commuting processes}. In fact, we will later prove a stronger statement, namely that the work distribution is Gaussian \textit{if and only if}~$\mathcal{Q}=0$:
 \begin{align}\label{eq:gaussiff}
 p(w_{\rm diss})\propto e^{-\beta\frac{\left(w_{\rm diss} -\average{w_{\text{diss}}}{}\right){}^2}{4 \average{w_{\text{diss}}}{}  }}\Longleftrightarrow \mathcal{Q}=0.
 \end{align}	
 In this way we see that in the slow driving regime  non Gaussianity of the work distribution provides a direct witness of quantum coherence.  In order to illustrate this point we will first give a closed expression for the CGF, study higher cumulants, and then give a characterisation of the probability distribution in the quasistatic regime.


 \section{The cumulant generating function}\label{sec:CGF}
 
 In this section we discuss the slow driving approximation  ($N\gg 1$) of Eq.~\eqref{eq:CGFdissdef}. Due to the technicality of the derivations, in order not to over complicate the exposition, we defer the main proofs to Appendix~\ref{app:Renyi} and~\ref{app:CGF}, limiting ourselves here to the qualitative discussion of the results.
 
 The main tool we are going to use is the expansion of the $\lambda$-R\'enyi divergence, given by (Appendix~\ref{app:Renyi}):
 \begin{align}\label{eq:Renyientropy}
 &S_\lambda(\varrho+\varepsilon\sigma|| \varrho) =\nonumber\\
 & = \frac{-\varepsilon^2}{2(\lambda -1)}\int_0^\lambda {\rm d}x \int_x^{1-x}{\rm d}y \, \text{cov}_\varrho^y(\J_\varrho^{-1}[\sigma], \J_\varrho^{-1}[\sigma])+\mathcal{O}(\epsilon^3) .
 \end{align}
 where we have defined the $y$-covariance as:
 \begin{align}\label{eq:covy}
 \text{cov}_\rho^y(A, B) := \Tr\sqrbra{\rho^{1-y}A \rho^{y} B} - \Tr\sqrbra{A \rho}\Tr\sqrbra{B \rho}.
 \end{align}
 The $y$-covariance represents a non-commutative generalisation of the classical covariance, reducing to the usual form $\langle A B \rangle-\langle A\rangle\langle B\rangle$ for commuting observables. 
 Using Eq.~\eqref{eq:Renyientropy} to expand the sum present in the CGF in Eq.~\eqref{eq:CGFdissdef}, and  passing to the continuous limit through the definition of Riemann sums, we finally obtain {\ms at first order}:
 \begin{align}
 K^{\text{diss}}(\lambda) = &-\frac{\beta^2}{2N} \int_\gamma\int_0^\lambda {\rm d}x \int_x^{1-x}{\rm d}y \, \text{cov}_t^y(\dot H_t,\dot H_t)\label{eq:CGFslowdriving},
 \end{align}
 where we used the simplified notation $\text{cov}_t^y \equiv \text{cov}_{\pi_t}^y $.
{\ms It is interesting to point out that the $y$-covariance can be rewritten as the connected two point correlation function between $\dot H_t (0)$ and its time evolved counterpart in the imaginary time $\dot H_t (i \beta y)$, with evolution in the Heisenberg picture denoted by $\dot H_t(\nu)=:e^{i\nu H_t}\dot H_t e^{-i\nu H_t}$  (Appendix~\ref{app:CGF}). This identification shows the connection between our approach and linear response theory~\cite{Kubo1957,parisiStatisticalFieldTheory1988}. Note however that the generalised correlation function $\int_0^\lambda {\rm d}x \int_x^{1-x}{\rm d}y \, \text{cov}_t^y(A,B)$ is necessary to characterise the higher order work cumulants, rather than just the usual Kubo correlation function $\int_0^1 {\rm d}y \  \text{cov}_t^y(A,B)$  that determines linear response perturbations only for $\average{w_{\rm diss}}{}$ \cite{Campisi2012geometric,sivakThermodynamicMetricsOptimal2012}. }

 As a sanity check for the validity of the approximation, we show in Appendix~\ref{app:CGF} that Eq.~\eqref{eq:CGFslowdriving} satisfies both the normalisation condition ($K^{\text{diss}}(0) = \log\average{1}{} = 0$) and the Jarzynski equality ($K^{\text{diss}}(1) =  \log \average{e^{-\beta(w-\Delta F)}}{}= 0$). For the derivation of the latter condition, one has to explicitly use the identity:
 \begin{align}\label{eq:covysymmetry}
 \text{cov}_t^y(A, B) = \text{cov}_t^{1-y}(B, A),
 \end{align}
 which can be linked in a precise manner with the KMS condition. In this way, one can understand the necessity of a thermal initial state for the Jarzynski equality to hold (Appendix~\ref{app:CGF}).
 

 The symmetry in the $y$-covariance Eq.~\eqref{eq:covysymmetry} also implies that the CGF satisfies the relation $K^{\text{diss}}(\lambda) = K^{\text{diss}}(1-\lambda)$. This should be contrasted with the general case, in which $K^{\text{diss}}(1-\lambda) = K^{\text{diss}}_{\text{rev}}(\lambda)$,  
 where we implicitly defined the CGF for the time reversed process (see Appendix D for its definition). Putting the two conditions together, we can deduce that the probability of having a dissipation $w_{\rm diss}$ is the same both for the forward and the backwards protocol. Then, using Crooks fluctuation theorem, we see that (details in Appendix \ref{app:CGF}):
 \begin{align}\label{eq:modifiedCrooks}
 \frac{p(w_{\rm diss})}{p(-w_{\rm diss})} = e^{\beta w_{\rm diss}},
 \end{align}
 meaning that negative values of the fluctuations are exponentially suppressed. This relation takes the name of Evans-Searles fluctuation theorem \cite{Evans2002} and it will be further analysed in section~\ref{sec:evansearles}.
 
 In the case in which the protocol does not create coherences ($[H_t, \dot H_t] = 0$), the $y$-covariance reduces to the usual variance, $\text{cov}_t^y(\dot H_t,\dot H_t) \equiv \text{Var}_t [\dot H_t]$. Then, one can explicitly carry out the $x$ and $y$ integral in the CGF Eq.~\eqref{eq:CGFslowdriving}, which gives:
 \begin{align}\label{eq:CGFpolynomial}
 K^{\text{diss}}_{\text{comm}}(\lambda)=  &\frac{\beta^2(\lambda^2-\lambda)}{2N}\int_\gamma\text{Var}_t [\dot H_t].
 \end{align}
 Both Gaussianity and the classical FDR can be directly inferred  from this form of the cumulant generating function. In general, since the appearance of finite cumulants of order three or higher is a quantum witness, we can expect that their expression can be directly connected with a measure of coherence, similarly to Eq.~\eqref{eq:Qdef}. This is indeed true: in fact, the CGF can be split in the form:
 \begin{align}
 K^{\text{diss}}(\lambda)= K^{\text{diss}}_{\text{comm}}(\lambda) + \frac{\beta^2}{2N} \int_\gamma\int_0^\lambda {\rm d}x \int_x^{1-x}{\rm d}y\,  I^y(\pi_t, \dot H_t);\label{eq:CGFcommsplit}
 \end{align}
 As a consequence, from \eqref{eq:cumulants}  we see that all cumulants decouple into a classical (i.e., commuting) and a quantum contribution \textcolor{ms}{in the slow driving regime}. 
 The connection between this expression and the creation of asymmetry across the protocol will be investigated in section~\ref{sec:asymmetry}. We can now proceed to the study of the functional form of higher cumulants.

 \subsection{Characterisation of higher cumulants}\label{sec:highercumulants}
 
 Eq.~\eqref{eq:CGFcommsplit} can be used to give a particularly simple expression for the cumulants. In particular, as it was anticipated in section~\ref{sec:QFDR}, the first two cumulants are given by:
 \begin{align}
 &	\average{w_{\rm diss}}{}  = \frac{\beta}{2N}\norbra{\int_\gamma\text{Var}_t [\dot H_t] - \int_\gamma \int_0^1 {\rm d}y \, I^y(\pi_t, \dot H_t)},\label{eq:wdiss2}\\
 &	\frac{1}{2} \beta\, \sigma^2_{\rm diss} = \frac{\beta}{2N}\int_\gamma\text{Var}_t [\dot H_t],
 \end{align}
 in which we see how the quantum correction $\mathcal{Q}$ naturally appears in Eq.~\eqref{eq:noncommFDR}. On the other hand, since the commuting CGF contributes only to the first two cumulants, Eq.~\eqref{eq:CGFcommsplit} highlights how higher cumulants depends only on the Wigner-Yanase-Dyson skew information $I^y(\pi_t, \dot H_t)$. In fact, by differentiation we obtain the formula
 \begin{align}
 \kappa_{w}^{(n>2)} = \frac{\beta}{N(-\beta)^{n-1}}\,\int_\gamma\norbra{ \frac{\de^{(n-2)}}{\de\lambda^{(n-2)}}\,I^\lambda(\pi_t, \dot H_t)\Big|_{\lambda=0}}.\label{eq:highercumulants}
 \end{align} 
 This shows that all higher {\ms cumulants} of $p(w)$ {\ms for slow processes} can be directly inferred by taking derivatives of the quantum skew information, a measure of purely quantum uncertainty. 
 
 Using the definition of $I^\lambda(\pi_t, \dot H_t)$ given in Eq.~\eqref{eq:skewinfoDef} we can also express the cumulants in a compact form in terms of nested commutators between the Hamiltonian and $\dot H_t$ (Appendix~\ref{app:cumulants}): 
 \begin{align}
 \kappa_{w}^{(2n+1)} &= \frac{1}{N } \int_\gamma \Tr[\pi_t\, C_{n-1}^\dagger C_n],\label{eq:oddcumulants}\\
 \kappa_{w}^{(2n+2)} &= \frac{1}{N } \int_\gamma \Tr[\pi_t\, C_{n}^\dagger C_n] ,\label{eq:evencumulants}
 \end{align}
 where $n\in \mathbb{N}^*$, and we recursively defined the family of operators $C_n$ by the two properties: $C_0 := \dot H_t$, and ${C_n := [H_t, C_{n-1}]}$. For example, the first {\ms two} higher cumulants take the simple form:
 \begin{align}
 \kappa_{w}^{(3)} &= \frac{1}{N } \int_\gamma \Tr\sqrbra{  \pi_t [\dot{H}_t,H_t] \dot{H}_t},\label{eq:thirdcumulant}\\
 \kappa_{w}^{(4)} &= -\frac{1}{N } \int_\gamma \Tr\sqrbra{  \pi_t\, [\dot{H}_t,H_t]^2 }.\label{eq:fourthcumulant}
 \end{align}

 In Appendix~\ref{app:cumulants},  we show that all the cumulants (odd and even) are positive in the presence of coherence. That is, for protocols with quantum coherence ($[H_t, \dot{H}_t] \neq 0$ for some $t$), we have
 \begin{align}
 \kappa_{w}^{(n>2)} > 0,
 \label{positivity}
 \end{align}
 whereas $\kappa_{w}^{(n>2)} = 0$ for commuting protocols in which $[H_t, \dot{H}_t] = 0$. The general proof of \eqref{positivity} is quite {\ms cumbersome} being based on a coordinate expression for the CGF. Yet note that from Eq.~\eqref{eq:evencumulants} one can immediately deduce the positivity of all even cumulants. In the case $n=3$, the positivity can be deduced by noting that  the skew information is positive for $\lambda \in (0,1)$, but identically zero for $\lambda=0$. Hence the first derivative, which gives $\kappa_{w}^{(3)}$, must be positive. 
 From \eqref{positivity} we can infer some information about the shape of the distribution: indeed, the skewness $\gamma_1$ and the excess kurtosis $\gamma_2$ are connected to the cumulants by the relations:
 \begin{align}
 &\gamma_1 = \frac{\kappa_w^{(3)}}{(\kappa_w^{(2)})^{3/2}},\\
 &\gamma_2 = \frac{\kappa_w^{(4)}}{(\kappa_w^{(2)})^{2}}.
 \end{align}
 The positivity of $\kappa_w^{(3)}$ and $\kappa_w^{(4)}$ then means that the probability distribution has a fat tail on the right of the average $\average{w_{\rm diss}}{}$. That is, compared to a normal distribution, values of the dissipation which are bigger than the average by five or more standard deviations are more likely to occur due to quantum fluctuations.

 \subsection{Explicit form of the CGF and numerical verifications}
 
 Before going further with the analysis, in order to gain some intuition on the {\ms specific }form of the CGF  for {\ms particular} physical systems we present here the CGF for a two level system and a quantum Ising chain in a transverse field.

 We parametrise the Hamiltonian of the two level system by spherical coordinates:
 \begin{align}
 H(r, \theta, \phi) = r \cos \phi \sin \theta\,\hat\sigma^x +  r \sin \phi \sin \theta\,\hat\sigma^y+ r \cos \theta\,\hat\sigma^z;\label{eq:qubitHamiltonian}
 \end{align}
 notice that we can neglect any term proportional to the identity since this would only correspond to a shift in the ground state energy. In this setting, assuming external control over the parameters $(r, \theta, \phi)$, we can write the CGF as (Appendix~\ref{app:qubit}):
 \begin{align}\label{eq:CGFqubit}
 K^{\text{diss}}(\lambda) = \frac{\beta^2}{N}\int_\gamma \sqrbra{\dot r^2_t e^{\rm c}_t(\lambda) + r^2_t(\dot \theta^2_t +\sin^2(\theta_t) \dot \phi^2_t) e^{\rm q}_t(\lambda)},
 \end{align}
 where we separated the classical contribution and the quantum one, which respectively read:
 \begin{align}\label{eq:Eclassical}
 e^{\rm c}_t(\lambda) &= \frac{1}{2} \left(\lambda ^2-\lambda \right) \text{sech}^2(\beta  r_t),\\
 e^{\rm q}_t(\lambda) &=\frac{\text{sech}(\beta  r_t) \cosh (\beta  r_t-2 \beta  \lambda  r_t)-1}{4 \beta ^2 r^2_t}.\label{eq:Equantum}
 \end{align}
 As it can be noticed, the first eigenvalue is associated to a change in the energy spacing only, while the second one corresponds to a change in the eigenbasis orientation. The similarity between Eq.~\eqref{eq:CGFpolynomial} and Eq.~\eqref{eq:Eclassical} is not a coincidence: since changing the energy levels does not create coherence, this part of the protocol will contribute only to the first two cumulants, and will ultimately behave classically.

 The second model we consider is an Ising chain in a transverse field, whose Hamiltonian is given by:
 \begin{align}
 H(h)= -J \sum_{i=1}^L(\hat \sigma_i^x\hat\sigma_{i+1}^x + h \hat\sigma_i^z) ,
 \end{align}
 where we assume control only on the magnetic field $h$. Since this model can be mapped into free fermions via a Jordan-Wigner transformation (Appendix~\ref{app:Ising}), it is possible to exactly diagonalise it. This allows us to give a closed expression of the $y$-covariance close to the thermodynamic limit $L\gg1$, which reads:
 \begin{align}\label{eq:covyIsing}
 \text{cov}_{h_t}^y(\dot H_t, \dot H_t) = \dot h^2 L \int_0^\pi \de k\, C(k, y, h) + \bigo{1},
 \end{align}
 where the explicit definition of the function $C(k, y, h)$ is provided in Appendix~\ref{app:Ising}. Plugging this expression in Eq.~\eqref{eq:CGFslowdriving} gives the CGF of the model. It should be noticed that {\ms due} to the factor $L$ in Eq.~\eqref{eq:covyIsing}, both the $y$-covariance and the CGF are extensive.

 At this point we can numerically investigate the validity of the slow driving approximation we used to obtain Eq.~\eqref{eq:CGFslowdriving} from  Eq.~\eqref{eq:CGFdissdef}. For this reason, in Fig.~\ref{fig:fig1} we compare the first four cumulants obtained differentiating Eq.~\eqref{eq:CGFslowdriving} and the ones coming from the exact evolution. It can be seen that the approximation behaves well already for a small number of steps (of the order $N\sim10$). Fast convergences of higher cumulants is also guaranteed by the plot of $\mathcal{Q}$. 	
 
 \begin{figure*}[t]
 	\centering
 	\includegraphics[width=0.494\linewidth]{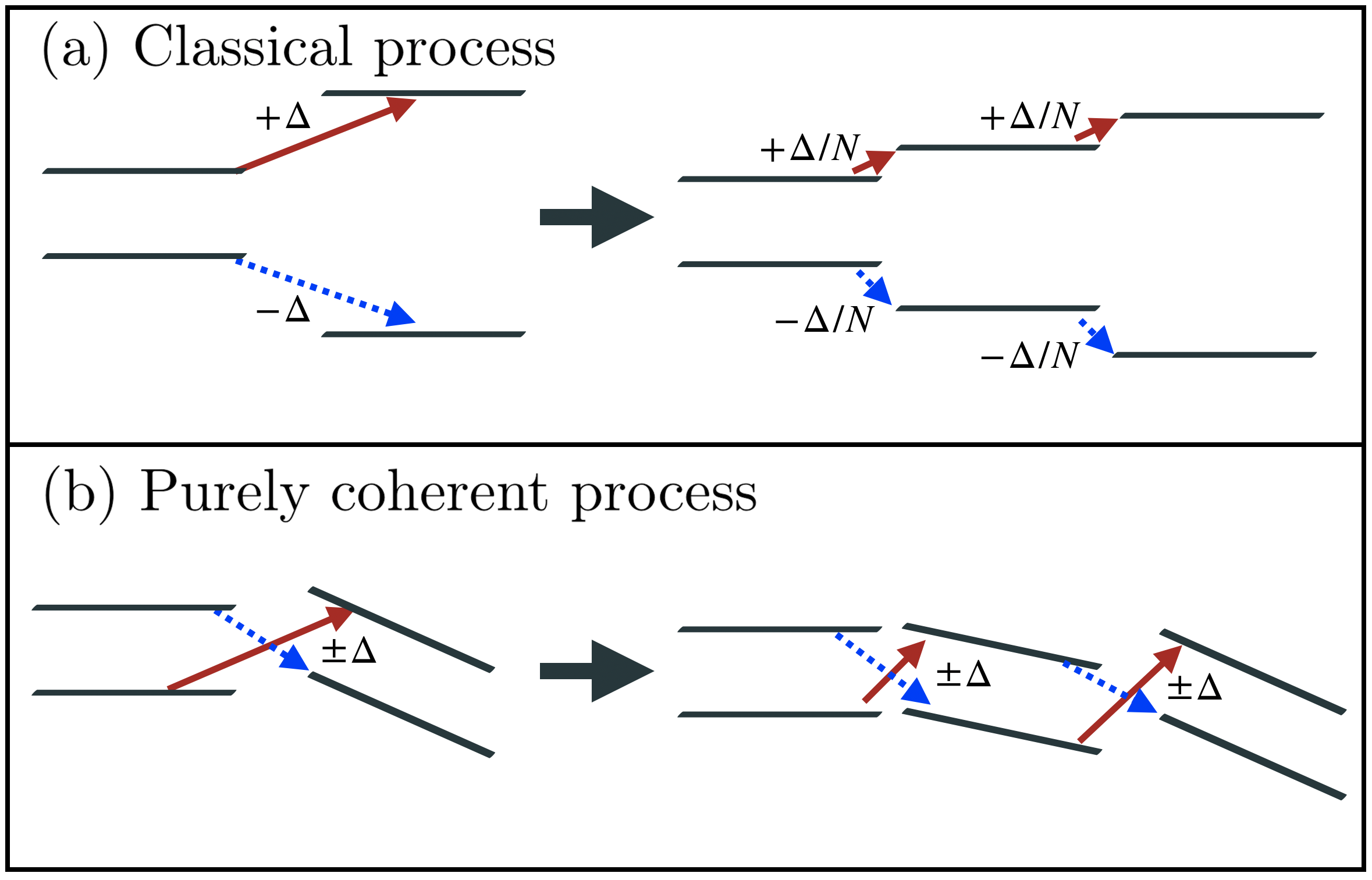}
 	\includegraphics[width=0.499\linewidth]{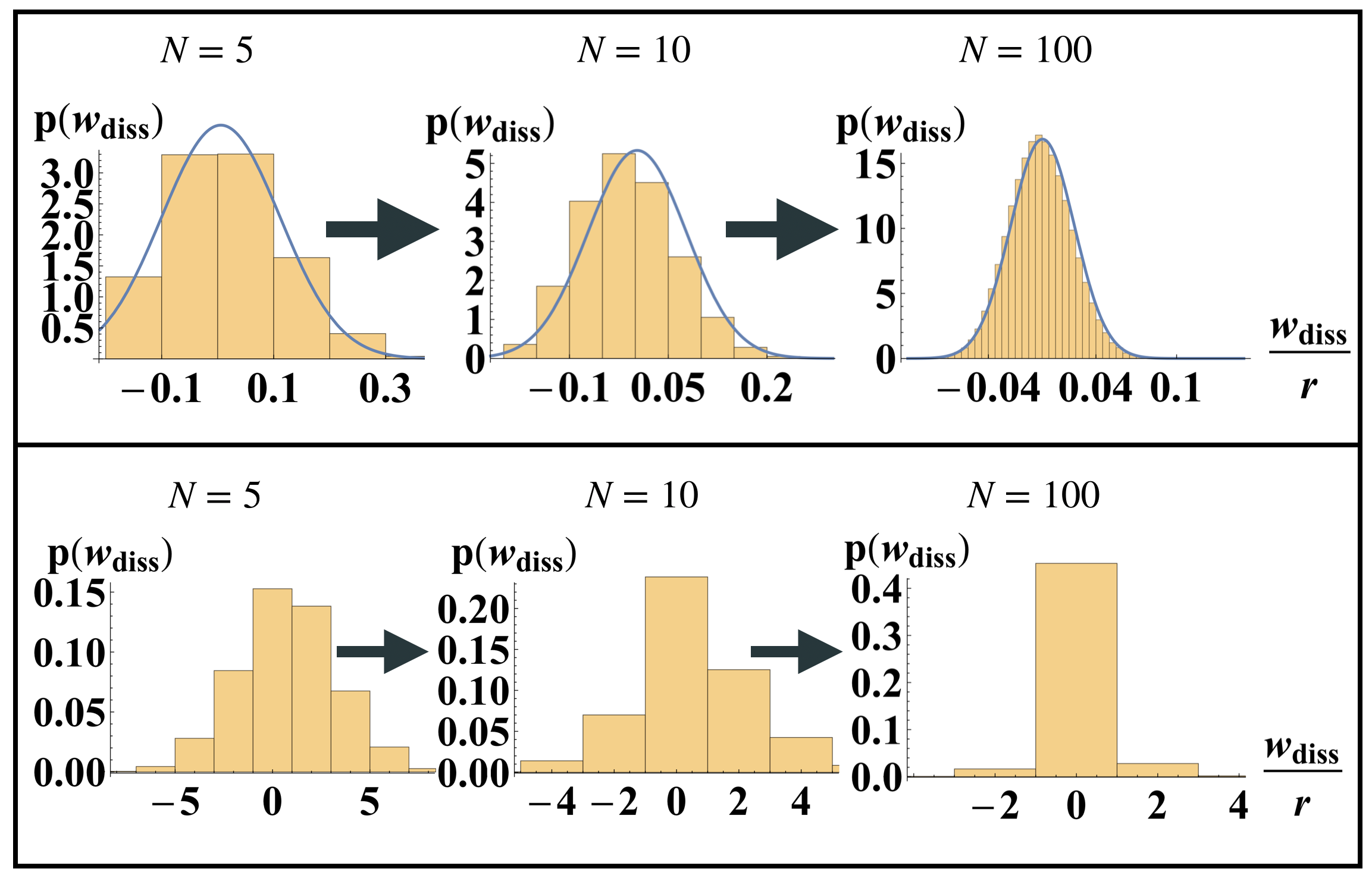}
 	\caption{Illustration of how the slow driving limit affects a process for a classical protocol and a purely coherent one. While in the first case the work output becomes infinitesimal, giving rise to a continuous distribution in the quasistatic limit, for a purely coherent process only the transition probability from one state to the other is affected, while the work output at each step will be either $\pm \Delta$ or $0$. The two protocols are respectively: (a) $H(t) = (1+t) \,\hat\sigma^z/4$, and (b) $H(t) =( \cos(2\pi t) \,\hat\sigma^z + \sin(2\pi t)\, \hat\sigma^x)/4$, for $t\in[0,1]$.}
 	\label{fig:fig3}
 \end{figure*}
 
 \section{Reconstruction of the probability distribution}\label{sec:propDensity}
 
 The expression for the CGF obtained in Eq.~\eqref{eq:CGFslowdriving} yields a tool to qualitatively characterise the entropy production distribution, both as a consequence of the symmetries of $K^{\text{diss}}(\lambda)$ (e.g., as we {\ms have shown} with the modified Crooks relations Eq.~\eqref{eq:modifiedCrooks}), and by providing an algorithm to compute the cumulants, which in turn are used to characterise the shape of the probability density.

 Additionally to these qualitative results one can  use the expression for the CGF to obtain the probability distribution via an inverse Fourier transform. Analytically continuing $K^{\text{diss}}(\lambda)$ to imaginary values ($\lambda \equiv i \nu/\beta$) gives the identity 
 \begin{align}\label{eq:CGFFourier}
 \text{exp}(K^{-\beta w}(i\nu/\beta )) = \int \de w\, p(w) e^{-i \nu w} = \hat p(\nu) .
 \end{align}
 
 Hence, in order  to reconstruct the probability distribution it is sufficient to inverse Fourier transform the relation just obtained. 
 We illustrate this procedure in the two opposite limits of a protocol in which no coherence between the energy levels is created and one in which only a change of basis is performed. The qualitatively different results we obtain will be connected with the different origin of the dissipation.

 \subsection{Gaussianity of the distribution: commuting protocols}
 
 We first consider the simple case in which the protocol does not produce coherence between energy levels, see Fig.~\ref{fig:fig3}(a). This means that only the commuting part of the CGF Eq.~\eqref{eq:CGFpolynomial} has to be considered. Thanks to the quadratic structure of $K^{\text{diss}}_{\text{comm}}(\lambda)$ it is straightforward to perform the inverse Fourier transform, which simply gives:
 \begin{align}
 p(w_{\rm diss})  = \sqrt{\frac{\beta}{4\pi \average{w_{\text{diss}}}{} }} \, e^{-\beta\frac{\left(w_{\rm diss} -\average{w_{\text{diss}}}{}\right){}^2}{4 \average{w_{\text{diss}}}{}  }},
 \end{align}
 where we defined the average dissipated work as: $\average{w_{\text{diss}}}{} \equiv\frac{\beta}{2N}\int_\gamma\text{Var}_t [\dot H_t] $. From this expression one can also directly read off the work fluctuation dissipation relation Eq.~\eqref{eq:commFDR}.

 \subsection{Non Gaussianity of the distribution: purely coherent protocol}\label{sec:CGFcohe}
 
 In the opposite limit, we now study the case in which the process does not affect the spectrum, but it only creates coherences between different energy levels, see Fig.~\ref{fig:fig3}(b). In order to illustrate this point we will consider here the example of a qubit in Eq.~\eqref{eq:qubitHamiltonian} for which $\dot r \equiv 0$. Since the eigenvalue $e^{\rm q}(\lambda)$ only depends on $r$ and not on the coordinates $\phi$ and $\theta$, the CGF Eq.~\eqref{eq:CGFqubit} takes the form:
 \begin{align}
 K^{\text{diss}}(\lambda) = \frac{\beta^2r^2 e^{\rm q}(\lambda)}{N}\int_\gamma \sqrbra{ (\dot \theta^2_t +\sin^2(\theta_t) \dot \phi^2_t)}.
 \end{align}

 In this way the time integration reduces to a path dependent constant. For concreteness we choose here to study a protocol in which $\phi$ is constant and $\theta$ changes as $\theta = 2\pi t$, for $t\in [0,1]$. In this case the integral reduces to $4\pi^2$, and the shape of the distribution is completely characterised by $e^{\rm q}(\lambda)$ {\ms alone (defined in Eq.~\eqref{eq:Equantum})}. Writing down the explicit formula for the probability distribution one obtains
 \begin{align}\label{eq:probDistPurelycoherentPreliminary}
 p(w_{\rm diss}= x) = \frac{1}{2\pi} \int {\rm d}\nu \, e^{i \nu x} \text{exp}(K^{\rm diss}(i\nu/\beta )),
 \end{align}
 where the CGF reads
 \begin{align}
 K^{\rm diss}(i\nu/\beta ) = \frac{1}{4} (\cos (2 \nu r)-i \tanh (\beta r) \sin (2 \nu r)-1).
 \end{align}
 
 Before passing to actually work out the integral in Eq.~\eqref{eq:probDistPurelycoherentPreliminary}, it is interesting to perform the change of variables $\nu \rightarrow \nu + \pi/r$. This gives the condition
 \begin{align}\label{eq:simmetryPurelycoherent}
 p(w_{\rm diss}) = e^{i \frac{w_{\rm diss}}{r}\pi} p(w_{\rm diss}).
 \end{align}
 {\ms Since} $p(w_{\rm diss})$ is real by definition this {\ms relation} tells us that a non-zero probability is possible only for ${w_{\rm diss}\sim 2 k r}$ for $k \in \mathbb{N}$. In fact, since $e^{\rm q}(i \nu/\beta)$ is a periodic function in $\nu$, we can express the exponential in Eq.~\eqref{eq:probDistPurelycoherentPreliminary} in terms of the Fourier series:
 \begin{align}
 \text{exp}(K^{\rm diss}(i\nu/\beta ))=: \sum_{k =-\infty}^{\infty} c_k e^{i\nu r k},
 \end{align}
 where the factors $c_k$ are given by:
 \begin{align}
 c_k  =\frac{1}{2\pi} \int_{-\pi}^\pi {\rm d} (\nu r) \text{exp}(K^{\rm diss}(i\nu/\beta )) e^{-i k \nu r}.
 \end{align}
 
 Plugging this decomposition into Eq.~\eqref{eq:probDistPurelycoherentPreliminary},  we see that the probability distribution is simply given by a sum of Dirac deltas of the form:
 \begin{align}\label{eq:deltaWork}
 p(w_{\rm diss}) &= \sum_{k =-\infty}^{\infty}  \frac{c_k}{2\pi} \int {\rm d}\nu \, e^{i \nu (w_{\rm diss}+r k)}  \nonumber \\
 &= \sum_{k =-\infty}^{\infty}  c_k\delta(w_{\rm diss}+r k);
 \end{align}
 this result is compatible with Eq.~\eqref{eq:simmetryPurelycoherent}. The distribution is presented in Fig.~\ref{fig:fig3}(b) while \ref{fig:fig3}(a) shows the result for the purely commutative protocol, which leads to a Gaussian.

 The persistence of a discrete distribution in the large $N$ limit is a purely quantum effect: in fact, in classical systems there is no way to produce work (or dissipation) without affecting the energy levels. At each step one will produce an energy output of the form $\curbra{\Delta_i/N}$, resulting in a continuous distribution in the regime $N\gg1$ (see Fig.~\ref{fig:fig3} for a illustrative depiction). Contrary, for quantum systems there also exists the freedom of manipulating the system without changing the energy levels, by creating coherence between different eigenvectors.  It should be noticed that in this case, at each step, there is a finite probability of producing an energy output given by $\pm \Delta$, where $\Delta$ is the spectral gap between the two levels. This quantity does not scale with $N$, so that even in the limit of infinite number of steps one can only obtain a distribution concentrated on a discrete set. This is illustrated in Fig. \ref{fig:fig3}.  This discrete behaviour of the work distribution in slow processes is purely quantum in nature, and the presence of $\delta$-{\ms peaks} in the work distribution of a slowly driven system {\ms is therefore} a quantum witness.

 \begin{figure}
 	\centering
 	\includegraphics[width=.45\linewidth]{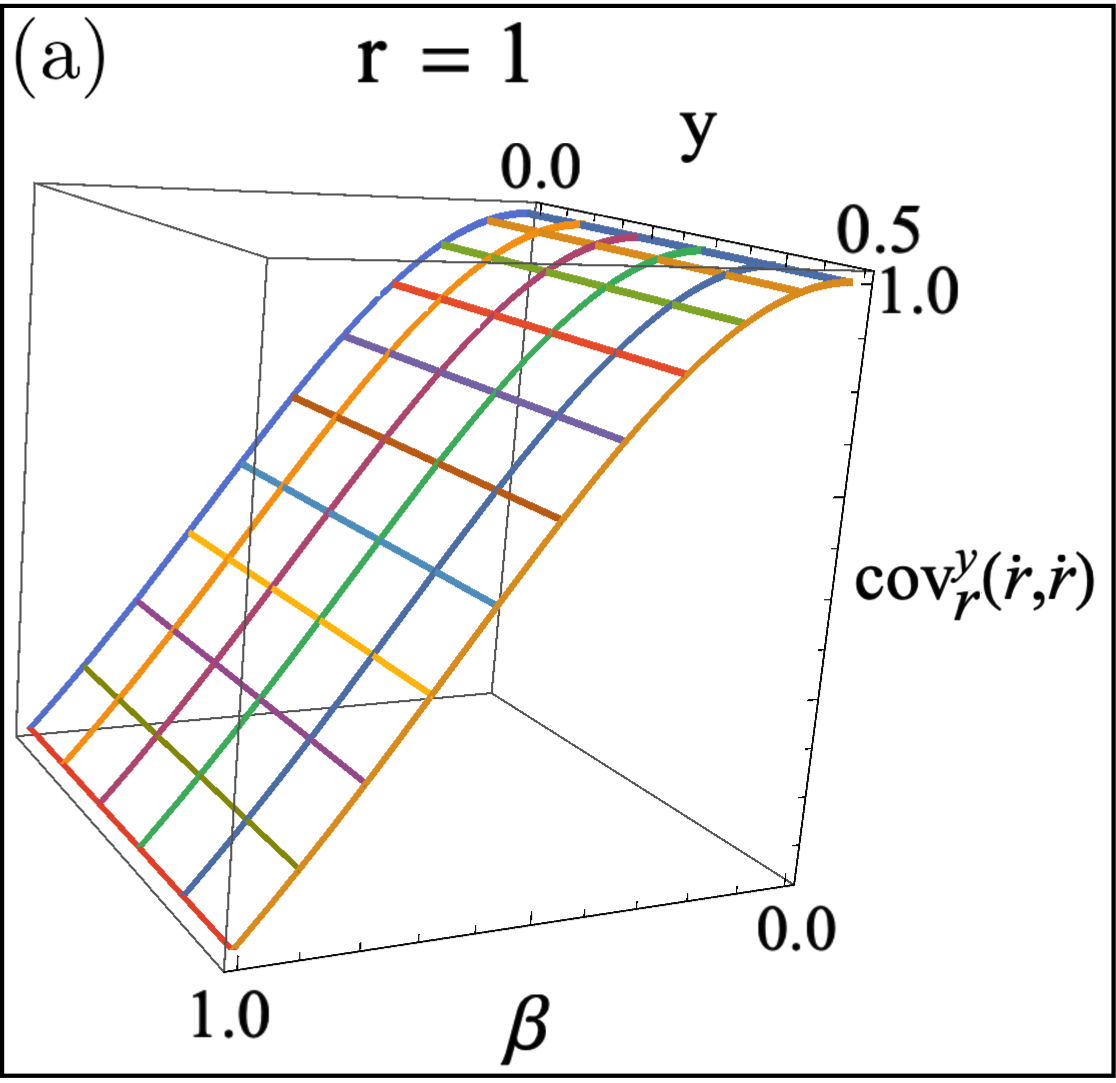}\,
 	\includegraphics[width=.45\linewidth]{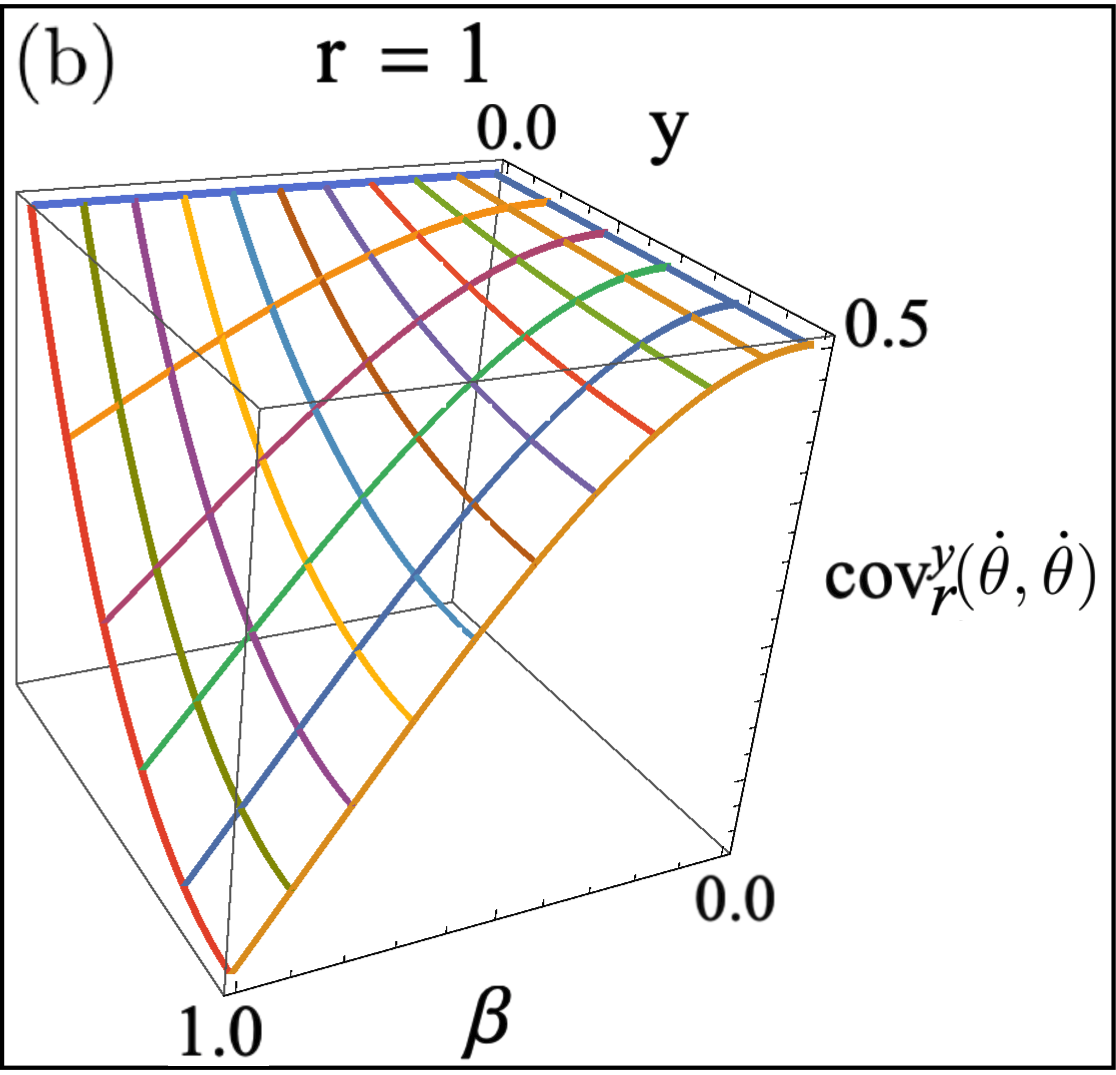}\\
 	\includegraphics[width=.45\linewidth]{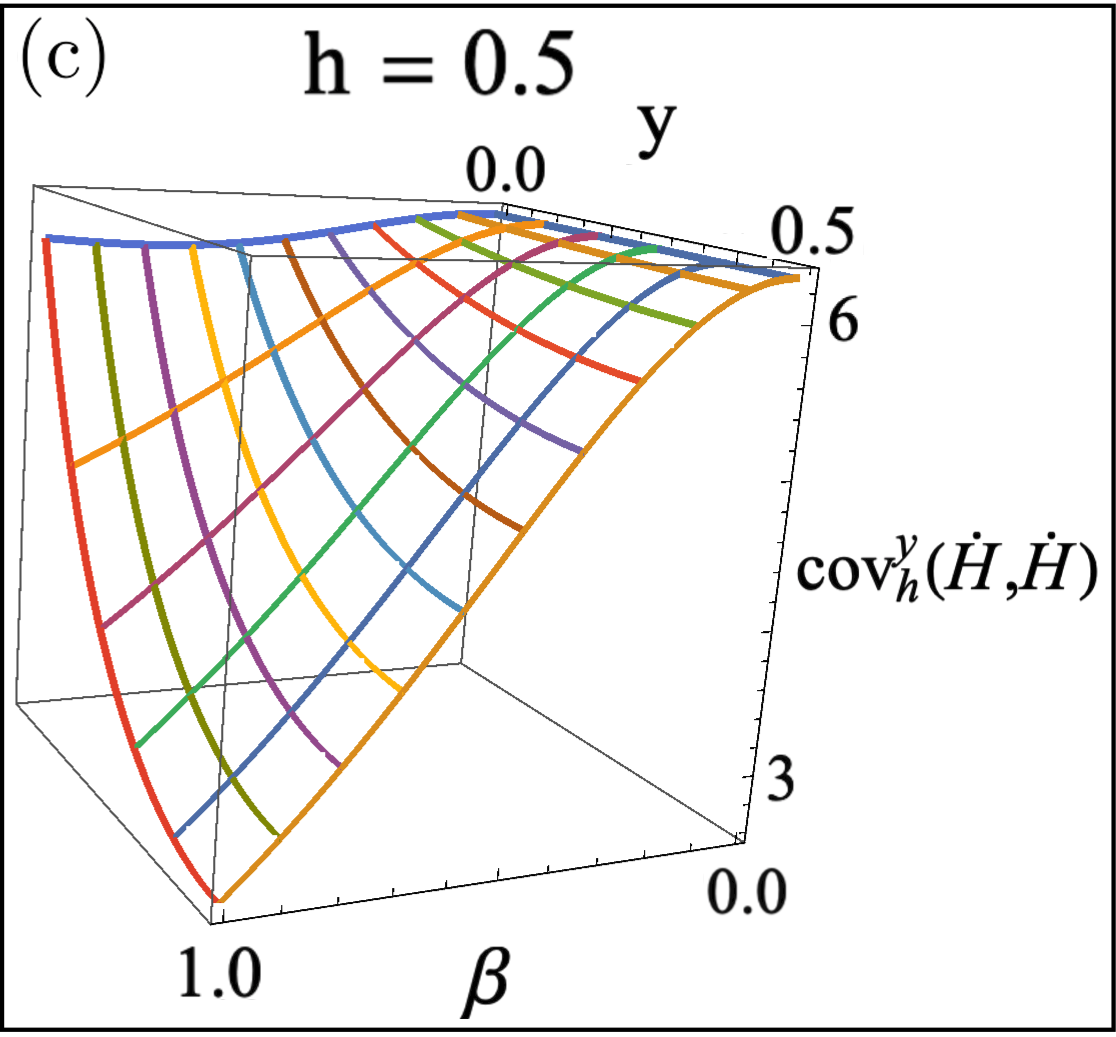}\,
 	\includegraphics[width=.45\linewidth]{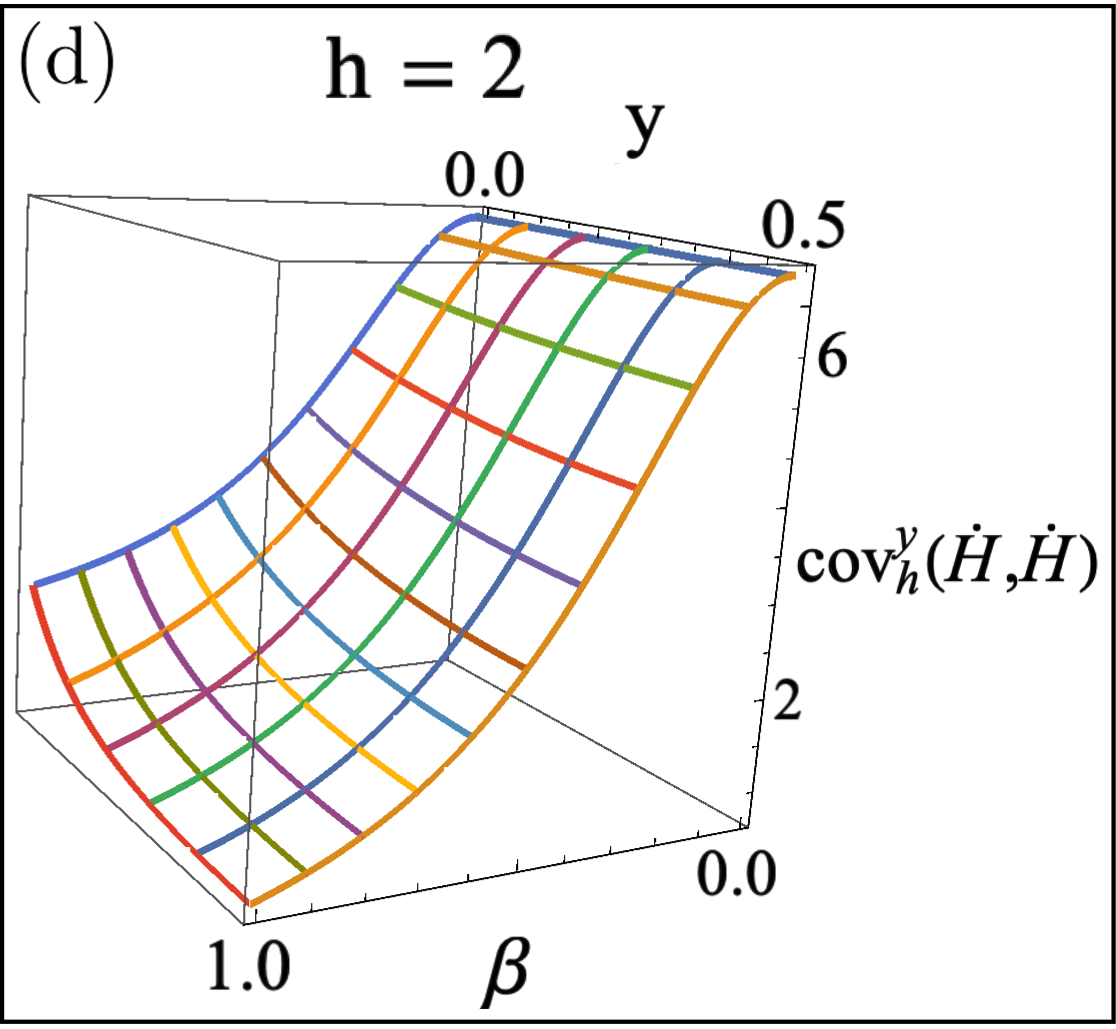}
 	\caption{ Generalised $y$-covariance as a function of the inverse temperature, for the qubit (figures (a) and (b)) and for the Ising chain (figures (c) and (d)). Figure (a) shows the case in which the $y$-covariance does not depend on $y$ for any temperatures. Any deviation from this behaviour signals the emergence of non Gaussianity in the probability distribution. Thanks to the symmetry of the $y$-covariance Eq.~\eqref{eq:covysymmetry}, it is sufficient to study $y\in[0,0.5]$. In figure (c) and (d) we normalised the $y$-covariance by the number of sites.}
 	\label{fig:fig5}
 \end{figure}

 \subsection{Non Gaussianity of the distribution: qualitative behaviour}
 
 For generic transformations of the Hamiltonian in which both the eigenvalues and the eigenvectors are modified, one will obtain work distributions which interpolate between the two extreme cases described above: indeed, $p(w)$ will {\ms in general} tend to a continuous {\ms non Gaussian} distribution, with ${\ms \delta}${\ms -}peaks {\ms which are a manifestation of the quantum adiabatic theorem}.  
 
 {\ms In order to get the full probability distribution of the dissipated work one has to perform an inverse Fourier transform, a task which could be numerically challenging, and certainly not analytically appealing. Nonetheless, if one is interested only in the deviation from Gaussianity, it is sufficient} to {\ms study} the sensitivity of the $y$-covariance (Eq.~\eqref{eq:covy}) on $y$. We know in fact, from Eq.~\eqref{eq:CGFpolynomial}, that if the $y$-covariance does not depend on $y$ at all, the resulting CGF will be quadratic and, consequently, the distribution will be Gaussian. On the opposite side, we have seen how a non-polynomial dependence of the CGF on $\lambda$ can lead to {\ms qualitative deviations from Gaussianity, as it is exemplified in Eq.~\eqref{eq:deltaWork}}. We will illustrate here how this intuition can be applied to qualitatively explore the high and low temperature limit for the qubit and the Ising chain. 
 

 First, notice that if we plot the $y$-covariance for the qubit and Ising chain as a function of $\beta$ and $y$ (Fig.~\ref{fig:fig5}) we can see how the non Gaussianity of the distribution is affected by the temperature. In fact for both models, in the high temperature limit the $y$ dependence becomes flat. In this way we regain the expected result that at high temperature the system will {\ms generically} behave classically. On the other hand, for higher values of $\beta$  a non trivial dependence on $y$ manifests, signalling the appearance of quantum effects. 
 
 In particular, it is interesting to notice how for the Ising chain the system is more responsive to changes in $y$ for values of the transverse field $h\leq1$. This phenomenon can be understood as a signal of the zero temperature phase transition to a long ordered phase.
 
 Let us now discuss the dependence on the temperature of the qubit terms $e^{\rm c/q}(\lambda)$ (Eq.~\eqref{eq:Eclassical} and~\eqref{eq:Equantum}), which characterise the work distribution for the driven qubit. First notice that in the high temperature limit we have:
 \begin{align}
 \lim_{\beta\rightarrow 0} e^{\rm c/q}(\lambda) = \frac{1}{2} \left(\lambda ^2-\lambda \right),
 \end{align}
 witnessing the emergence of a Gaussian behaviour. The fact that both the classical and the quantum contribution converges to the same limit is a consequence of the fact that for $\beta\sim 0$ regardless of the orientation of the eigenbasis or the energy spacing between the two levels the Gibbs state will be given by $\id/2$. For this reason, $[H, \dot H] \sim 0$ for any change of parameters, making the classical and quantum contribution equal.
 
 If we look at the opposite limit, the case in which $\beta\rightarrow\infty$, we can first notice that:
 \begin{align}
 \lim_{\beta\rightarrow \infty} \beta^2\,e^{\rm c}(\lambda) = 0,
 \end{align}
 meaning that in the low temperature limit changing the energy spacing will not affect the work distribution. Indeed, {\ms since} most of the population of the system lives in the ground state, any manipulation of the excited states will leave the system mostly unaffected. On the other hand, if we look at the same regime for $e^{\rm q}$ a more exotic behaviour emerges. Taking the limit along the imaginary axis we obtain:
 \begin{align}
 \lim_{\beta\rightarrow \infty} r^2\beta^2\,e^{\rm q}(i \nu/\beta) =\frac{1}{2}\, e^{-i \left(r \nu +\frac{3 \pi }{2}\right)} \sin (r \nu ),
 \end{align}
 where the periodicity of the function signals {\ms how} the distribution becomes {\ms more and more} concentrated on a discrete set of points, {\ms reproducing} what happens for purely coherent protocols.
 
 The analysis just presented shows how $y$-covariances can be used not only to infer statistical properties of a distribution with a level of detail higher than the averages alone, but it can also provide a tool to infer the physics of the underlying system.
 
 \subsection{Gaussianity of the distribution: central limit theorem}\label{sec:CGFcentral}
 
 The results obtained in the previous sections could seem in contradiction with the central limit theorem: looking at the definition of $\mathcal{Q}$, which witnesses a breakdown of the Gaussianity of the distribution,  it is easy to prove that this quantity is extensive. This means that if one considers a system $\varrho$ made up of $L$ non interacting copies of the same subsystem $\eta$ ($\varrho \equiv \eta^{\otimes L}$), the quantum correction will behave as:
 \begin{align}
 \mathcal{Q}(\varrho) = L\mathcal{Q}(\eta).
 \end{align}
 Similarly, $\langle w_{\rm diss}\rangle $ and $\sigma_w^2$, so that all terms in the FDR are extensive. 
 This condition, together with the Jarzynski equality, implies that the probability distribution of the work output for any finite $L$, however big,  will deviate from a Gaussian distribution.
 
 At the same time though, the central limit theorem says that the standardised sum $\Sigma$ of $L$ i.i.d. random variables $X_i$, defined by $\Sigma := \sum_{i=1}^L X_i/\sqrt{L}$, converges in distribution to a Gaussian as $L\rightarrow\infty$. For this reason one could be lead to think that $\mathcal{Q}$ should converge to zero, due to considerations in the spirit of Eq.~\eqref{eq:qzero}.
 
 This apparent contradiction comes from an erroneous interpretation of Jarzynski equality: in fact, it should be noticed that it applies only to the work distribution, and not to its rescaled version, which is the one treated by the central limit theorem. For this reason, for any $L$ the work distribution output by $\varrho$ will deviate from Gaussianity whenever coherences are present. On the other hand, since cumulants of order  $n$ are homogeneous of degree $n$, for the rescaled work output $w(\varrho)/\sqrt{L}$ we have:
 \begin{align}
 \kappa_{\frac{w(\varrho)}{\sqrt{L}}}^{(n)}  = L^{1-n/2} \, \kappa_{w(\eta)}^{(n)},
 \end{align}
 which is a simple demonstration of the central limit theorem. Using this formula, we have that the rescaled distribution will converge to a Gaussian
 \begin{align}
 p_{\frac{w(\varrho)}{\sqrt{L}}} \stackrel{L\rightarrow\infty}{\longrightarrow }{\ms \G} \left(\sqrt{L} \average{w_{\text{diss}}(\eta)}{}, \frac{2}{\beta}\left(\average{w_{\text{diss}}(\eta)}{}+\mathcal{Q}(\eta)\right) \right),
 \end{align}
 where ${\ms \G}(\mu,\sigma^2)=e^{-(w-\mu)^2/2\sigma^2}/\sqrt{2\pi \sigma^2}$.
 In this way we see that, even if the Gaussianity of the standardised sum still holds, one can deduce the underlying production of coherence by the breakdown of the classical FDR. In other words, the \textcolor{ms}{rescaled} work distribution of large macroscopic quantum systems ($L\gg 1 $) which are slowly driven will tend to a Gaussian distribution with a larger variance than {\ms the one} predicted by the classical FDR. 
 
 In this context, it is also interesting to study what happens to the work distribution of the Ising model in the thermodynamic limit. In fact, if we consider scales larger than the correlation length, the system will behave as the sum of $L_{\text{eff}}$ independent copies, meaning that the central limit theorem should hold. Indeed, the dissipative CGF for the rescaled work $w/L$ (where the scaling is chosen so to make the average dissipation finite for $L\rightarrow\infty$) takes the form (Appendix~\ref{app:Ising}):
 \begin{align}\label{eq:IsingTL}
 K^{\text{diss}}_{\text{resc}}(\lambda) = -\frac{\beta^2 }{2N}  \int_\gamma \dot h^2& \Big( \lambda\Tr\sqrbra{\partial_h H\, \J_{\pi_t} [\Delta_{\pi_t}\partial_h H]} \nonumber\\ 
 &- \frac{\lambda^2}{L} \Tr\sqrbra{\partial_h H\, \operatorS_{\pi_t} [\Delta_{\pi_t}\partial_h H]}\Big),
 \end{align}
 up to corrections of order $\bigo{1/L^2}$. We can recognise from this formula the average dissipation and the fluctuations defined in Eq.~\eqref{eq:avWork} and Eq.~\eqref{eq:fluctuations}. We also recognise the CFG of a Gaussian distribution, with  mean independent of $L$ and variance $\propto L^{-1/2}$.
 If we now take the thermodynamic limit $L\rightarrow \infty$, we see that the fluctuations term goes to zero, leaving us with a $\delta$-distribution centred in $\average{w_{\text{diss}}}{}$. This result is a consequence of the equivalence between the canonical and the microcanonical ensemble in the thermodynamic limit~\cite{ruelleStatisticalMechanicsRigorous1999}.

 \section{Consequences of the slow driving regime}
 As it was pointed out in the introduction, the characterisation of the entropy production for processes arbitrary out of equilibrium is in general difficult, and it is foreseeable that any universal result won't be sufficient to constrain the statistical properties of the dissipation. On the other hand, we have seen that in the linear response regime the probability distribution of the entropy production can be characterised with relative ease, making reference only to the instantaneous thermal state and to the driving speed. The simplicity of the expressions obtained can be partially ascribed to the decoupling of different channels of the entropy production (section~\ref{sec:asymmetry}), and to the time reversal symmetry which arises for slow driving protocols (section~\ref{sec:evansearles}). These effects and their consequences will be described in the following sections.
 
 \subsection{Channels of entropy production: non Gaussianity and asymmetry}\label{sec:asymmetry}
 
 The second law of thermodynamics, together with Eq.~\eqref{eq:equivalence}, implies that the entropy production can be interpreted as the deterioration of the ability of a system to perform work. This motivates the study of thermodynamics as a resource theory \cite{Lostaglio2019}, where in particular one interprets a system out of equilibrium as a resource that can be expended to generate work. 
 This intuition was first investigated by Lieb and Yngvason in~\cite{liebPhysicsMathematicsSecond1999}, where the uniqueness of the entropy functional was proved for equilibrium states.

 \begin{figure}
	\centering
	\includegraphics[width=\linewidth]{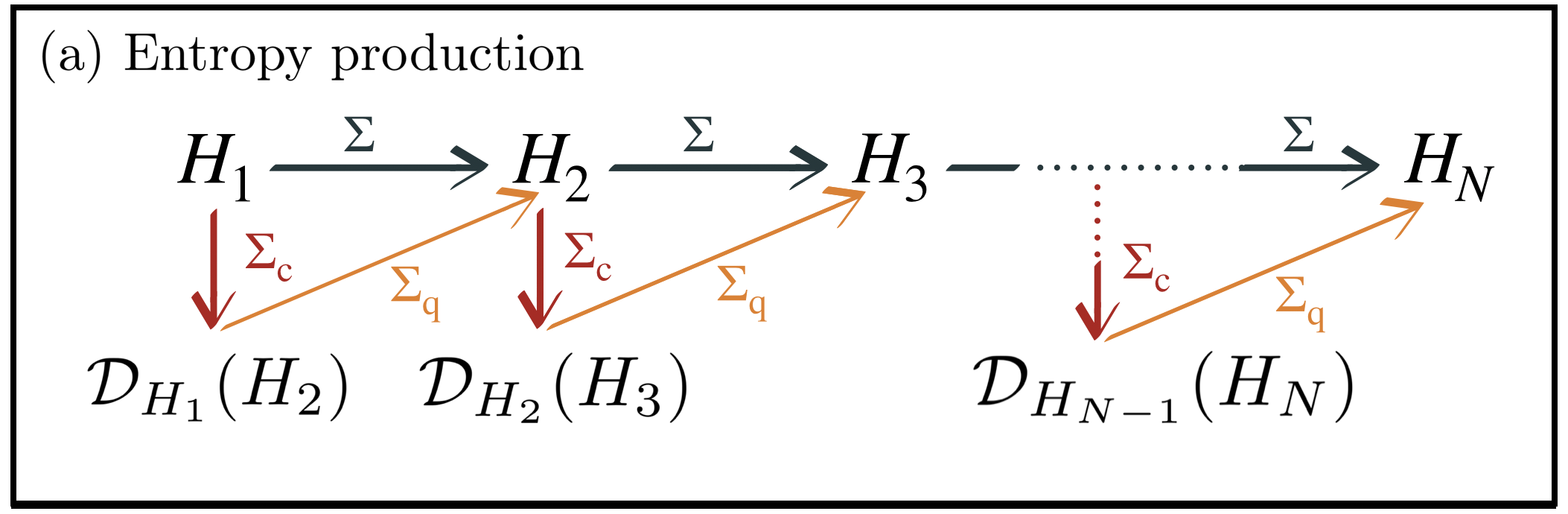}\,
	\caption{In the quasistatic regime the entropy production $\Sigma$ splits in two additive contribution, one which accounts only for the dissipation associated with the athermality created at each step ($\Sigma_{\rm c}$, red line), and a part coming solely from a change in the energy basis ($\Sigma_{\rm q}$, orange line).}
	\label{fig:fig6}
\end{figure}

 In the context of the resource theory of thermodynamics, a single law is not sufficient to characterise the irreversibility of a thermodynamic process. In fact, it has been shown in~\cite{brandaoSecondLawsQuantum2015b} that a necessary condition for the transition {\ms between two diagonal states} $\varrho\rightarrow\sigma$ to happen through thermal operations is that the following family of second laws is satisfied:
 \begin{align}
 \forall \lambda\geq 0 \hspace{0.5cm} S_\lambda (\varrho || \pi(H)))\geq S_\lambda (\sigma || \pi(H)).\label{eq:athermalityconstraint}
 \end{align}
 The R\'enyi divergences measure how statistically different a state is from the {\ms Gibbs ensemble, considered as the zero of the theory since no work can be extracted from it}. The corresponding resource is called \emph{athermality} and it is progressively lost under thermodynamic evolution.
 
 Moreover, if the state presents off-diagonal terms in the energy eigenbasis an additional family of constraints have to be satisfied by any thermal operations~\cite{lostaglioDescriptionQuantumCoherence2015, lostaglioQuantumCoherenceTimeTranslation2015,PhysRevLett.115.210403}:
 \begin{align}
 \forall \lambda\geq 0 \hspace{0.5cm} S_\lambda (\varrho || \mathcal{D}_H (\varrho))\geq S_\lambda (\sigma || \mathcal{D}_H (\sigma)),\label{eq:asymmetrycostraints}
 \end{align}
 where $\mathcal{D}_H$ is the dephasing map in the $H$ eigenbasis{\ms , so that the R\'enyi divergences quantify how different the state is from a diagonal one}. The corresponding resource is called \emph{asymmetry}, as it is connected with the breaking of the time translation symmetry of the state.  From this set of operations it appears that one cannot increase the coherence between different energy levels with thermal operations alone. Coherence does not come for free, as it could have been guessed from the fact that it can be converted into work in the presence of a coherent bath~\cite{abergCatalyticCoherence2014}.
 
 We can now pass to analyse the work extraction protocol in this framework. Focussing on a step of the process only, we see that the system starts in a thermal state (which is automatically time symmetric). The quench in the Hamiltonian provides work to the system, which brings the state out of equilibrium and, at the same time, breaks its symmetry {\ms by introducing off-diagonal terms}. Hence, part of the work is converted in athermality, part in asymmetry. Right after the quench a perfectly thermalising operation is applied, which dissipates both resources, bringing the system back to a symmetric equilibrium state.
 
 In general the entropy production $\Sigma$ can be split in a classical contribution $\Sigma_{\rm c}$ and a purely coherent one $\Sigma_{\rm q}$ only on average~\cite{lostaglioDescriptionQuantumCoherence2015, lostaglioQuantumCoherenceTimeTranslation2015}. In this case, the second law is measured by the relative entropy, which corresponds to the limit $\lambda\rightarrow1$ of the $\lambda$-R\'enyi divergence in Eq.~(\ref{eq:athermalityconstraint},\ref{eq:asymmetrycostraints}), where in Eq.~\eqref{eq:athermalityconstraint} $\varrho$ is substituted by $\mathcal{D}_H (\varrho)$ \cite{lostaglioDescriptionQuantumCoherence2015}. Notably this situation changes in the quasistatic regime, and the two channels of entropy production decouple at all levels of statistics. In fact, the $y$-covariance, which arises in the expansion of the R\'enyi divergences Eq.~\eqref{eq:Renyientropy}, can be split in a dephased and a coherent contribution:
 \begin{align}\label{eq:covysplitting}
 \text{cov}_t^y(\dot H_t,\dot H_t) = \text{cov}_t^y(\mathcal{D}_t(\dot H_t),\mathcal{D}_t(\dot H_t))+ \text{cov}_t^y(\dot H_t^{\text{c}},\dot H_t^{\text{c}}),
 \end{align}
 where we introduced the bookkeeping notation $\dot H_t^{\text{c}} := \dot H_t-\mathcal{D}_t(\dot H_t)$. This also leads to naturally define the dephased and coherent CGF in the quasistatic regime as:
 \begin{align}\label{eq:splitCGF}
 K^{\text{diss}}(\lambda) = K^{\text{diss}}_{\text{deph}}(\lambda) +K^{\text{diss}}_{\text{asymm}}(\lambda).
 \end{align}
 In Appendix~\ref{app:asymmetry} we show how each contribution can be linked with the expansion of terms akin to Eq.~(\ref{eq:athermalityconstraint}, \ref{eq:asymmetrycostraints}). We can then interpret Eq.~\eqref{eq:splitCGF} as the fact that in the slow driving regime $\Sigma_{\rm c}$ and $\Sigma_{\rm q}$ become independent random variables. This situation is exemplified in Fig.~\ref{fig:fig6}.
 
 In this context, since $\text{cov}_t^y(\mathcal{D}_t(\dot H_t),\mathcal{D}_t(\dot H_t)) = \text{Var}[\mathcal{D}_t(\dot H_t)]$, we can see that the diagonal family of second laws in Eq.~\eqref{eq:athermalityconstraint} collapse into a single constraint, in the spirit of~\cite{liebPhysicsMathematicsSecond1999}. This result can be thought as a consequence of the adiabatic theorem: in the slow driving regime, in fact, the most probable transitions are of the form $\ket{E^{(k)}_i}\rightarrow\ket{ E^{(k)}_{i+1}}$, so that the work output becomes quasi-continuous (this discussion should be compared with the one in section~\ref{sec:CGFcohe}). Since the only difference between a diagonal quantum state and a classical one is the discreteness of the spectrum, it can be intuitively understood that when this discreteness is smeared out one regains the classical result.

 The CGF corresponding to the degradation of athermality takes the form:
 \begin{align}\label{eq:CGFdeph}
 K^{\text{diss}}_{\text{deph}}(\lambda) =  &\frac{\beta^2(\lambda^2-\lambda)}{2N}\int_\gamma\text{Var}_t [\mathcal{D}_t(\dot H_t)].
 \end{align}
 On the other hand, the CGF for the dissipation of coherence is given by: 
 \begin{align}\label{eq:CGFasymm}
 K^{\text{diss}}_{\text{asymm}} = & 	\frac{\beta^2(\lambda^2-\lambda)}{2N}\int_\gamma\text{Var}_t [\dot H_t^{\text{c}}]+\nonumber\\
 &+ \frac{\beta^2}{2N} \int_\gamma\int_0^\lambda {\rm d}x \int_x^{1-x}{\rm d}y\,  I^y(\pi_t, \dot H_t^{\text{c}}).
 \end{align}

 The splitting in Eq.~\eqref{eq:splitCGF} implies that all the cumulants decompose as $\kappa_{w}^{(n)} = \kappa_{\text{deph}}^{(n)}  + \kappa_{\text{asymm}}^{(n)} $, which, in particular, means:
 \begin{align}
 \average{w_{\text{diss}}}{} \geq \langle w_{\text{diss}}^{\text{deph}}\rangle .
 \end{align} 
 This result confirms the intuition that the additional channel of entropy production provided by the dissipation of asymmetry raises the average dissipation. Indeed, thanks to the fact that all the cumulants derived from Eq.~\eqref{eq:CGFasymm} are positive (section~\ref{sec:CGF}), it also holds that $\kappa_{w}^{(n)} \geq \kappa_{\text{deph}}^{(n)} $, for any $n$.

 More interestingly, the two terms in Eq.~\eqref{eq:splitCGF} independently satisfy the Jarzynski equality, as it can be verified by evaluating Eq.~\eqref{eq:CGFdeph} and Eq.~\eqref{eq:CGFasymm} for $\lambda=1$, in analogy with what is discussed in section~\ref{sec:CGF}. This means that both $K^{\text{diss}}_{\text{deph}}(\lambda)$ and  $K^{\text{diss}}_{\text{asymm}}(\lambda)$ can be considered as arising from two independent thermal processes. The probability distribution of the total dissipated work will then be given by the convolution between the Gaussian coming from the dissipation of athermality resources with the probability distribution coming from the degradation of asymmetry. This observation also implies that the two extremal regimes studied in section~\ref{sec:propDensity} can be considered as the cornerstone of any quasistatic thermodynamic process.
 
 The FDR are also satisfied independently by the two terms. We can then write the inequality:
 \begin{align}\label{eq:tradeofTUR}
 \frac{\langle w_{\text{diss}}^{\text{deph}}\rangle }{ \sigma^2_{\text{deph}}} = \frac{\beta}{2}\geq \frac{\langle w_{\text{diss}}^{\text{asymm}}\rangle}{\sigma^2_{\text{asymm}}} = \frac{\beta}{2} -\frac{ \int_\gamma \int_0^1 {\rm d}y \, I^y(\pi_t, \dot H^{\text{c}}_t) }{\frac{2}{\beta}\int_\gamma\text{Var}_t [\dot H^{\text{c}}_t]}  .
 \end{align}
 This equation can be read off as the fact that a coherent process dissipates less for the same unit of fluctuation. {\ms Moreover}, thanks to Jarzynski equality, the presence of a negative correction allows for positive higher cumulants. Therefore, the second law of thermodynamics manifests itself not as a higher dissipation on average, but rather with a fat tail for positive dissipation, that is a tendency of the system to {\ms fluctuate above} \textcolor{ms}{$\average{w_{\rm diss}}{}$}. This means that the entropy production associated with the degradation of asymmetry is inherently different than the one associated ot the dissipation of athermality, {\ms having bigger fluctuations, arising partly from the thermal disorder, partly from the genuinely quantum uncertainty in the state}. Notice that the disordered nature of the work output from systems coupled to a coherent bath was already noticed in~\cite{abergCatalyticCoherence2014}.

 \subsection{Time reversal symmetry: the Evans-Searles Fluctuation theorem}\label{sec:evansearles}
 
 As it was pointed out in section~\ref{sec:CGF}, from the relation $K^{\text{diss}}(\lambda) = K^{\text{diss}}(1-\lambda)$ we can deduce the fluctuation relation $p(w_{\text{diss}})=p(-w_{\text{diss}})e^{\beta w_{\text{diss}}}$, which {\ms is }typically referred to as the \emph{Evans-Searles} relation \cite{Crooks,Evans2002}. {\ms It} places a considerable constraint on the fluctuations in entropy production, with negative values exponentially suppressed. In fact, one may derive as a direct consequence of  this fluctuation theorem the following:
 \begin{align}
 p(w_{\text{diss}}\leq 0)\leq \frac{1}{2}\norbra{1-\frac{\langle \Sigma \rangle_{+}}{s_{+}}\bigg(e^{s_{+}/\langle \Sigma \rangle_{+}}-1\bigg)},
 \end{align}
 where $\langle . \rangle_+$ denotes an average over the positive values of entropy production $\Sigma=\beta w_{\text{diss}}$ and $s_+=\langle \Sigma^2 \rangle_{+} $ \cite{Merhav2010}. One may also obtain a lower bound on the likelihood of negative dissipation:
 \begin{align}
 p(w_{\text{diss}}\leq 0)\geq \frac{1+\psi(\langle \Sigma \rangle_{+})-\langle \Sigma \rangle_{+}}{e^{\psi(\langle \Sigma \rangle_{+})}+\psi(\langle \Sigma \rangle_{+})+1},
 \end{align}
 where $\psi(.)$ is the inverse of the function $\phi(z)=z/(1+e^{-z})$. These constraints are tighter than those imposed by applying general concentration bounds that do not make use of the additional information provided by the fluctuation theorem. 
 
 The Evans-Searles relation should be compared with the weaker Crooks fluctuation theorem that relates the entropy production to a hypothetical reverse process:
 \begin{align}
 \frac{p(w_{\text{diss}})}{p_{\text{rev}}(-w_{\text{diss}})}=e^{\beta w_{\text{diss}}}.
 \end{align}
 Here $p_{\text{rev}}(-w_{\text{diss}})$ represents the probability induced by the time reversed driving $H^{\text{rev}}_i=H_{1-t}$. Comparing the two fluctuation relations we see that the work statistics for the forwards and reverse protocols are indistinguishable:
 \begin{align}
 p_{\text{rev}}(w_{\text{diss}})=p(w_{\text{diss}}).
 \end{align}
 This result has a straightforward interpretation: in fact, we are expanding the dissipated work around its minimum, i.e., the manifold of equilibrium states. This implies that the second laws in the CGF (Eq.~\eqref{eq:CGFdissdef}) become quadratic forms, so that the system does not differentiate between the driving $\dot H_t$ and its reverse $\dot{H}^{\text{rev}}_t=-\dot H_{1-t}$.

 If we again consider the splitting of the entropy production $\Sigma=\Sigma_c+\Sigma_q$, we see from the symmetry of the $y$-covariances in Eq.~\eqref{eq:covysplitting} that the dephased and coherent contributions actually satisfy the Evans-Searles fluctuation theorem independently:
 \begin{align}
 \frac{p(\Sigma_{x})}{p(-\Sigma_{x})} = e^{\Sigma_{x}}, \ \ x=\lbrace c,q \rbrace.
 \end{align}
 As these terms are independent random variables, we also have
 \begin{align}\label{eq:evansearles}
 \frac{p(\Sigma_q,\Sigma_c)}{p(-\Sigma_q,-\Sigma_c)} = e^{\Sigma},
 \end{align}
 The fact that Eq.~\eqref{eq:evansearles} holds true in this regime provides an interesting connection to the so-called thermodynamic uncertainty relation (TUR) \cite{Barato,Pietzonka,Pietzonka2018}.  The TUR imposes a trade-off between the noise-to-signal ratio $\mathcal{S}(x)=\Delta x /\langle x\rangle$ of a given time-integrated current and the average entropy production $\langle \Sigma \rangle$ along a process. For small average currents $x\ll1$, it has been shown that any distribution of the form Eq.~\eqref{eq:evansearles} leads to the TUR \cite{Guarnieri2019}:
 \begin{align}
 \langle \Sigma \rangle \ \mathcal{S}^2(x)  \geq 2,
 \end{align}
 The TUR implies that reducing the noise-to-signal ratio associated with a given current comes at a price of increased entropy production. This is consistent with the trade-off relation Eq.~\eqref{eq:tradeofTUR} for currents $x=\lbrace w_{\text{diss}}^{\text{deph}},w_{\text{diss}}^{\text{asymm}} \rbrace$. Notably the TUR is saturated by a Gaussian distribution, which is satisfied for the dephased work $x=w_{\text{diss}}^{\text{deph}}$. On the other hand, any non-zero quantum contribution $x=w_{\text{diss}}^{\text{asymm}}$ cannot saturate the TUR, as clearly seen in Eq.~\eqref{eq:tradeofTUR}. This highlights the fact that coherence provides a fundamental limitation to the trade-off between reversibility (small $\langle \Sigma \rangle$) and constancy (small $\mathcal{S}$).

 \section{Entropy production for continuous protocols}\label{sec:contProtocol}
 Historically, the motivation to analyse step equilibration processes was to construct a framework in which effects arising from the slow driving regime could be isolated from the particular details of the relaxing dynamics~\cite{nultonQuasistaticProcessesStep1985b}. In fact, a process constituted by a discrete sequence of quenches and thermalisations steps can be thought as the simplification of a continuous open-system process in which the equilibration {\ms dynamics} is trivial. We will here motivate this claim and connect the CGF for a continuous protocol with the one we obtained for a discrete one.
 
 In particular, we focus on the regime in which the state of the system can be approximated at all times as ${\varrho_t = \pi_t +\delta\varrho_t}$, where $\delta\varrho_t$ is of order $\bigo{1/\tau}$, and $\tau$ is the duration of the protocol. One can expect this kind of behaviour whenever the dynamics is relaxing and the parameters are changed in a sufficiently slow manner.
 
 The cumulant generating function for a continuous process which initially starts at equilibrium is {\ms exactly} given by~\cite{weiRelationsDissipatedWork2017, guarnieriQuantumWorkStatistics2019}:
 \begin{align}
 K^{-\beta W}(\lambda) &=   \log\Tr\sqrbra{ e^{-\beta \lambda H_{\tau}} U_{\tau} e^{\beta \lambda H_{0}} \pi_0 U^{\dagger}_{\tau}}  \nonumber\\
 &=-\beta \lambda \Delta F + \log\Tr\sqrbra{ \pi_\tau^{\lambda} U_{\tau}  \pi_0^{1-\lambda}U^{\dagger}_{\tau}}.
 \end{align}
 For what follows, we denote the time evolved state as ${\varrho_\tau :=U_{\tau}  \pi_0U^{\dagger}_{\tau}}$. Using this notation, we can give a compact expression of the dissipative CGF, which takes the form:
 \begin{align}\label{eq:CGFcontinuous}
 K^{\text{diss}}(\lambda) =(\lambda - 1) S_\lambda(\pi_\tau || \varrho_\tau).
 \end{align}
 This equation shows the close relation between the cumulants of the dissipated work and the statistical difference of the evolved state $\varrho_\tau$ from the thermal state~\cite{weiRelationsDissipatedWork2017}. In the case of a quench the state is given by $\varrho_\tau = \pi_0$, and Eq.~\eqref{eq:CGFcontinuous} reduces to the expression for a single step of {\ms the discrete process in} Eq.~\eqref{eq:CGFdissdef}. In this sense, in a discrete process the dynamics, which can be assumed to be given by a generic thermalising map, is not taken into account by the CGF in any other way than in the information about the initial conditions $\varrho_i \equiv \pi_i$.
 
 The dependency of $\varrho_\tau$ on the particular protocol is implicit in this expression. Knowing that the dissipation is path dependent, though, it is useful to highlight this dependency by rewriting Eq.~\eqref{eq:CGFcontinuous} as:
 \begin{align}\label{eq:CGFcontinuous1}
 K^{\text{diss}}(\lambda)  = \int_0^\tau \dt \norbra{\frac{\de}{\dt} \log\Tr\sqrbra{ \pi_t^{\lambda} \varrho_t^{1-\lambda}}}.
 \end{align}
 We show in Appendix~\ref{app:contEv} that this takes the form:
 \begin{align}
 &K^{\text{diss}}(\lambda) = 
 -\beta \int_\gamma \int_0^\lambda {\rm d}x  \frac{\Tr\sqrbra{ \pi_t^{x}\Delta_t \dot H  \varrho_t^{1-x}}}{\Tr\sqrbra{ \pi_t^{\lambda} \varrho_t^{1-\lambda}}} \; +\nonumber\\
 &-\beta \int_\gamma \int_0^\lambda {\rm d}x\int_0^x\de y \frac{\Tr\sqrbra{ \pi_t^{y}\Delta_t \dot H \pi_t^{x-y}(\log \pi_t - \log \varrho_t ) \varrho_t^{1-x}}}{\Tr\sqrbra{ \pi_t^{\lambda} \varrho_t^{1-\lambda}}}\label{eq:CGFcontinuous2}
 \end{align}
 where we can already see that both terms {\ms will be} of order $\bigo{1/\tau}$ in the quasistatic limit. Notice{\ms, however,} that this expression is exact and it directly connects the cumulant generating function with the particular trajectory in the parameter space taken during the protocol.
 
 We can now pass to the slow driving limit of Eq.~\eqref{eq:CGFcontinuous2}. This means that we assume the state to be given by the approximation $\varrho_t = \pi_t +\delta\varrho_t$, and that we can neglect terms of order $\bigo{\delta \varrho^2}$. Then, Eq.~\eqref{eq:CGFcontinuous2} can be simplified to the form (Appendix~\ref{app:contEv}):
 \begin{align}
 K^{\text{diss}}(\lambda) =  -\beta \int_\gamma  \int_0^\lambda {\rm d}x\int_x^{1-x} \de y \,\text{cov}_t^y(\dot H_t, \J_t^{-1}[\delta \varrho_t ]).\label{eq:CGFquasistaticcontinuous}
 \end{align}
 Note that in the above derivation, the adiabatic term $\pi_t$ need not be thermal. 
 
 Let us now imagine that the system is embedded in a larger Hilbert space $\mathcal{H}\to\mathcal{H}\otimes \mathcal{H}_R$ through coupling to a reservoir at inverse temperature $\beta$, while only the local system Hamiltonian is varied in time. We restrict {\ms our} attention to situations where the coupling is weak enough so that we may approximate the global state as a tensor product $\rho_t\otimes \pi_R\simeq (\pi_t +\delta \rho_t)\otimes \pi_R$, where $\pi_R$ is a fixed equilibrium state of the reservoir. The $y$-covariance is then invariant under partial trace over the reservoir, and we end up with the expression in Eq.~\eqref{eq:CGFquasistaticcontinuous} for the open system. We note that while neglecting the influence of correlations on the $y$-covariance is a rather strong assumption, this approximation is sufficient for illustrating the connection between the discrete protocols adopted in the previous sections and continuous-time dynamics. We leave a more rigorous treatment of the CGF for general open quantum systems for future work.

 From here we can consider Lindblad evolutions; if the reduced dynamics of the system is given by a time dependent relaxing Lindbladian $\dot{\rho}_t=\lind_t(\rho_t)$, then the correction term in the slow driving regime takes the form $\delta\varrho_t\equiv -\beta\lind_t^+[\J_t[\Delta_t \dot H]]$, where the cross denotes the Drazin inverse~\cite{mandalAnalysisSlowTransitions2016,cavinaSlowDynamicsThermodynamics2017, scandiThermodynamicLengthOpen2019}. Under the assumption of quantum detailed balance \cite{Fagnola2010a}, we show in Appendix~\ref{app:contEv} that the CGF in Eq.~\eqref{eq:CGFquasistaticcontinuous} becomes
 \begin{align}\label{eq:CGFquasistaticcontinuous2}
 K^{\text{diss}}(\lambda) =   -\beta^2 \int_\gamma  \int_0^\lambda {\rm d}x\int_x^{1-x} \de y \ \tau_{eq}^y(t) \ \text{cov}_t^y(\dot H_t ,\dot H_t),
 \end{align}
 where 
 \begin{align}
 \tau_{eq}^y(t):= \int^\infty_0 d\nu \ \frac{\text{cov}_t^y(\dot H_t(\nu) ,\dot H_t(0))}{\text{cov}_t^y(\dot H_t (0) ,\dot H_t(0))}\geq 0,
 \end{align}
 represents a quantum generalisation of the \textit{integral relaxation timescale} introduced in \cite{sivakThermodynamicMetricsOptimal2012}. Here we denote the operator $\dot H_t(\nu)=e^{\nu \lind_t^\dagger}[\dot H_t]$ evolved in the Heisenberg picture. This timescale quantifies the time over which the fluctuations in power, as quantified by the $y$-covariance, decay to their equilibrium values. As an example, for a simple Lindbladian of the form $\lind_t(\rho_t)=(\pi_t-\rho_t)/\Gamma_{eq}(t)$, the integral relaxation time reduces to a single timescale $\tau_{eq}^y(t)=\Gamma_{eq}(t)$.
 
 From this formula the expression for the average dissipated work and the work fluctuations presented in~\cite{millerWorkFluctuationsSlow2019} can be obtained in a straightforward manner, extending the work therein to arbitrary cumulants. Furthermore, from Eq.~\eqref{eq:CGFquasistaticcontinuous2} we now find a connection between the continuous protocol approach and the discrete protocol described by the CGF in Eq.~\eqref{eq:CGFslowdriving}. If we identify the ratio between the integral relaxation time and total time with a uniform step size $N$, namely $\tau_{eq}^y=1/2N$, we find that the two protocols are described by the same statistics. In essence, we may therefore view the discrete protocol as indistinguishable from that of a continuous process $\gamma$ described by the trivial relaxing Lindbladian $\lind_t(\rho_t)=2N(\pi_t-\rho_t)$.

 \section{Discussion, conclusions and outlook}
 
 In this paper, we have characterised the fluctuations of work, dissipation, and entropy production in quantum quasi-isothermal processes, where the system of interest stays close to equilibrium along the thermodynamic transformation. 
 In this regime, all cumulants of the work distribution decouple into a classical (non-coherent) and a quantum (coherent) contribution, hence extending previous considerations for average quantities \cite{Janzing2006,lostaglioDescriptionQuantumCoherence2015,Santos2019,Francica2019}. In fact, all cumulants beyond the second one have a purely quantum origin, and they can be obtained by differentiating the quantum skew information (Eq. \eqref{eq:highercumulants}), a measure of quantum uncertainities  \cite{Hansen2008,Frerot2016}. Such quantum  fluctuations  lead to  positive skewness and excess kurtosis, witnessing a tendency of the system for extreme deviations above the average dissipation. These results shed new light on  our understanding of quantum features in the work distribution \cite{Allahverdyan2014,baumer2018fluctuating,Talkner2016,Solinas2017,Miller2018,Solinas2015,Lostaglio2018,Xu2018,PatrickPotts2019,Wu2019,mingo2018superpositions}. 
 
 It is also important to comment on the deviations from Gaussianity in the tails of the work distribution that have been observed for classical processes \cite{hoppenauWorkDistributionQuasistatic2013a}. These deviations can be understood here by the impossibility of expanding in terms of $1/N$ work values of order~$\sim \bigo{N}$ (in a single step); these contributions will certainly appear in unbounded spectra, albeit being extremely unlikely. This problem  does not appear for quantum systems with bounded spectra (hence there is always an $N$ sufficiently large as compared to all possible work values during a single step).   In this case 
 we can build  a strong relation between non Gaussianity and quantumness: a work distribution becomes non Gaussian  in the quasi-isothermal regime if and only if quantum fluctuations appear. For unbounded spectra, this statement remains true for typical values of the work, while on the tails the ``only if'' ceases to be valid. 
 
 The thermodynamic quantum signatures reported in this article (breakdown of the classical fluctuation-dissipation relation Eq. \eqref{eq:FDRclassical}, non Gaussianity of the distribution, with positive skewness and kurtosis)  appear to be measurable with state-of-the-art technologies. Indeed, it is sufficient to have an experimental platform where the following three operations can be implemented: (i) projective measurement of the energy for non commuting Hamiltonians of the system $\curbra{H_i}$; (ii) quenching of the system Hamiltonian from $H_i$ to $H_{i+1}$, and (iii) thermalisation of the system for the Hamiltonians $H_{i+1}$. Quantum signatures will then be observed whenever $[H_i, H_{i+1}] \neq 0$ for at least some time step $i$. Ion traps provide excellent controllability and have previously been used to measure quantum work distributions~\cite{An2014,Lindenfels2019}, 
 and similarly for NMR systems~\cite{Tiago2014}. 
 Other promising platforms for measuring such quantum signatures in the work distribution are quantum dots and superconducting qubits~\cite{Pekola2015,Cottet2018,Naghiloo2018}.   

 From the observation that the entropy production distribution for a particular process equals the one for its time reverse, we have also proven  the Evans-Searles relation \cite{Evans2002}, a stronger form of Crooks fluctuation theorem \cite{Crooks}. This relation enables us to set strong constrains on the fluctuations in entropy production,  with negative values exponentially supressed.  This result has also enabled us to make a connection with (quantum) Thermodynamic Uncertainty Relations \cite{Barato,Pietzonka,Pietzonka2018,Guarnieri2019}, which in this context set a tradeoff between fluctuations and average entropy production. We have then shown that quantum coherence prevents the saturation of the TUR in the slow driving regime. 
 
 Our results have also implications for a seemingly unrelated question, namely the interconvertibility of states within the resource theory of thermodynamics \cite{Brandao2013,Lostaglio2019}. Indeed, as it is also observed in \cite{guarnieriQuantumWorkStatistics2019}, there is a close connection  between work statistics and the second laws of thermodynamics of \cite{brandaoSecondLawsQuantum2015b}, both being expressed through R\'enyi divergences.  The expansions of R\'enyi divergences close to thermal equilibrium developed here imply that the continuous family of second laws of \cite{brandaoSecondLawsQuantum2015b} for the interconvertibility of diagonal states reduces to a single one (the second law of thermodynamics) close to thermal equilibrium. This is in spirit similar to the well-known fact that in the many-copies limit such second laws converge to a single one \cite{Brandao2013}, but this result holds at the single-copy level. On the other hand, the additional constraints from quantum coherence~\cite{lostaglioDescriptionQuantumCoherence2015,lostaglioQuantumCoherenceTimeTranslation2015} remain untouched close to thermal equilibrium.   In other words, the fact that the work distribution becomes Gaussian close to equilibrium for diagonal states translates into the many laws of~\cite{brandaoSecondLawsQuantum2015b} reducing to a single one; whereas the fact that the work distribution is non-trivial for quantum coherent states  implies that the asymmetry restrictions of~\cite{lostaglioDescriptionQuantumCoherence2015,lostaglioQuantumCoherenceTimeTranslation2015} do not simplify close to equilibrium. We believe our results might find other implications in the resource theory of thermodynamics; for example, from the expansions of the relative entropy and the relative entropy of variance it appears that  the  conversions of thermal  resources close to equilibrium will  always be resonant, in the sense developed in \cite{Korzekwa2019}. 
 
 We have also discussed how the quantum signatures in the work distribution behave for macroscopic systems. The first important observation is that the quantum correction $\mathcal{Q}$ in the quantum fluctuation dissipation relation (FDR), $\frac{\beta}{2}\sigma_w^2=\average{w_{\rm diss}}{}+\mathcal{Q}$, is extensive (and so are the dissipation and the variance) which means that no matter how large is the system under observation, a correction to the classical FDR in Eq.~\eqref{eq:FDRclassical} will appear  for protocols where $[\dot{H}_t,H_t] \neq 0$, hence causing the work distribution to become non Gaussian. Note that non-commutativity  is in fact ubiquitous in many-body quantum systems, where usually the (local) control does not commute with the global Hamiltonian.  We also discussed  how this observation relates with the central limit theorem, which implies that the rescaled work distribution  will converge to a Gaussian, at least for non-interacting or locally interacting many-body systems away from a phase transition  (\textcolor{ms}{note that there is no contradiction with our considerations: the full work distribution remains non-Gaussian, and only the rescaled version becomes Gaussian}). 
 For such a rescaled distribution, the quantumness in the distribution is encoded in a larger variance of the distribution due to the presence of $\mathcal{Q}$.  We have illustrated these considerations in a  Ising chain in a driven transverse field, where we have computed the average and variance of the quantum work distribution. These considerations contribute to recent efforts to characterise the work distribution  of many-body systems \cite{Silva2008,Dorner2012,Gambassi2012,Fusco2014,Zongping2014,Goold2018,Wang2018,zawadzki2019work,arrais2019work}. 
 
 While most of our results have been derived through a discrete model of quasi-isothermal processes, we have shown in Sec. \ref{sec:contProtocol} that our considerations can be naturally extended to more complex continuous dynamics. 
 It remains as an interesting future question to derive these same results by means of a quantum jump approach, or quantum trajectories, given a Lindblad master equation \cite{Horowitz2013a,Manzano2018}, a direction that we are currently exploring. 
 Another interesting complementary  question is to derive similar slow-driving expansions through linear response theory for higher cumulants of the work distribution, hence extending previous results for average dissipation \cite{Campisi2012geometric,Bonana2014,acconcia2015shortcuts,Ludovico2016} (see also \cite{suomelaMomentsWorkTwopoint2014a} for a discussion of the work moments beyond weak coupling).  In all such extensions, we expect that the results reported for the discrete case should be recovered in the simplest model of an exponential relaxation to equilibrium with a well-defined time-scale, as argued in Sec. \ref{sec:contProtocol}.


 \emph{Acknowledgements.} We thank M. Lostaglio and J. Eisert for the useful comments. This project has received funding from the European Union’s Horizon 2020 research and innovation programme under the Marie Skłodowska-Curie grant agreement No 713729, and from Spanish MINECO (QIBEQI FIS2016-80773-P, Severo Ochoa SEV-2015-0522), Fundacio Cellex, Generalitat de Catalunya (SGR 1381 and CERCA Programme). J.A. acknowledges support from EPSRC (grant EP/R045577/1) and the Royal Society.

 \bibliographystyle{apsrev4-1}

 \bibliography{qwf.bib}

\begin{thebibliography}{93}%
\makeatletter
\providecommand \@ifxundefined [1]{%
 \@ifx{#1\undefined}
}%
\providecommand \@ifnum [1]{%
 \ifnum #1\expandafter \@firstoftwo
 \else \expandafter \@secondoftwo
 \fi
}%
\providecommand \@ifx [1]{%
 \ifx #1\expandafter \@firstoftwo
 \else \expandafter \@secondoftwo
 \fi
}%
\providecommand \natexlab [1]{#1}%
\providecommand \enquote  [1]{``#1''}%
\providecommand \bibnamefont  [1]{#1}%
\providecommand \bibfnamefont [1]{#1}%
\providecommand \citenamefont [1]{#1}%
\providecommand \href@noop [0]{\@secondoftwo}%
\providecommand \href [0]{\begingroup \@sanitize@url \@href}%
\providecommand \@href[1]{\@@startlink{#1}\@@href}%
\providecommand \@@href[1]{\endgroup#1\@@endlink}%
\providecommand \@sanitize@url [0]{\catcode `\\12\catcode `\$12\catcode
  `\&12\catcode `\#12\catcode `\^12\catcode `\_12\catcode `\%12\relax}%
\providecommand \@@startlink[1]{}%
\providecommand \@@endlink[0]{}%
\providecommand \url  [0]{\begingroup\@sanitize@url \@url }%
\providecommand \@url [1]{\endgroup\@href {#1}{\urlprefix }}%
\providecommand \urlprefix  [0]{URL }%
\providecommand \Eprint [0]{\href }%
\providecommand \doibase [0]{http://dx.doi.org/}%
\providecommand \selectlanguage [0]{\@gobble}%
\providecommand \bibinfo  [0]{\@secondoftwo}%
\providecommand \bibfield  [0]{\@secondoftwo}%
\providecommand \translation [1]{[#1]}%
\providecommand \BibitemOpen [0]{}%
\providecommand \bibitemStop [0]{}%
\providecommand \bibitemNoStop [0]{.\EOS\space}%
\providecommand \EOS [0]{\spacefactor3000\relax}%
\providecommand \BibitemShut  [1]{\csname bibitem#1\endcsname}%
\let\auto@bib@innerbib\@empty
\bibitem [{\citenamefont {Jarzynski}(2011{\natexlab{a}})}]{Jarzynski2011}%
  \BibitemOpen
  \bibfield  {author} {\bibinfo {author} {\bibfnamefont {C.}~\bibnamefont
  {Jarzynski}},\ }\href {\doibase 10.1146/annurev-conmatphys-062910-140506}
  {\bibfield  {journal} {\bibinfo  {journal} {Annual Review of Condensed Matter
  Physics}\ }\textbf {\bibinfo {volume} {2}},\ \bibinfo {pages} {329} (\bibinfo
  {year} {2011}{\natexlab{a}})}\BibitemShut {NoStop}%
\bibitem [{\citenamefont {Seifert}(2012)}]{Seifert_2012}%
  \BibitemOpen
  \bibfield  {author} {\bibinfo {author} {\bibfnamefont {U.}~\bibnamefont
  {Seifert}},\ }\href {\doibase 10.1088/0034-4885/75/12/126001} {\bibfield
  {journal} {\bibinfo  {journal} {Reports on Progress in Physics}\ }\textbf
  {\bibinfo {volume} {75}},\ \bibinfo {pages} {126001} (\bibinfo {year}
  {2012})}\BibitemShut {NoStop}%
\bibitem [{\citenamefont {Esposito}\ \emph {et~al.}(2009)\citenamefont
  {Esposito}, \citenamefont {Harbola},\ and\ \citenamefont
  {Mukamel}}]{Esposito2009}%
  \BibitemOpen
  \bibfield  {author} {\bibinfo {author} {\bibfnamefont {M.}~\bibnamefont
  {Esposito}}, \bibinfo {author} {\bibfnamefont {U.}~\bibnamefont {Harbola}}, \
  and\ \bibinfo {author} {\bibfnamefont {S.}~\bibnamefont {Mukamel}},\ }\href
  {\doibase 10.1103/revmodphys.81.1665} {\bibfield  {journal} {\bibinfo
  {journal} {Reviews of Modern Physics}\ }\textbf {\bibinfo {volume} {81}},\
  \bibinfo {pages} {1665} (\bibinfo {year} {2009})}\BibitemShut {NoStop}%
\bibitem [{\citenamefont {Campisi}\ \emph {et~al.}(2011)\citenamefont
  {Campisi}, \citenamefont {H{\"{a}}nggi},\ and\ \citenamefont
  {Talkner}}]{Campisi2011c}%
  \BibitemOpen
  \bibfield  {author} {\bibinfo {author} {\bibfnamefont {M.}~\bibnamefont
  {Campisi}}, \bibinfo {author} {\bibfnamefont {P.}~\bibnamefont
  {H{\"{a}}nggi}}, \ and\ \bibinfo {author} {\bibfnamefont {P.}~\bibnamefont
  {Talkner}},\ }\href {\doibase 10.1103/RevModPhys.83.771} {\bibfield
  {journal} {\bibinfo  {journal} {Rev. Mod. Phys.}\ }\textbf {\bibinfo {volume}
  {83}},\ \bibinfo {pages} {771} (\bibinfo {year} {2011})}\BibitemShut
  {NoStop}%
\bibitem [{\citenamefont {Goold}\ \emph {et~al.}(2016)\citenamefont {Goold},
  \citenamefont {Huber}, \citenamefont {Riera}, \citenamefont {del Rio},\ and\
  \citenamefont {Skrzypczyk}}]{Goold_2016}%
  \BibitemOpen
  \bibfield  {author} {\bibinfo {author} {\bibfnamefont {J.}~\bibnamefont
  {Goold}}, \bibinfo {author} {\bibfnamefont {M.}~\bibnamefont {Huber}},
  \bibinfo {author} {\bibfnamefont {A.}~\bibnamefont {Riera}}, \bibinfo
  {author} {\bibfnamefont {L.}~\bibnamefont {del Rio}}, \ and\ \bibinfo
  {author} {\bibfnamefont {P.}~\bibnamefont {Skrzypczyk}},\ }\href {\doibase
  10.1088/1751-8113/49/14/143001} {\bibfield  {journal} {\bibinfo  {journal}
  {Journal of Physics A: Mathematical and Theoretical}\ }\textbf {\bibinfo
  {volume} {49}},\ \bibinfo {pages} {143001} (\bibinfo {year}
  {2016})}\BibitemShut {NoStop}%
\bibitem [{\citenamefont {Jarzynski}(1997)}]{Jarzynski1997d}%
  \BibitemOpen
  \bibfield  {author} {\bibinfo {author} {\bibfnamefont {C.}~\bibnamefont
  {Jarzynski}},\ }\href {\doibase 10.1103/PhysRevLett.78.2690} {\bibfield
  {journal} {\bibinfo  {journal} {Phys. Rev. Lett.}\ }\textbf {\bibinfo
  {volume} {78}},\ \bibinfo {pages} {2690} (\bibinfo {year}
  {1997})}\BibitemShut {NoStop}%
\bibitem [{\citenamefont {Crooks}(1999{\natexlab{a}})}]{Crooks1999}%
  \BibitemOpen
  \bibfield  {author} {\bibinfo {author} {\bibfnamefont {G.~E.}\ \bibnamefont
  {Crooks}},\ }\href {\doibase 10.1103/PhysRevE.60.2721} {\bibfield  {journal}
  {\bibinfo  {journal} {Phys. Rev. E}\ }\textbf {\bibinfo {volume} {60}},\
  \bibinfo {pages} {2721} (\bibinfo {year} {1999}{\natexlab{a}})}\BibitemShut
  {NoStop}%
\bibitem [{\citenamefont {H\"{a}nggi}\ and\ \citenamefont
  {Talkner}(2015)}]{Hnggi2015}%
  \BibitemOpen
  \bibfield  {author} {\bibinfo {author} {\bibfnamefont {P.}~\bibnamefont
  {H\"{a}nggi}}\ and\ \bibinfo {author} {\bibfnamefont {P.}~\bibnamefont
  {Talkner}},\ }\href {\doibase 10.1038/nphys3167} {\bibfield  {journal}
  {\bibinfo  {journal} {Nature Physics}\ }\textbf {\bibinfo {volume} {11}},\
  \bibinfo {pages} {108} (\bibinfo {year} {2015})}\BibitemShut {NoStop}%
\bibitem [{\citenamefont {Funo}\ \emph {et~al.}(2018)\citenamefont {Funo},
  \citenamefont {Ueda},\ and\ \citenamefont {Sagawa}}]{Funo2018}%
  \BibitemOpen
  \bibfield  {author} {\bibinfo {author} {\bibfnamefont {K.}~\bibnamefont
  {Funo}}, \bibinfo {author} {\bibfnamefont {M.}~\bibnamefont {Ueda}}, \ and\
  \bibinfo {author} {\bibfnamefont {T.}~\bibnamefont {Sagawa}},\ }in\ \href
  {\doibase 10.1007/978-3-319-99046-0_10} {\emph {\bibinfo {booktitle}
  {Fundamental Theories of Physics}}}\ (\bibinfo  {publisher} {Springer
  International Publishing},\ \bibinfo {year} {2018})\ pp.\ \bibinfo {pages}
  {249--273}\BibitemShut {NoStop}%
\bibitem [{\citenamefont {Nulton}\ \emph {et~al.}(1985)\citenamefont {Nulton},
  \citenamefont {Salamon}, \citenamefont {Andresen},\ and\ \citenamefont
  {Anmin}}]{nultonQuasistaticProcessesStep1985b}%
  \BibitemOpen
  \bibfield  {author} {\bibinfo {author} {\bibfnamefont {J.}~\bibnamefont
  {Nulton}}, \bibinfo {author} {\bibfnamefont {P.}~\bibnamefont {Salamon}},
  \bibinfo {author} {\bibfnamefont {B.}~\bibnamefont {Andresen}}, \ and\
  \bibinfo {author} {\bibfnamefont {Q.}~\bibnamefont {Anmin}},\ }\href
  {\doibase 10.1063/1.449774} {\bibfield  {journal} {\bibinfo  {journal} {The
  Journal of Chemical Physics}\ }\textbf {\bibinfo {volume} {83}},\ \bibinfo
  {pages} {334} (\bibinfo {year} {1985})}\BibitemShut {NoStop}%
\bibitem [{\citenamefont {Speck}\ and\ \citenamefont
  {Seifert}(2004)}]{speckDistributionWorkIsothermal2004}%
  \BibitemOpen
  \bibfield  {author} {\bibinfo {author} {\bibfnamefont {T.}~\bibnamefont
  {Speck}}\ and\ \bibinfo {author} {\bibfnamefont {U.}~\bibnamefont
  {Seifert}},\ }\href {\doibase 10.1103/PhysRevE.70.066112} {\bibfield
  {journal} {\bibinfo  {journal} {Physical Review E}\ }\textbf {\bibinfo
  {volume} {70}},\ \bibinfo {pages} {066112} (\bibinfo {year}
  {2004})}\BibitemShut {NoStop}%
\bibitem [{\citenamefont {Crooks}(2007)}]{Crooks2007}%
  \BibitemOpen
  \bibfield  {author} {\bibinfo {author} {\bibfnamefont {G.~E.}\ \bibnamefont
  {Crooks}},\ }\href {\doibase 10.1103/PhysRevLett.99.100602} {\bibfield
  {journal} {\bibinfo  {journal} {Phys. Rev. Lett.}\ }\textbf {\bibinfo
  {volume} {99}},\ \bibinfo {pages} {100602} (\bibinfo {year}
  {2007})}\BibitemShut {NoStop}%
\bibitem [{\citenamefont {Hoppenau}\ and\ \citenamefont
  {Engel}(2013)}]{hoppenauWorkDistributionQuasistatic2013a}%
  \BibitemOpen
  \bibfield  {author} {\bibinfo {author} {\bibfnamefont {J.}~\bibnamefont
  {Hoppenau}}\ and\ \bibinfo {author} {\bibfnamefont {A.}~\bibnamefont
  {Engel}},\ }\href {\doibase 10.1088/1742-5468/2013/06/P06004} {\bibfield
  {journal} {\bibinfo  {journal} {Journal of Statistical Mechanics: Theory and
  Experiment}\ }\textbf {\bibinfo {volume} {2013}},\ \bibinfo {pages} {P06004}
  (\bibinfo {year} {2013})}\BibitemShut {NoStop}%
\bibitem [{\citenamefont {Kwon}\ \emph {et~al.}(2013)\citenamefont {Kwon},
  \citenamefont {Noh},\ and\ \citenamefont {Park}}]{Kwon2013}%
  \BibitemOpen
  \bibfield  {author} {\bibinfo {author} {\bibfnamefont {C.}~\bibnamefont
  {Kwon}}, \bibinfo {author} {\bibfnamefont {J.~D.}\ \bibnamefont {Noh}}, \
  and\ \bibinfo {author} {\bibfnamefont {H.}~\bibnamefont {Park}},\ }\href
  {\doibase 10.1103/PhysRevE.88.062102} {\bibfield  {journal} {\bibinfo
  {journal} {Phys. Rev. E}\ }\textbf {\bibinfo {volume} {88}},\ \bibinfo
  {pages} {062102} (\bibinfo {year} {2013})}\BibitemShut {NoStop}%
\bibitem [{\citenamefont {Mandal}\ and\ \citenamefont
  {Jarzynski}(2016{\natexlab{a}})}]{Mandal2016a}%
  \BibitemOpen
  \bibfield  {author} {\bibinfo {author} {\bibfnamefont {D.}~\bibnamefont
  {Mandal}}\ and\ \bibinfo {author} {\bibfnamefont {C.}~\bibnamefont
  {Jarzynski}},\ }\href {\doibase 10.1088/1742-5468/2016/06/063204} {\bibfield
  {journal} {\bibinfo  {journal} {J. Stat. Mech.}\ }\textbf {\bibinfo {volume}
  {2016}},\ \bibinfo {pages} {063204} (\bibinfo {year}
  {2016}{\natexlab{a}})}\BibitemShut {NoStop}%
\bibitem [{\citenamefont {Miller}\ \emph {et~al.}(2019)\citenamefont {Miller},
  \citenamefont {Scandi}, \citenamefont {Anders},\ and\ \citenamefont
  {{Perarnau-Llobet}}}]{millerWorkFluctuationsSlow2019}%
  \BibitemOpen
  \bibfield  {author} {\bibinfo {author} {\bibfnamefont {H.~J.~D.}\
  \bibnamefont {Miller}}, \bibinfo {author} {\bibfnamefont {M.}~\bibnamefont
  {Scandi}}, \bibinfo {author} {\bibfnamefont {J.}~\bibnamefont {Anders}}, \
  and\ \bibinfo {author} {\bibfnamefont {M.}~\bibnamefont
  {{Perarnau-Llobet}}},\ }\href {\doibase 10.1103/PhysRevLett.123.230603}
  {\bibfield  {journal} {\bibinfo  {journal} {Physical Review Letters}\
  }\textbf {\bibinfo {volume} {123}},\ \bibinfo {pages} {230603} (\bibinfo
  {year} {2019})}\BibitemShut {NoStop}%
\bibitem [{\citenamefont {Anders}\ and\ \citenamefont
  {Giovannetti}(2013)}]{Anders2013}%
  \BibitemOpen
  \bibfield  {author} {\bibinfo {author} {\bibfnamefont {J.}~\bibnamefont
  {Anders}}\ and\ \bibinfo {author} {\bibfnamefont {V.}~\bibnamefont
  {Giovannetti}},\ }\href {\doibase 10.1088/1367-2630/15/3/033022} {\bibfield
  {journal} {\bibinfo  {journal} {New Journal of Physics}\ }\textbf {\bibinfo
  {volume} {15}},\ \bibinfo {pages} {033022} (\bibinfo {year}
  {2013})}\BibitemShut {NoStop}%
\bibitem [{\citenamefont {Gallego}\ \emph {et~al.}(2014)\citenamefont
  {Gallego}, \citenamefont {Riera},\ and\ \citenamefont
  {Eisert}}]{Gallego2014}%
  \BibitemOpen
  \bibfield  {author} {\bibinfo {author} {\bibfnamefont {R.}~\bibnamefont
  {Gallego}}, \bibinfo {author} {\bibfnamefont {A.}~\bibnamefont {Riera}}, \
  and\ \bibinfo {author} {\bibfnamefont {J.}~\bibnamefont {Eisert}},\ }\href
  {\doibase 10.1088/1367-2630/16/12/125009} {\bibfield  {journal} {\bibinfo
  {journal} {New Journal of Physics}\ }\textbf {\bibinfo {volume} {16}},\
  \bibinfo {pages} {125009} (\bibinfo {year} {2014})}\BibitemShut {NoStop}%
\bibitem [{\citenamefont {B\"{a}umer}\ \emph {et~al.}(2019)\citenamefont
  {B\"{a}umer}, \citenamefont {Perarnau-Llobet}, \citenamefont {Kammerlander},
  \citenamefont {Wilming},\ and\ \citenamefont {Renner}}]{Bumer2019}%
  \BibitemOpen
  \bibfield  {author} {\bibinfo {author} {\bibfnamefont {E.}~\bibnamefont
  {B\"{a}umer}}, \bibinfo {author} {\bibfnamefont {M.}~\bibnamefont
  {Perarnau-Llobet}}, \bibinfo {author} {\bibfnamefont {P.}~\bibnamefont
  {Kammerlander}}, \bibinfo {author} {\bibfnamefont {H.}~\bibnamefont
  {Wilming}}, \ and\ \bibinfo {author} {\bibfnamefont {R.}~\bibnamefont
  {Renner}},\ }\href {\doibase 10.22331/q-2019-06-24-153} {\bibfield  {journal}
  {\bibinfo  {journal} {Quantum}\ }\textbf {\bibinfo {volume} {3}},\ \bibinfo
  {pages} {153} (\bibinfo {year} {2019})}\BibitemShut {NoStop}%
\bibitem [{\citenamefont {Evans}\ and\ \citenamefont
  {Searles}(2002)}]{Evans2002}%
  \BibitemOpen
  \bibfield  {author} {\bibinfo {author} {\bibfnamefont {D.~J.}\ \bibnamefont
  {Evans}}\ and\ \bibinfo {author} {\bibfnamefont {D.~J.}\ \bibnamefont
  {Searles}},\ }\href {\doibase 10.1080/00018730210155133} {\bibfield
  {journal} {\bibinfo  {journal} {Advances in Physics}\ }\textbf {\bibinfo
  {volume} {51}},\ \bibinfo {pages} {1529} (\bibinfo {year} {2002})},\ \Eprint
  {http://arxiv.org/abs/https://doi.org/10.1080/00018730210155133}
  {https://doi.org/10.1080/00018730210155133} \BibitemShut {NoStop}%
\bibitem [{\citenamefont {Mohammady}\ \emph {et~al.}(2019)\citenamefont
  {Mohammady}, \citenamefont {Auff{\'e}ves},\ and\ \citenamefont
  {Anders}}]{mohammady2019energetic}%
  \BibitemOpen
  \bibfield  {author} {\bibinfo {author} {\bibfnamefont {M.}~\bibnamefont
  {Mohammady}}, \bibinfo {author} {\bibfnamefont {A.}~\bibnamefont
  {Auff{\'e}ves}}, \ and\ \bibinfo {author} {\bibfnamefont {J.}~\bibnamefont
  {Anders}},\ }\href {http://arxiv.org/abs/1907.06559} {\bibfield  {journal}
  {\bibinfo  {journal} {arXiv:1907.06559}\ } (\bibinfo {year}
  {2019})}\BibitemShut {NoStop}%
\bibitem [{\citenamefont {Lostaglio}(2019)}]{Lostaglio2019}%
  \BibitemOpen
  \bibfield  {author} {\bibinfo {author} {\bibfnamefont {M.}~\bibnamefont
  {Lostaglio}},\ }\href {\doibase 10.1088/1361-6633/ab46e5} {\bibfield
  {journal} {\bibinfo  {journal} {Reports on Progress in Physics}\ }\textbf
  {\bibinfo {volume} {82}},\ \bibinfo {pages} {114001} (\bibinfo {year}
  {2019})}\BibitemShut {NoStop}%
\bibitem [{\citenamefont {Brand{\~a}o}\ \emph {et~al.}(2015)\citenamefont
  {Brand{\~a}o}, \citenamefont {Horodecki}, \citenamefont {Ng}, \citenamefont
  {Oppenheim},\ and\ \citenamefont {Wehner}}]{brandaoSecondLawsQuantum2015b}%
  \BibitemOpen
  \bibfield  {author} {\bibinfo {author} {\bibfnamefont {F.}~\bibnamefont
  {Brand{\~a}o}}, \bibinfo {author} {\bibfnamefont {M.}~\bibnamefont
  {Horodecki}}, \bibinfo {author} {\bibfnamefont {N.}~\bibnamefont {Ng}},
  \bibinfo {author} {\bibfnamefont {J.}~\bibnamefont {Oppenheim}}, \ and\
  \bibinfo {author} {\bibfnamefont {S.}~\bibnamefont {Wehner}},\ }\href
  {\doibase 10.1073/pnas.1411728112} {\bibfield  {journal} {\bibinfo  {journal}
  {Proceedings of the National Academy of Sciences}\ }\textbf {\bibinfo
  {volume} {112}},\ \bibinfo {pages} {3275} (\bibinfo {year}
  {2015})}\BibitemShut {NoStop}%
\bibitem [{\citenamefont {Lostaglio}\ \emph
  {et~al.}(2015{\natexlab{a}})\citenamefont {Lostaglio}, \citenamefont
  {Jennings},\ and\ \citenamefont
  {Rudolph}}]{lostaglioDescriptionQuantumCoherence2015}%
  \BibitemOpen
  \bibfield  {author} {\bibinfo {author} {\bibfnamefont {M.}~\bibnamefont
  {Lostaglio}}, \bibinfo {author} {\bibfnamefont {D.}~\bibnamefont {Jennings}},
  \ and\ \bibinfo {author} {\bibfnamefont {T.}~\bibnamefont {Rudolph}},\ }\href
  {\doibase 10.1038/ncomms7383} {\bibfield  {journal} {\bibinfo  {journal}
  {Nature Communications}\ }\textbf {\bibinfo {volume} {6}},\ \bibinfo {pages}
  {6383} (\bibinfo {year} {2015}{\natexlab{a}})}\BibitemShut {NoStop}%
\bibitem [{\citenamefont {Lostaglio}\ \emph
  {et~al.}(2015{\natexlab{b}})\citenamefont {Lostaglio}, \citenamefont
  {Korzekwa}, \citenamefont {Jennings},\ and\ \citenamefont
  {Rudolph}}]{lostaglioQuantumCoherenceTimeTranslation2015}%
  \BibitemOpen
  \bibfield  {author} {\bibinfo {author} {\bibfnamefont {M.}~\bibnamefont
  {Lostaglio}}, \bibinfo {author} {\bibfnamefont {K.}~\bibnamefont {Korzekwa}},
  \bibinfo {author} {\bibfnamefont {D.}~\bibnamefont {Jennings}}, \ and\
  \bibinfo {author} {\bibfnamefont {T.}~\bibnamefont {Rudolph}},\ }\href
  {\doibase 10.1103/PhysRevX.5.021001} {\bibfield  {journal} {\bibinfo
  {journal} {Physical Review X}\ }\textbf {\bibinfo {volume} {5}},\ \bibinfo
  {pages} {021001} (\bibinfo {year} {2015}{\natexlab{b}})}\BibitemShut
  {NoStop}%
\bibitem [{\citenamefont {Guarnieri}\ \emph {et~al.}(2019)\citenamefont
  {Guarnieri}, \citenamefont {Ng}, \citenamefont {Modi}, \citenamefont
  {Eisert}, \citenamefont {Paternostro},\ and\ \citenamefont
  {Goold}}]{guarnieriQuantumWorkStatistics2019}%
  \BibitemOpen
  \bibfield  {author} {\bibinfo {author} {\bibfnamefont {G.}~\bibnamefont
  {Guarnieri}}, \bibinfo {author} {\bibfnamefont {N.~H.~Y.}\ \bibnamefont
  {Ng}}, \bibinfo {author} {\bibfnamefont {K.}~\bibnamefont {Modi}}, \bibinfo
  {author} {\bibfnamefont {J.}~\bibnamefont {Eisert}}, \bibinfo {author}
  {\bibfnamefont {M.}~\bibnamefont {Paternostro}}, \ and\ \bibinfo {author}
  {\bibfnamefont {J.}~\bibnamefont {Goold}},\ }\href {\doibase
  10.1103/PhysRevE.99.050101} {\bibfield  {journal} {\bibinfo  {journal}
  {Physical Review E}\ }\textbf {\bibinfo {volume} {99}},\ \bibinfo {pages}
  {050101} (\bibinfo {year} {2019})},\ \Eprint
  {http://arxiv.org/abs/1804.09962} {arXiv:1804.09962} \BibitemShut {NoStop}%
\bibitem [{\citenamefont {Talkner}\ and\ \citenamefont
  {H\"anggi}(2016)}]{Talkner2016}%
  \BibitemOpen
  \bibfield  {author} {\bibinfo {author} {\bibfnamefont {P.}~\bibnamefont
  {Talkner}}\ and\ \bibinfo {author} {\bibfnamefont {P.}~\bibnamefont
  {H\"anggi}},\ }\href {\doibase 10.1103/PhysRevE.93.022131} {\bibfield
  {journal} {\bibinfo  {journal} {Phys. Rev. E}\ }\textbf {\bibinfo {volume}
  {93}},\ \bibinfo {pages} {022131} (\bibinfo {year} {2016})}\BibitemShut
  {NoStop}%
\bibitem [{\citenamefont {B\"{a}umer}\ \emph {et~al.}(2018)\citenamefont
  {B\"{a}umer}, \citenamefont {Lostaglio}, \citenamefont {Perarnau-Llobet},\
  and\ \citenamefont {Sampaio}}]{baumer2018fluctuating}%
  \BibitemOpen
  \bibfield  {author} {\bibinfo {author} {\bibfnamefont {E.}~\bibnamefont
  {B\"{a}umer}}, \bibinfo {author} {\bibfnamefont {M.}~\bibnamefont
  {Lostaglio}}, \bibinfo {author} {\bibfnamefont {M.}~\bibnamefont
  {Perarnau-Llobet}}, \ and\ \bibinfo {author} {\bibfnamefont {R.}~\bibnamefont
  {Sampaio}},\ }in\ \href {\doibase 10.1007/978-3-319-99046-0_11} {\emph
  {\bibinfo {booktitle} {Fundamental Theories of Physics}}}\ (\bibinfo
  {publisher} {Springer International Publishing},\ \bibinfo {year} {2018})\
  pp.\ \bibinfo {pages} {275--300}\BibitemShut {NoStop}%
\bibitem [{\citenamefont {Debarba}\ \emph {et~al.}(2019)\citenamefont
  {Debarba}, \citenamefont {Manzano}, \citenamefont {Guryanova}, \citenamefont
  {Huber},\ and\ \citenamefont {Friis}}]{Debarba2019}%
  \BibitemOpen
  \bibfield  {author} {\bibinfo {author} {\bibfnamefont {T.}~\bibnamefont
  {Debarba}}, \bibinfo {author} {\bibfnamefont {G.}~\bibnamefont {Manzano}},
  \bibinfo {author} {\bibfnamefont {Y.}~\bibnamefont {Guryanova}}, \bibinfo
  {author} {\bibfnamefont {M.}~\bibnamefont {Huber}}, \ and\ \bibinfo {author}
  {\bibfnamefont {N.}~\bibnamefont {Friis}},\ }\href {\doibase
  10.1088/1367-2630/ab4d9d} {\bibfield  {journal} {\bibinfo  {journal} {New
  Journal of Physics}\ }\textbf {\bibinfo {volume} {21}},\ \bibinfo {pages}
  {113002} (\bibinfo {year} {2019})}\BibitemShut {NoStop}%
\bibitem [{\citenamefont {Strasberg}(2019)}]{Strasberg2019}%
  \BibitemOpen
  \bibfield  {author} {\bibinfo {author} {\bibfnamefont {P.}~\bibnamefont
  {Strasberg}},\ }\href {\doibase 10.1103/PhysRevE.100.022127} {\bibfield
  {journal} {\bibinfo  {journal} {Phys. Rev. E}\ }\textbf {\bibinfo {volume}
  {100}},\ \bibinfo {pages} {022127} (\bibinfo {year} {2019})}\BibitemShut
  {NoStop}%
\bibitem [{\citenamefont {Talkner}\ \emph {et~al.}(2007)\citenamefont
  {Talkner}, \citenamefont {Lutz},\ and\ \citenamefont
  {H{\"{a}}nggi}}]{Talkner2007c}%
  \BibitemOpen
  \bibfield  {author} {\bibinfo {author} {\bibfnamefont {P.}~\bibnamefont
  {Talkner}}, \bibinfo {author} {\bibfnamefont {E.}~\bibnamefont {Lutz}}, \
  and\ \bibinfo {author} {\bibfnamefont {P.}~\bibnamefont {H{\"{a}}nggi}},\
  }\href {\doibase 10.1103/PhysRevE.75.050102} {\bibfield  {journal} {\bibinfo
  {journal} {Phys. Rev. E}\ }\textbf {\bibinfo {volume} {75}},\ \bibinfo
  {pages} {050102} (\bibinfo {year} {2007})}\BibitemShut {NoStop}%
\bibitem [{\citenamefont {Wei}\ and\ \citenamefont
  {Plenio}(2017)}]{weiRelationsDissipatedWork2017}%
  \BibitemOpen
  \bibfield  {author} {\bibinfo {author} {\bibfnamefont {B.-B.}\ \bibnamefont
  {Wei}}\ and\ \bibinfo {author} {\bibfnamefont {M.~B.}\ \bibnamefont
  {Plenio}},\ }\href {\doibase 10.1088/1367-2630/aa579e} {\bibfield  {journal}
  {\bibinfo  {journal} {New Journal of Physics}\ }\textbf {\bibinfo {volume}
  {19}},\ \bibinfo {pages} {023002} (\bibinfo {year} {2017})}\BibitemShut
  {NoStop}%
\bibitem [{\citenamefont {Campisi}\ \emph {et~al.}(2012)\citenamefont
  {Campisi}, \citenamefont {Denisov},\ and\ \citenamefont
  {H\"anggi}}]{Campisi2012geometric}%
  \BibitemOpen
  \bibfield  {author} {\bibinfo {author} {\bibfnamefont {M.}~\bibnamefont
  {Campisi}}, \bibinfo {author} {\bibfnamefont {S.}~\bibnamefont {Denisov}}, \
  and\ \bibinfo {author} {\bibfnamefont {P.}~\bibnamefont {H\"anggi}},\ }\href
  {\doibase 10.1103/PhysRevA.86.032114} {\bibfield  {journal} {\bibinfo
  {journal} {Phys. Rev. A}\ }\textbf {\bibinfo {volume} {86}},\ \bibinfo
  {pages} {032114} (\bibinfo {year} {2012})}\BibitemShut {NoStop}%
\bibitem [{\citenamefont {Sivak}\ and\ \citenamefont
  {Crooks}(2012)}]{sivakThermodynamicMetricsOptimal2012}%
  \BibitemOpen
  \bibfield  {author} {\bibinfo {author} {\bibfnamefont {D.~A.}\ \bibnamefont
  {Sivak}}\ and\ \bibinfo {author} {\bibfnamefont {G.~E.}\ \bibnamefont
  {Crooks}},\ }\href {\doibase 10.1103/PhysRevLett.108.190602} {\bibfield
  {journal} {\bibinfo  {journal} {Physical Review Letters}\ }\textbf {\bibinfo
  {volume} {108}} (\bibinfo {year} {2012}),\
  10.1103/PhysRevLett.108.190602}\BibitemShut {NoStop}%
\bibitem [{\citenamefont {Bonan{\c{c}}a}\ and\ \citenamefont
  {Deffner}(2014)}]{Bonana2014}%
  \BibitemOpen
  \bibfield  {author} {\bibinfo {author} {\bibfnamefont {M.~V.~S.}\
  \bibnamefont {Bonan{\c{c}}a}}\ and\ \bibinfo {author} {\bibfnamefont
  {S.}~\bibnamefont {Deffner}},\ }\href {\doibase 10.1063/1.4885277} {\bibfield
   {journal} {\bibinfo  {journal} {The Journal of Chemical Physics}\ }\textbf
  {\bibinfo {volume} {140}},\ \bibinfo {pages} {244119} (\bibinfo {year}
  {2014})}\BibitemShut {NoStop}%
\bibitem [{\citenamefont {Acconcia}\ \emph {et~al.}(2015)\citenamefont
  {Acconcia}, \citenamefont {Bonan{\c{c}}a},\ and\ \citenamefont
  {Deffner}}]{acconcia2015shortcuts}%
  \BibitemOpen
  \bibfield  {author} {\bibinfo {author} {\bibfnamefont {T.~V.}\ \bibnamefont
  {Acconcia}}, \bibinfo {author} {\bibfnamefont {M.~V.~S.}\ \bibnamefont
  {Bonan{\c{c}}a}}, \ and\ \bibinfo {author} {\bibfnamefont {S.}~\bibnamefont
  {Deffner}},\ }\href
  {https://journals.aps.org/pre/abstract/10.1103/PhysRevE.92.042148} {\bibfield
   {journal} {\bibinfo  {journal} {Physical Review E}\ }\textbf {\bibinfo
  {volume} {92}},\ \bibinfo {pages} {042148} (\bibinfo {year}
  {2015})}\BibitemShut {NoStop}%
\bibitem [{\citenamefont {Ludovico}\ \emph {et~al.}(2016)\citenamefont
  {Ludovico}, \citenamefont {Battista}, \citenamefont {von Oppen},\ and\
  \citenamefont {Arrachea}}]{Ludovico2016}%
  \BibitemOpen
  \bibfield  {author} {\bibinfo {author} {\bibfnamefont {M.~F.}\ \bibnamefont
  {Ludovico}}, \bibinfo {author} {\bibfnamefont {F.}~\bibnamefont {Battista}},
  \bibinfo {author} {\bibfnamefont {F.}~\bibnamefont {von Oppen}}, \ and\
  \bibinfo {author} {\bibfnamefont {L.}~\bibnamefont {Arrachea}},\ }\href
  {\doibase 10.1103/PhysRevB.93.075136} {\bibfield  {journal} {\bibinfo
  {journal} {Phys. Rev. B}\ }\textbf {\bibinfo {volume} {93}},\ \bibinfo
  {pages} {075136} (\bibinfo {year} {2016})}\BibitemShut {NoStop}%
\bibitem [{\citenamefont {Allahverdyan}(2014)}]{Allahverdyan2014}%
  \BibitemOpen
  \bibfield  {author} {\bibinfo {author} {\bibfnamefont {A.~E.}\ \bibnamefont
  {Allahverdyan}},\ }\href {\doibase 10.1103/PhysRevE.90.032137} {\bibfield
  {journal} {\bibinfo  {journal} {Phys. Rev. E}\ }\textbf {\bibinfo {volume}
  {90}},\ \bibinfo {pages} {032137} (\bibinfo {year} {2014})}\BibitemShut
  {NoStop}%
\bibitem [{\citenamefont {Solinas}\ and\ \citenamefont
  {Gasparinetti}(2015)}]{Solinas2015}%
  \BibitemOpen
  \bibfield  {author} {\bibinfo {author} {\bibfnamefont {P.}~\bibnamefont
  {Solinas}}\ and\ \bibinfo {author} {\bibfnamefont {S.}~\bibnamefont
  {Gasparinetti}},\ }\href {\doibase 10.1103/PhysRevE.92.042150} {\bibfield
  {journal} {\bibinfo  {journal} {Phys. Rev. E}\ }\textbf {\bibinfo {volume}
  {92}},\ \bibinfo {pages} {042150} (\bibinfo {year} {2015})}\BibitemShut
  {NoStop}%
\bibitem [{\citenamefont {Hofer}(2017)}]{Hofer2017}%
  \BibitemOpen
  \bibfield  {author} {\bibinfo {author} {\bibfnamefont {P.~P.}\ \bibnamefont
  {Hofer}},\ }\href {\doibase 10.22331/q-2017-10-12-32} {\bibfield  {journal}
  {\bibinfo  {journal} {Quantum}\ }\textbf {\bibinfo {volume} {1}},\ \bibinfo
  {pages} {32} (\bibinfo {year} {2017})}\BibitemShut {NoStop}%
\bibitem [{\citenamefont {Scandi}\ and\ \citenamefont
  {{Perarnau-Llobet}}(2019)}]{scandiThermodynamicLengthOpen2019}%
  \BibitemOpen
  \bibfield  {author} {\bibinfo {author} {\bibfnamefont {M.}~\bibnamefont
  {Scandi}}\ and\ \bibinfo {author} {\bibfnamefont {M.}~\bibnamefont
  {{Perarnau-Llobet}}},\ }\href {\doibase 10.22331/q-2019-10-24-197} {\bibfield
   {journal} {\bibinfo  {journal} {Quantum}\ }\textbf {\bibinfo {volume} {3}},\
  \bibinfo {pages} {197} (\bibinfo {year} {2019})}\BibitemShut {NoStop}%
\bibitem [{\citenamefont {Marshall}\ and\ \citenamefont
  {Olkin}(1985)}]{marshallInequalitiesTraceFunction1985}%
  \BibitemOpen
  \bibfield  {author} {\bibinfo {author} {\bibfnamefont {A.~W.}\ \bibnamefont
  {Marshall}}\ and\ \bibinfo {author} {\bibfnamefont {I.}~\bibnamefont
  {Olkin}},\ }\href {\doibase 10.1007/BF02189811} {\bibfield  {journal}
  {\bibinfo  {journal} {aequationes mathematicae}\ }\textbf {\bibinfo {volume}
  {29}},\ \bibinfo {pages} {36} (\bibinfo {year} {1985})}\BibitemShut {NoStop}%
\bibitem [{\citenamefont {Wigner}\ and\ \citenamefont
  {Yanase}(1963)}]{wignerINFORMATIONCONTENTSDISTRIBUTIONS1963}%
  \BibitemOpen
  \bibfield  {author} {\bibinfo {author} {\bibfnamefont {E.~P.}\ \bibnamefont
  {Wigner}}\ and\ \bibinfo {author} {\bibfnamefont {M.~M.}\ \bibnamefont
  {Yanase}},\ }\href {\doibase 10.1073/pnas.49.6.910} {\bibfield  {journal}
  {\bibinfo  {journal} {Proceedings of the National Academy of Sciences of the
  United States of America}\ }\textbf {\bibinfo {volume} {49}},\ \bibinfo
  {pages} {910} (\bibinfo {year} {1963})}\BibitemShut {NoStop}%
\bibitem [{\citenamefont {Kubo}(1957)}]{Kubo1957}%
  \BibitemOpen
  \bibfield  {author} {\bibinfo {author} {\bibfnamefont {R.}~\bibnamefont
  {Kubo}},\ }\href@noop {} {\bibfield  {journal} {\bibinfo  {journal} {J. Phys.
  Soc. Jap.}\ }\textbf {\bibinfo {volume} {12}},\ \bibinfo {pages} {570}
  (\bibinfo {year} {1957})}\BibitemShut {NoStop}%
\bibitem [{\citenamefont {Parisi}(1988)}]{parisiStatisticalFieldTheory1988}%
  \BibitemOpen
  \bibfield  {author} {\bibinfo {author} {\bibfnamefont {G.}~\bibnamefont
  {Parisi}},\ }\href@noop {} {\emph {\bibinfo {title} {Statistical Field
  Theory}}}\ (\bibinfo  {publisher} {{Addison-Wesley}},\ \bibinfo {year}
  {1988})\BibitemShut {NoStop}%
\bibitem [{\citenamefont
  {Ruelle}(1999)}]{ruelleStatisticalMechanicsRigorous1999}%
  \BibitemOpen
  \bibfield  {author} {\bibinfo {author} {\bibfnamefont {D.}~\bibnamefont
  {Ruelle}},\ }\href@noop {} {\emph {\bibinfo {title} {Statistical
  {{Mechanics}}: {{Rigorous Results}}}}}\ (\bibinfo  {publisher} {{World
  Scientific}},\ \bibinfo {year} {1999})\BibitemShut {NoStop}%
\bibitem [{\citenamefont {Lieb}\ and\ \citenamefont
  {Yngvason}(1999)}]{liebPhysicsMathematicsSecond1999}%
  \BibitemOpen
  \bibfield  {author} {\bibinfo {author} {\bibfnamefont {E.~H.}\ \bibnamefont
  {Lieb}}\ and\ \bibinfo {author} {\bibfnamefont {J.}~\bibnamefont
  {Yngvason}},\ }\href {\doibase 10.1016/S0370-1573(98)00082-9} {\bibfield
  {journal} {\bibinfo  {journal} {Physics Reports}\ }\textbf {\bibinfo {volume}
  {310}},\ \bibinfo {pages} {1} (\bibinfo {year} {1999})}\BibitemShut {NoStop}%
\bibitem [{\citenamefont {\ifmmode \acute{C}\else
  \'{C}\fi{}wikli\ifmmode~\acute{n}\else \'{n}\fi{}ski}\ \emph
  {et~al.}(2015)\citenamefont {\ifmmode \acute{C}\else
  \'{C}\fi{}wikli\ifmmode~\acute{n}\else \'{n}\fi{}ski}, \citenamefont
  {Studzi\ifmmode~\acute{n}\else \'{n}\fi{}ski}, \citenamefont {Horodecki},\
  and\ \citenamefont {Oppenheim}}]{PhysRevLett.115.210403}%
  \BibitemOpen
  \bibfield  {author} {\bibinfo {author} {\bibfnamefont {P.}~\bibnamefont
  {\ifmmode \acute{C}\else \'{C}\fi{}wikli\ifmmode~\acute{n}\else
  \'{n}\fi{}ski}}, \bibinfo {author} {\bibfnamefont {M.}~\bibnamefont
  {Studzi\ifmmode~\acute{n}\else \'{n}\fi{}ski}}, \bibinfo {author}
  {\bibfnamefont {M.}~\bibnamefont {Horodecki}}, \ and\ \bibinfo {author}
  {\bibfnamefont {J.}~\bibnamefont {Oppenheim}},\ }\href {\doibase
  10.1103/PhysRevLett.115.210403} {\bibfield  {journal} {\bibinfo  {journal}
  {Phys. Rev. Lett.}\ }\textbf {\bibinfo {volume} {115}},\ \bibinfo {pages}
  {210403} (\bibinfo {year} {2015})}\BibitemShut {NoStop}%
\bibitem [{\citenamefont {Aberg}(2014)}]{abergCatalyticCoherence2014}%
  \BibitemOpen
  \bibfield  {author} {\bibinfo {author} {\bibfnamefont {J.}~\bibnamefont
  {Aberg}},\ }\href {\doibase 10.1103/PhysRevLett.113.150402} {\bibfield
  {journal} {\bibinfo  {journal} {Physical Review Letters}\ }\textbf {\bibinfo
  {volume} {113}},\ \bibinfo {pages} {150402} (\bibinfo {year}
  {2014})}\BibitemShut {NoStop}%
\bibitem [{\citenamefont {Crooks}(1999{\natexlab{b}})}]{Crooks}%
  \BibitemOpen
  \bibfield  {author} {\bibinfo {author} {\bibfnamefont {G.}~\bibnamefont
  {Crooks}},\ }\href {\doibase 10.1103/PhysRevE.60.2721} {\bibfield  {journal}
  {\bibinfo  {journal} {Phys. Rev. E}\ }\textbf {\bibinfo {volume} {60}},\
  \bibinfo {pages} {2721} (\bibinfo {year} {1999}{\natexlab{b}})}\BibitemShut
  {NoStop}%
\bibitem [{\citenamefont {Merhav}\ and\ \citenamefont
  {Kafri}(2010)}]{Merhav2010}%
  \BibitemOpen
  \bibfield  {author} {\bibinfo {author} {\bibfnamefont {N.}~\bibnamefont
  {Merhav}}\ and\ \bibinfo {author} {\bibfnamefont {Y.}~\bibnamefont {Kafri}},\
  }\href {\doibase 10.1088/1742-5468/2010/12/P12022} {\bibfield  {journal}
  {\bibinfo  {journal} {J. Stat. Mech.}\ }\textbf {\bibinfo {volume} {2010}},\
  \bibinfo {pages} {P12022} (\bibinfo {year} {2010})}\BibitemShut {NoStop}%
\bibitem [{\citenamefont {Barato}\ and\ \citenamefont
  {Seifert}(2015)}]{Barato}%
  \BibitemOpen
  \bibfield  {author} {\bibinfo {author} {\bibfnamefont {A.~C.}\ \bibnamefont
  {Barato}}\ and\ \bibinfo {author} {\bibfnamefont {U.}~\bibnamefont
  {Seifert}},\ }\href {https://doi.org/10.1103/PhysRevLett.114.158101}
  {\bibfield  {journal} {\bibinfo  {journal} {Phys. Rev. Lett.}\ }\textbf
  {\bibinfo {volume} {114}},\ \bibinfo {pages} {158101} (\bibinfo {year}
  {2015})}\BibitemShut {NoStop}%
\bibitem [{\citenamefont {Pietzonka}\ \emph {et~al.}(2016)\citenamefont
  {Pietzonka}, \citenamefont {Barato},\ and\ \citenamefont
  {Seifert}}]{Pietzonka}%
  \BibitemOpen
  \bibfield  {author} {\bibinfo {author} {\bibfnamefont {P.}~\bibnamefont
  {Pietzonka}}, \bibinfo {author} {\bibfnamefont {A.~C.}\ \bibnamefont
  {Barato}}, \ and\ \bibinfo {author} {\bibfnamefont {U.}~\bibnamefont
  {Seifert}},\ }\href {https://doi.org/10.1103/PhysRevE.93.052145} {\bibfield
  {journal} {\bibinfo  {journal} {Phys. Rev. E}\ }\textbf {\bibinfo {volume}
  {93}},\ \bibinfo {pages} {052145} (\bibinfo {year} {2016})}\BibitemShut
  {NoStop}%
\bibitem [{\citenamefont {Pietzonka}\ and\ \citenamefont
  {Seifert}(2018)}]{Pietzonka2018}%
  \BibitemOpen
  \bibfield  {author} {\bibinfo {author} {\bibfnamefont {P.}~\bibnamefont
  {Pietzonka}}\ and\ \bibinfo {author} {\bibfnamefont {U.}~\bibnamefont
  {Seifert}},\ }\href {https://doi.org/10.1103/PhysRevLett.120.190602}
  {\bibfield  {journal} {\bibinfo  {journal} {Phys. Rev. Lett.}\ }\textbf
  {\bibinfo {volume} {120}},\ \bibinfo {pages} {190602} (\bibinfo {year}
  {2018})}\BibitemShut {NoStop}%
\bibitem [{\citenamefont {Timpanaro}\ \emph {et~al.}(2019)\citenamefont
  {Timpanaro}, \citenamefont {Guarnieri}, \citenamefont {Goold},\ and\
  \citenamefont {Landi}}]{Guarnieri2019}%
  \BibitemOpen
  \bibfield  {author} {\bibinfo {author} {\bibfnamefont {A.~M.}\ \bibnamefont
  {Timpanaro}}, \bibinfo {author} {\bibfnamefont {G.}~\bibnamefont
  {Guarnieri}}, \bibinfo {author} {\bibfnamefont {J.}~\bibnamefont {Goold}}, \
  and\ \bibinfo {author} {\bibfnamefont {G.~T.}\ \bibnamefont {Landi}},\ }\href
  {\doibase 10.1103/PhysRevLett.123.090604} {\bibfield  {journal} {\bibinfo
  {journal} {Phys. Rev. Lett.}\ }\textbf {\bibinfo {volume} {123}},\ \bibinfo
  {pages} {090604} (\bibinfo {year} {2019})}\BibitemShut {NoStop}%
\bibitem [{\citenamefont {Mandal}\ and\ \citenamefont
  {Jarzynski}(2016{\natexlab{b}})}]{mandalAnalysisSlowTransitions2016}%
  \BibitemOpen
  \bibfield  {author} {\bibinfo {author} {\bibfnamefont {D.}~\bibnamefont
  {Mandal}}\ and\ \bibinfo {author} {\bibfnamefont {C.}~\bibnamefont
  {Jarzynski}},\ }\href {\doibase 10.1088/1742-5468/2016/06/063204} {\bibfield
  {journal} {\bibinfo  {journal} {Journal of Statistical Mechanics: Theory and
  Experiment}\ }\textbf {\bibinfo {volume} {2016}},\ \bibinfo {pages} {063204}
  (\bibinfo {year} {2016}{\natexlab{b}})},\ \Eprint
  {http://arxiv.org/abs/1507.06269} {arXiv:1507.06269} \BibitemShut {NoStop}%
\bibitem [{\citenamefont {Cavina}\ \emph {et~al.}(2017)\citenamefont {Cavina},
  \citenamefont {Mari},\ and\ \citenamefont
  {Giovannetti}}]{cavinaSlowDynamicsThermodynamics2017}%
  \BibitemOpen
  \bibfield  {author} {\bibinfo {author} {\bibfnamefont {V.}~\bibnamefont
  {Cavina}}, \bibinfo {author} {\bibfnamefont {A.}~\bibnamefont {Mari}}, \ and\
  \bibinfo {author} {\bibfnamefont {V.}~\bibnamefont {Giovannetti}},\ }\href
  {\doibase 10.1103/PhysRevLett.119.050601} {\bibfield  {journal} {\bibinfo
  {journal} {Physical Review Letters}\ }\textbf {\bibinfo {volume} {119}}
  (\bibinfo {year} {2017}),\ 10.1103/PhysRevLett.119.050601}\BibitemShut
  {NoStop}%
\bibitem [{\citenamefont {Fagnola}\ and\ \citenamefont
  {Umanit{\`{a}}}(2010)}]{Fagnola2010a}%
  \BibitemOpen
  \bibfield  {author} {\bibinfo {author} {\bibfnamefont {F.}~\bibnamefont
  {Fagnola}}\ and\ \bibinfo {author} {\bibfnamefont {V.}~\bibnamefont
  {Umanit{\`{a}}}},\ }\href {\doibase 10.1007/s00220-010-1011-1} {\bibfield
  {journal} {\bibinfo  {journal} {Communications in Mathematical Physics}\
  }\textbf {\bibinfo {volume} {298}},\ \bibinfo {pages} {523} (\bibinfo {year}
  {2010})}\BibitemShut {NoStop}%
\bibitem [{\citenamefont {Janzing}(2006)}]{Janzing2006}%
  \BibitemOpen
  \bibfield  {author} {\bibinfo {author} {\bibfnamefont {D.}~\bibnamefont
  {Janzing}},\ }\href {\doibase 10.1007/s10955-006-9220-x} {\bibfield
  {journal} {\bibinfo  {journal} {Journal of Statistical Physics}\ }\textbf
  {\bibinfo {volume} {125}},\ \bibinfo {pages} {761} (\bibinfo {year}
  {2006})}\BibitemShut {NoStop}%
\bibitem [{\citenamefont {Santos}\ \emph {et~al.}(2019)\citenamefont {Santos},
  \citenamefont {C{\'{e}}leri}, \citenamefont {Landi},\ and\ \citenamefont
  {Paternostro}}]{Santos2019}%
  \BibitemOpen
  \bibfield  {author} {\bibinfo {author} {\bibfnamefont {J.~P.}\ \bibnamefont
  {Santos}}, \bibinfo {author} {\bibfnamefont {L.~C.}\ \bibnamefont
  {C{\'{e}}leri}}, \bibinfo {author} {\bibfnamefont {G.~T.}\ \bibnamefont
  {Landi}}, \ and\ \bibinfo {author} {\bibfnamefont {M.}~\bibnamefont
  {Paternostro}},\ }\href {\doibase 10.1038/s41534-019-0138-y} {\bibfield
  {journal} {\bibinfo  {journal} {npj Quantum Information}\ }\textbf {\bibinfo
  {volume} {5}} (\bibinfo {year} {2019}),\
  10.1038/s41534-019-0138-y}\BibitemShut {NoStop}%
\bibitem [{\citenamefont {Francica}\ \emph {et~al.}(2019)\citenamefont
  {Francica}, \citenamefont {Goold},\ and\ \citenamefont
  {Plastina}}]{Francica2019}%
  \BibitemOpen
  \bibfield  {author} {\bibinfo {author} {\bibfnamefont {G.}~\bibnamefont
  {Francica}}, \bibinfo {author} {\bibfnamefont {J.}~\bibnamefont {Goold}}, \
  and\ \bibinfo {author} {\bibfnamefont {F.}~\bibnamefont {Plastina}},\ }\href
  {\doibase 10.1103/physreve.99.042105} {\bibfield  {journal} {\bibinfo
  {journal} {Physical Review E}\ }\textbf {\bibinfo {volume} {99}} (\bibinfo
  {year} {2019}),\ 10.1103/physreve.99.042105}\BibitemShut {NoStop}%
\bibitem [{\citenamefont {Hansen}(2008)}]{Hansen2008}%
  \BibitemOpen
  \bibfield  {author} {\bibinfo {author} {\bibfnamefont {F.}~\bibnamefont
  {Hansen}},\ }\href {\doibase 10.1073/pnas.0803323105} {\bibfield  {journal}
  {\bibinfo  {journal} {Proc. Natl. Acad. Sci. USA}\ }\textbf {\bibinfo
  {volume} {105}},\ \bibinfo {pages} {9909} (\bibinfo {year}
  {2008})}\BibitemShut {NoStop}%
\bibitem [{\citenamefont {Fr{\'{e}}rot}\ and\ \citenamefont
  {Roscilde}(2016)}]{Frerot2016}%
  \BibitemOpen
  \bibfield  {author} {\bibinfo {author} {\bibfnamefont {I.}~\bibnamefont
  {Fr{\'{e}}rot}}\ and\ \bibinfo {author} {\bibfnamefont {T.}~\bibnamefont
  {Roscilde}},\ }\href {http://arxiv.org/abs/1509.06741} {\bibfield  {journal}
  {\bibinfo  {journal} {Phys. Rev. A}\ }\textbf {\bibinfo {volume} {94}},\
  \bibinfo {pages} {075121} (\bibinfo {year} {2016})}\BibitemShut {NoStop}%
\bibitem [{\citenamefont {Solinas}\ \emph {et~al.}(2017)\citenamefont
  {Solinas}, \citenamefont {Miller},\ and\ \citenamefont
  {Anders}}]{Solinas2017}%
  \BibitemOpen
  \bibfield  {author} {\bibinfo {author} {\bibfnamefont {P.}~\bibnamefont
  {Solinas}}, \bibinfo {author} {\bibfnamefont {H.~J.~D.}\ \bibnamefont
  {Miller}}, \ and\ \bibinfo {author} {\bibfnamefont {J.}~\bibnamefont
  {Anders}},\ }\href {http://arxiv.org/abs/1705.10296} {\bibfield  {journal}
  {\bibinfo  {journal} {Phys. Rev. A}\ }\textbf {\bibinfo {volume} {96}},\
  \bibinfo {pages} {052115} (\bibinfo {year} {2017})}\BibitemShut {NoStop}%
\bibitem [{\citenamefont {Miller}\ and\ \citenamefont
  {Anders}(2018)}]{Miller2018}%
  \BibitemOpen
  \bibfield  {author} {\bibinfo {author} {\bibfnamefont {H.~J.~D.}\
  \bibnamefont {Miller}}\ and\ \bibinfo {author} {\bibfnamefont
  {J.}~\bibnamefont {Anders}},\ }\href {http://arxiv.org/abs/1801.08057}
  {\bibfield  {journal} {\bibinfo  {journal} {Nat. Comm.}\ }\textbf {\bibinfo
  {volume} {9}},\ \bibinfo {pages} {2203} (\bibinfo {year} {2018})}\BibitemShut
  {NoStop}%
\bibitem [{\citenamefont {Lostaglio}(2018)}]{Lostaglio2018}%
  \BibitemOpen
  \bibfield  {author} {\bibinfo {author} {\bibfnamefont {M.}~\bibnamefont
  {Lostaglio}},\ }\href {\doibase 10.1103/PhysRevLett.120.040602} {\bibfield
  {journal} {\bibinfo  {journal} {Phys. Rev. Lett.}\ }\textbf {\bibinfo
  {volume} {120}},\ \bibinfo {pages} {040602} (\bibinfo {year}
  {2018})}\BibitemShut {NoStop}%
\bibitem [{\citenamefont {Xu}\ \emph {et~al.}(2018)\citenamefont {Xu},
  \citenamefont {Zou}, \citenamefont {Guo},\ and\ \citenamefont
  {Kong}}]{Xu2018}%
  \BibitemOpen
  \bibfield  {author} {\bibinfo {author} {\bibfnamefont {B.-M.}\ \bibnamefont
  {Xu}}, \bibinfo {author} {\bibfnamefont {J.}~\bibnamefont {Zou}}, \bibinfo
  {author} {\bibfnamefont {L.-S.}\ \bibnamefont {Guo}}, \ and\ \bibinfo
  {author} {\bibfnamefont {X.-M.}\ \bibnamefont {Kong}},\ }\href {\doibase
  10.1103/PhysRevA.97.052122} {\bibfield  {journal} {\bibinfo  {journal} {Phys.
  Rev. A}\ }\textbf {\bibinfo {volume} {97}},\ \bibinfo {pages} {052122}
  (\bibinfo {year} {2018})}\BibitemShut {NoStop}%
\bibitem [{\citenamefont {Potts}(2019)}]{PatrickPotts2019}%
  \BibitemOpen
  \bibfield  {author} {\bibinfo {author} {\bibfnamefont {P.~P.}\ \bibnamefont
  {Potts}},\ }\href {\doibase 10.1103/PhysRevLett.122.110401} {\bibfield
  {journal} {\bibinfo  {journal} {Phys. Rev. Lett.}\ }\textbf {\bibinfo
  {volume} {122}},\ \bibinfo {pages} {110401} (\bibinfo {year}
  {2019})}\BibitemShut {NoStop}%
\bibitem [{\citenamefont {Wu}\ \emph {et~al.}(2019)\citenamefont {Wu},
  \citenamefont {Yuan}, \citenamefont {Xiang}, \citenamefont {Li},
  \citenamefont {Guo},\ and\ \citenamefont {Perarnau-Llobet}}]{Wu2019}%
  \BibitemOpen
  \bibfield  {author} {\bibinfo {author} {\bibfnamefont {K.-D.}\ \bibnamefont
  {Wu}}, \bibinfo {author} {\bibfnamefont {Y.}~\bibnamefont {Yuan}}, \bibinfo
  {author} {\bibfnamefont {G.-Y.}\ \bibnamefont {Xiang}}, \bibinfo {author}
  {\bibfnamefont {C.-F.}\ \bibnamefont {Li}}, \bibinfo {author} {\bibfnamefont
  {G.-C.}\ \bibnamefont {Guo}}, \ and\ \bibinfo {author} {\bibfnamefont
  {M.}~\bibnamefont {Perarnau-Llobet}},\ }\href {\doibase
  10.1126/sciadv.aav4944} {\bibfield  {journal} {\bibinfo  {journal} {Science
  Advances}\ }\textbf {\bibinfo {volume} {5}},\ \bibinfo {pages} {eaav4944}
  (\bibinfo {year} {2019})}\BibitemShut {NoStop}%
\bibitem [{\citenamefont {Mingo}\ and\ \citenamefont
  {Jennings}(2018)}]{mingo2018superpositions}%
  \BibitemOpen
  \bibfield  {author} {\bibinfo {author} {\bibfnamefont {E.~H.}\ \bibnamefont
  {Mingo}}\ and\ \bibinfo {author} {\bibfnamefont {D.}~\bibnamefont
  {Jennings}},\ }\href {http://arxiv.org/abs/1812.08159} {\bibfield  {journal}
  {\bibinfo  {journal} {arXiv:1812.08159}\ } (\bibinfo {year}
  {2018})}\BibitemShut {NoStop}%
\bibitem [{\citenamefont {An}\ \emph {et~al.}(2014)\citenamefont {An},
  \citenamefont {Zhang}, \citenamefont {Um}, \citenamefont {Lv}, \citenamefont
  {Lu}, \citenamefont {Zhang}, \citenamefont {Yin}, \citenamefont {Quan},\ and\
  \citenamefont {Kim}}]{An2014}%
  \BibitemOpen
  \bibfield  {author} {\bibinfo {author} {\bibfnamefont {S.}~\bibnamefont
  {An}}, \bibinfo {author} {\bibfnamefont {J.-N.}\ \bibnamefont {Zhang}},
  \bibinfo {author} {\bibfnamefont {M.}~\bibnamefont {Um}}, \bibinfo {author}
  {\bibfnamefont {D.}~\bibnamefont {Lv}}, \bibinfo {author} {\bibfnamefont
  {Y.}~\bibnamefont {Lu}}, \bibinfo {author} {\bibfnamefont {J.}~\bibnamefont
  {Zhang}}, \bibinfo {author} {\bibfnamefont {Z.-Q.}\ \bibnamefont {Yin}},
  \bibinfo {author} {\bibfnamefont {H.~T.}\ \bibnamefont {Quan}}, \ and\
  \bibinfo {author} {\bibfnamefont {K.}~\bibnamefont {Kim}},\ }\href {\doibase
  10.1038/nphys3197} {\bibfield  {journal} {\bibinfo  {journal} {Nature
  Physics}\ }\textbf {\bibinfo {volume} {11}},\ \bibinfo {pages} {193}
  (\bibinfo {year} {2014})}\BibitemShut {NoStop}%
\bibitem [{\citenamefont {von Lindenfels}\ \emph {et~al.}(2019)\citenamefont
  {von Lindenfels}, \citenamefont {Gr\"ab}, \citenamefont {Schmiegelow},
  \citenamefont {Kaushal}, \citenamefont {Schulz}, \citenamefont {Mitchison},
  \citenamefont {Goold}, \citenamefont {Schmidt-Kaler},\ and\ \citenamefont
  {Poschinger}}]{Lindenfels2019}%
  \BibitemOpen
  \bibfield  {author} {\bibinfo {author} {\bibfnamefont {D.}~\bibnamefont {von
  Lindenfels}}, \bibinfo {author} {\bibfnamefont {O.}~\bibnamefont {Gr\"ab}},
  \bibinfo {author} {\bibfnamefont {C.~T.}\ \bibnamefont {Schmiegelow}},
  \bibinfo {author} {\bibfnamefont {V.}~\bibnamefont {Kaushal}}, \bibinfo
  {author} {\bibfnamefont {J.}~\bibnamefont {Schulz}}, \bibinfo {author}
  {\bibfnamefont {M.~T.}\ \bibnamefont {Mitchison}}, \bibinfo {author}
  {\bibfnamefont {J.}~\bibnamefont {Goold}}, \bibinfo {author} {\bibfnamefont
  {F.}~\bibnamefont {Schmidt-Kaler}}, \ and\ \bibinfo {author} {\bibfnamefont
  {U.~G.}\ \bibnamefont {Poschinger}},\ }\href {\doibase
  10.1103/PhysRevLett.123.080602} {\bibfield  {journal} {\bibinfo  {journal}
  {Phys. Rev. Lett.}\ }\textbf {\bibinfo {volume} {123}},\ \bibinfo {pages}
  {080602} (\bibinfo {year} {2019})}\BibitemShut {NoStop}%
\bibitem [{\citenamefont {Batalh\~ao}\ \emph {et~al.}(2014)\citenamefont
  {Batalh\~ao}, \citenamefont {Souza}, \citenamefont {Mazzola}, \citenamefont
  {Auccaise}, \citenamefont {Sarthour}, \citenamefont {Oliveira}, \citenamefont
  {Goold}, \citenamefont {De~Chiara}, \citenamefont {Paternostro},\ and\
  \citenamefont {Serra}}]{Tiago2014}%
  \BibitemOpen
  \bibfield  {author} {\bibinfo {author} {\bibfnamefont {T.~B.}\ \bibnamefont
  {Batalh\~ao}}, \bibinfo {author} {\bibfnamefont {A.~M.}\ \bibnamefont
  {Souza}}, \bibinfo {author} {\bibfnamefont {L.}~\bibnamefont {Mazzola}},
  \bibinfo {author} {\bibfnamefont {R.}~\bibnamefont {Auccaise}}, \bibinfo
  {author} {\bibfnamefont {R.~S.}\ \bibnamefont {Sarthour}}, \bibinfo {author}
  {\bibfnamefont {I.~S.}\ \bibnamefont {Oliveira}}, \bibinfo {author}
  {\bibfnamefont {J.}~\bibnamefont {Goold}}, \bibinfo {author} {\bibfnamefont
  {G.}~\bibnamefont {De~Chiara}}, \bibinfo {author} {\bibfnamefont
  {M.}~\bibnamefont {Paternostro}}, \ and\ \bibinfo {author} {\bibfnamefont
  {R.~M.}\ \bibnamefont {Serra}},\ }\href {\doibase
  10.1103/PhysRevLett.113.140601} {\bibfield  {journal} {\bibinfo  {journal}
  {Phys. Rev. Lett.}\ }\textbf {\bibinfo {volume} {113}},\ \bibinfo {pages}
  {140601} (\bibinfo {year} {2014})}\BibitemShut {NoStop}%
\bibitem [{\citenamefont {Pekola}(2015)}]{Pekola2015}%
  \BibitemOpen
  \bibfield  {author} {\bibinfo {author} {\bibfnamefont {J.~P.}\ \bibnamefont
  {Pekola}},\ }\href {\doibase 10.1038/nphys3169} {\bibfield  {journal}
  {\bibinfo  {journal} {Nature Physics}\ }\textbf {\bibinfo {volume} {11}},\
  \bibinfo {pages} {118} (\bibinfo {year} {2015})}\BibitemShut {NoStop}%
\bibitem [{\citenamefont {Cottet}\ and\ \citenamefont
  {Huard}(2018)}]{Cottet2018}%
  \BibitemOpen
  \bibfield  {author} {\bibinfo {author} {\bibfnamefont {N.}~\bibnamefont
  {Cottet}}\ and\ \bibinfo {author} {\bibfnamefont {B.}~\bibnamefont {Huard}},\
  }in\ \href {\doibase 10.1007/978-3-319-99046-0_40} {\emph {\bibinfo
  {booktitle} {Fundamental Theories of Physics}}}\ (\bibinfo  {publisher}
  {Springer International Publishing},\ \bibinfo {year} {2018})\ pp.\ \bibinfo
  {pages} {959--981}\BibitemShut {NoStop}%
\bibitem [{\citenamefont {Naghiloo}\ \emph {et~al.}(2018)\citenamefont
  {Naghiloo}, \citenamefont {Alonso}, \citenamefont {Romito}, \citenamefont
  {Lutz},\ and\ \citenamefont {Murch}}]{Naghiloo2018}%
  \BibitemOpen
  \bibfield  {author} {\bibinfo {author} {\bibfnamefont {M.}~\bibnamefont
  {Naghiloo}}, \bibinfo {author} {\bibfnamefont {J.~J.}\ \bibnamefont
  {Alonso}}, \bibinfo {author} {\bibfnamefont {A.}~\bibnamefont {Romito}},
  \bibinfo {author} {\bibfnamefont {E.}~\bibnamefont {Lutz}}, \ and\ \bibinfo
  {author} {\bibfnamefont {K.~W.}\ \bibnamefont {Murch}},\ }\href {\doibase
  10.1103/PhysRevLett.121.030604} {\bibfield  {journal} {\bibinfo  {journal}
  {Phys. Rev. Lett.}\ }\textbf {\bibinfo {volume} {121}},\ \bibinfo {pages}
  {030604} (\bibinfo {year} {2018})}\BibitemShut {NoStop}%
\bibitem [{\citenamefont {Brand\~ao}\ \emph {et~al.}(2013)\citenamefont
  {Brand\~ao}, \citenamefont {Horodecki}, \citenamefont {Oppenheim},
  \citenamefont {Renes},\ and\ \citenamefont {Spekkens}}]{Brandao2013}%
  \BibitemOpen
  \bibfield  {author} {\bibinfo {author} {\bibfnamefont {F.~G. S.~L.}\
  \bibnamefont {Brand\~ao}}, \bibinfo {author} {\bibfnamefont {M.}~\bibnamefont
  {Horodecki}}, \bibinfo {author} {\bibfnamefont {J.}~\bibnamefont
  {Oppenheim}}, \bibinfo {author} {\bibfnamefont {J.~M.}\ \bibnamefont
  {Renes}}, \ and\ \bibinfo {author} {\bibfnamefont {R.~W.}\ \bibnamefont
  {Spekkens}},\ }\href {\doibase 10.1103/PhysRevLett.111.250404} {\bibfield
  {journal} {\bibinfo  {journal} {Phys. Rev. Lett.}\ }\textbf {\bibinfo
  {volume} {111}},\ \bibinfo {pages} {250404} (\bibinfo {year}
  {2013})}\BibitemShut {NoStop}%
\bibitem [{\citenamefont {Korzekwa}\ \emph {et~al.}(2019)\citenamefont
  {Korzekwa}, \citenamefont {Chubb},\ and\ \citenamefont
  {Tomamichel}}]{Korzekwa2019}%
  \BibitemOpen
  \bibfield  {author} {\bibinfo {author} {\bibfnamefont {K.}~\bibnamefont
  {Korzekwa}}, \bibinfo {author} {\bibfnamefont {C.~T.}\ \bibnamefont {Chubb}},
  \ and\ \bibinfo {author} {\bibfnamefont {M.}~\bibnamefont {Tomamichel}},\
  }\href {\doibase 10.1103/PhysRevLett.122.110403} {\bibfield  {journal}
  {\bibinfo  {journal} {Phys. Rev. Lett.}\ }\textbf {\bibinfo {volume} {122}},\
  \bibinfo {pages} {110403} (\bibinfo {year} {2019})}\BibitemShut {NoStop}%
\bibitem [{\citenamefont {Silva}(2008)}]{Silva2008}%
  \BibitemOpen
  \bibfield  {author} {\bibinfo {author} {\bibfnamefont {A.}~\bibnamefont
  {Silva}},\ }\href {\doibase 10.1103/PhysRevLett.101.120603} {\bibfield
  {journal} {\bibinfo  {journal} {Phys. Rev. Lett.}\ }\textbf {\bibinfo
  {volume} {101}},\ \bibinfo {pages} {120603} (\bibinfo {year}
  {2008})}\BibitemShut {NoStop}%
\bibitem [{\citenamefont {Dorner}\ \emph {et~al.}(2012)\citenamefont {Dorner},
  \citenamefont {Goold}, \citenamefont {Cormick}, \citenamefont {Paternostro},\
  and\ \citenamefont {Vedral}}]{Dorner2012}%
  \BibitemOpen
  \bibfield  {author} {\bibinfo {author} {\bibfnamefont {R.}~\bibnamefont
  {Dorner}}, \bibinfo {author} {\bibfnamefont {J.}~\bibnamefont {Goold}},
  \bibinfo {author} {\bibfnamefont {C.}~\bibnamefont {Cormick}}, \bibinfo
  {author} {\bibfnamefont {M.}~\bibnamefont {Paternostro}}, \ and\ \bibinfo
  {author} {\bibfnamefont {V.}~\bibnamefont {Vedral}},\ }\href {\doibase
  10.1103/PhysRevLett.109.160601} {\bibfield  {journal} {\bibinfo  {journal}
  {Phys. Rev. Lett.}\ }\textbf {\bibinfo {volume} {109}},\ \bibinfo {pages}
  {160601} (\bibinfo {year} {2012})}\BibitemShut {NoStop}%
\bibitem [{\citenamefont {Gambassi}\ and\ \citenamefont
  {Silva}(2012)}]{Gambassi2012}%
  \BibitemOpen
  \bibfield  {author} {\bibinfo {author} {\bibfnamefont {A.}~\bibnamefont
  {Gambassi}}\ and\ \bibinfo {author} {\bibfnamefont {A.}~\bibnamefont
  {Silva}},\ }\href {\doibase 10.1103/PhysRevLett.109.250602} {\bibfield
  {journal} {\bibinfo  {journal} {Phys. Rev. Lett.}\ }\textbf {\bibinfo
  {volume} {109}},\ \bibinfo {pages} {250602} (\bibinfo {year}
  {2012})}\BibitemShut {NoStop}%
\bibitem [{\citenamefont {Fusco}\ \emph {et~al.}(2014)\citenamefont {Fusco},
  \citenamefont {Pigeon}, \citenamefont {Apollaro}, \citenamefont {Xuereb},
  \citenamefont {Mazzola}, \citenamefont {Campisi}, \citenamefont {Ferraro},
  \citenamefont {Paternostro},\ and\ \citenamefont {De~Chiara}}]{Fusco2014}%
  \BibitemOpen
  \bibfield  {author} {\bibinfo {author} {\bibfnamefont {L.}~\bibnamefont
  {Fusco}}, \bibinfo {author} {\bibfnamefont {S.}~\bibnamefont {Pigeon}},
  \bibinfo {author} {\bibfnamefont {T.~J.~G.}\ \bibnamefont {Apollaro}},
  \bibinfo {author} {\bibfnamefont {A.}~\bibnamefont {Xuereb}}, \bibinfo
  {author} {\bibfnamefont {L.}~\bibnamefont {Mazzola}}, \bibinfo {author}
  {\bibfnamefont {M.}~\bibnamefont {Campisi}}, \bibinfo {author} {\bibfnamefont
  {A.}~\bibnamefont {Ferraro}}, \bibinfo {author} {\bibfnamefont
  {M.}~\bibnamefont {Paternostro}}, \ and\ \bibinfo {author} {\bibfnamefont
  {G.}~\bibnamefont {De~Chiara}},\ }\href {\doibase 10.1103/PhysRevX.4.031029}
  {\bibfield  {journal} {\bibinfo  {journal} {Phys. Rev. X}\ }\textbf {\bibinfo
  {volume} {4}},\ \bibinfo {pages} {031029} (\bibinfo {year}
  {2014})}\BibitemShut {NoStop}%
\bibitem [{\citenamefont {Gong}\ \emph {et~al.}(2014)\citenamefont {Gong},
  \citenamefont {Deffner},\ and\ \citenamefont {Quan}}]{Zongping2014}%
  \BibitemOpen
  \bibfield  {author} {\bibinfo {author} {\bibfnamefont {Z.}~\bibnamefont
  {Gong}}, \bibinfo {author} {\bibfnamefont {S.}~\bibnamefont {Deffner}}, \
  and\ \bibinfo {author} {\bibfnamefont {H.~T.}\ \bibnamefont {Quan}},\ }\href
  {\doibase 10.1103/PhysRevE.90.062121} {\bibfield  {journal} {\bibinfo
  {journal} {Phys. Rev. E}\ }\textbf {\bibinfo {volume} {90}},\ \bibinfo
  {pages} {062121} (\bibinfo {year} {2014})}\BibitemShut {NoStop}%
\bibitem [{\citenamefont {Goold}\ \emph {et~al.}(2018)\citenamefont {Goold},
  \citenamefont {Plastina}, \citenamefont {Gambassi},\ and\ \citenamefont
  {Silva}}]{Goold2018}%
  \BibitemOpen
  \bibfield  {author} {\bibinfo {author} {\bibfnamefont {J.}~\bibnamefont
  {Goold}}, \bibinfo {author} {\bibfnamefont {F.}~\bibnamefont {Plastina}},
  \bibinfo {author} {\bibfnamefont {A.}~\bibnamefont {Gambassi}}, \ and\
  \bibinfo {author} {\bibfnamefont {A.}~\bibnamefont {Silva}},\ }in\ \href
  {\doibase 10.1007/978-3-319-99046-0_13} {\emph {\bibinfo {booktitle}
  {Fundamental Theories of Physics}}}\ (\bibinfo  {publisher} {Springer
  International Publishing},\ \bibinfo {year} {2018})\ pp.\ \bibinfo {pages}
  {317--336}\BibitemShut {NoStop}%
\bibitem [{\citenamefont {Wang}\ \emph {et~al.}(2018)\citenamefont {Wang},
  \citenamefont {Zhang},\ and\ \citenamefont {Quan}}]{Wang2018}%
  \BibitemOpen
  \bibfield  {author} {\bibinfo {author} {\bibfnamefont {B.}~\bibnamefont
  {Wang}}, \bibinfo {author} {\bibfnamefont {J.}~\bibnamefont {Zhang}}, \ and\
  \bibinfo {author} {\bibfnamefont {H.~T.}\ \bibnamefont {Quan}},\ }\href
  {\doibase 10.1103/PhysRevE.97.052136} {\bibfield  {journal} {\bibinfo
  {journal} {Phys. Rev. E}\ }\textbf {\bibinfo {volume} {97}},\ \bibinfo
  {pages} {052136} (\bibinfo {year} {2018})}\BibitemShut {NoStop}%
\bibitem [{\citenamefont {Zawadzki}\ \emph {et~al.}(2019)\citenamefont
  {Zawadzki}, \citenamefont {Serra},\ and\ \citenamefont
  {D'Amico}}]{zawadzki2019work}%
  \BibitemOpen
  \bibfield  {author} {\bibinfo {author} {\bibfnamefont {K.}~\bibnamefont
  {Zawadzki}}, \bibinfo {author} {\bibfnamefont {R.~M.}\ \bibnamefont {Serra}},
  \ and\ \bibinfo {author} {\bibfnamefont {I.}~\bibnamefont {D'Amico}},\ }\href
  {http://arxiv.org/abs/1908.06488} {\bibfield  {journal} {\bibinfo  {journal}
  {arXiv:1908.06488}\ } (\bibinfo {year} {2019})}\BibitemShut {NoStop}%
\bibitem [{\citenamefont {Arrais}\ \emph {et~al.}(2019)\citenamefont {Arrais},
  \citenamefont {Wisniacki}, \citenamefont {Roncaglia},\ and\ \citenamefont
  {Toscano}}]{arrais2019work}%
  \BibitemOpen
  \bibfield  {author} {\bibinfo {author} {\bibfnamefont {E.~G.}\ \bibnamefont
  {Arrais}}, \bibinfo {author} {\bibfnamefont {D.~A.}\ \bibnamefont
  {Wisniacki}}, \bibinfo {author} {\bibfnamefont {A.~J.}\ \bibnamefont
  {Roncaglia}}, \ and\ \bibinfo {author} {\bibfnamefont {F.}~\bibnamefont
  {Toscano}},\ }\href {http://arxiv.org/abs/1907.06285} {\bibfield  {journal}
  {\bibinfo  {journal} {arXiv:1907.06285}\ } (\bibinfo {year}
  {2019})}\BibitemShut {NoStop}%
\bibitem [{\citenamefont {Horowitz}\ and\ \citenamefont
  {Parrondo}(2013)}]{Horowitz2013a}%
  \BibitemOpen
  \bibfield  {author} {\bibinfo {author} {\bibfnamefont {J.~M.}\ \bibnamefont
  {Horowitz}}\ and\ \bibinfo {author} {\bibfnamefont {J.~M.~R.}\ \bibnamefont
  {Parrondo}},\ }\href {\doibase 10.1088/1367-2630/15/8/085028} {\bibfield
  {journal} {\bibinfo  {journal} {N. J. Phys}\ }\textbf {\bibinfo {volume}
  {15}},\ \bibinfo {pages} {085028} (\bibinfo {year} {2013})}\BibitemShut
  {NoStop}%
\bibitem [{\citenamefont {Manzano}\ \emph {et~al.}(2018)\citenamefont
  {Manzano}, \citenamefont {Horowitz},\ and\ \citenamefont
  {Parrondo}}]{Manzano2018}%
  \BibitemOpen
  \bibfield  {author} {\bibinfo {author} {\bibfnamefont {G.}~\bibnamefont
  {Manzano}}, \bibinfo {author} {\bibfnamefont {J.~M.}\ \bibnamefont
  {Horowitz}}, \ and\ \bibinfo {author} {\bibfnamefont {J.~M.~R.}\ \bibnamefont
  {Parrondo}},\ }\href {\doibase 10.1103/PhysRevX.8.031037} {\bibfield
  {journal} {\bibinfo  {journal} {Phys. Rev. X}\ }\textbf {\bibinfo {volume}
  {8}},\ \bibinfo {pages} {31037} (\bibinfo {year} {2018})}\BibitemShut
  {NoStop}%
\bibitem [{\citenamefont {Suomela}\ \emph {et~al.}(2014)\citenamefont
  {Suomela}, \citenamefont {Solinas}, \citenamefont {Pekola}, \citenamefont
  {Ankerhold},\ and\ \citenamefont
  {{Ala-Nissila}}}]{suomelaMomentsWorkTwopoint2014a}%
  \BibitemOpen
  \bibfield  {author} {\bibinfo {author} {\bibfnamefont {S.}~\bibnamefont
  {Suomela}}, \bibinfo {author} {\bibfnamefont {P.}~\bibnamefont {Solinas}},
  \bibinfo {author} {\bibfnamefont {J.~P.}\ \bibnamefont {Pekola}}, \bibinfo
  {author} {\bibfnamefont {J.}~\bibnamefont {Ankerhold}}, \ and\ \bibinfo
  {author} {\bibfnamefont {T.}~\bibnamefont {{Ala-Nissila}}},\ }\href {\doibase
  10.1103/PhysRevB.90.094304} {\bibfield  {journal} {\bibinfo  {journal}
  {Physical Review B}\ }\textbf {\bibinfo {volume} {90}} (\bibinfo {year}
  {2014}),\ 10.1103/PhysRevB.90.094304}\BibitemShut {NoStop}%
\bibitem [{\citenamefont {Hiai}\ and\ \citenamefont
  {Petz}(2014)}]{hiaiIntroductionMatrixAnalysis2014a}%
  \BibitemOpen
  \bibfield  {author} {\bibinfo {author} {\bibfnamefont {F.}~\bibnamefont
  {Hiai}}\ and\ \bibinfo {author} {\bibfnamefont {D.}~\bibnamefont {Petz}},\
  }\href@noop {} {\emph {\bibinfo {title} {Introduction to Matrix Analysis and
  Applications}}},\ \bibinfo {series} {Texts and Readings in Mathematics}\
  No.~\bibinfo {number} {70}\ (\bibinfo  {publisher} {{Hindustan Book
  Agency}},\ \bibinfo {address} {{New Dehli}},\ \bibinfo {year}
  {2014})\BibitemShut {NoStop}%
\bibitem [{\citenamefont
  {Scandi}(2018)}]{scandiQuantifyingDissipationThermodynamic2018}%
  \BibitemOpen
  \bibfield  {author} {\bibinfo {author} {\bibfnamefont {M.}~\bibnamefont
  {Scandi}},\ }\href@noop {} {\emph {\bibinfo {title} {Quantifying
  {{Dissipation}} via {{Thermodynamic Length}}}}}\ (\bibinfo  {publisher}
  {{Unpublished master's thesis}},\ \bibinfo {year} {2018})\BibitemShut
  {NoStop}%
\bibitem [{\citenamefont
  {Jarzynski}(2011{\natexlab{b}})}]{jarzynskiEqualitiesInequalitiesIrreversibility2011a}%
  \BibitemOpen
  \bibfield  {author} {\bibinfo {author} {\bibfnamefont {C.}~\bibnamefont
  {Jarzynski}},\ }\href {\doibase 10.1146/annurev-conmatphys-062910-140506}
  {\bibfield  {journal} {\bibinfo  {journal} {Annual Review of Condensed Matter
  Physics}\ }\textbf {\bibinfo {volume} {2}},\ \bibinfo {pages} {329} (\bibinfo
  {year} {2011}{\natexlab{b}})}\BibitemShut {NoStop}%
\end{thebibliography}%

 \newpage~
 \newpage
 \appendix
 
 \onecolumngrid
 
 \section{Work probability in the two point measurement scheme }\label{app:TPM}
 In this section we explain how one can arrive to the expression in Eq.~\eqref{eq:CGF1II} for the cumulant generating function of the work starting from the definition of the work probability in the two point measurement scheme (TPM) for discrete processes.
 
 In order to obtain the work output after $N$ steps, we would need to do a convolution between the probability distributions at each step, which is clearly an untreatable task. On the other hand, though, the cumulant generating function for the sum of independent variables factorises in the sum of the CGF of each variable. For this reason, it is useful to introduce the CGF for the full work as:
 \begin{align}
 K^{-\beta w}(\lambda)&:= \log\int_{-\infty}^{\infty} \de w\, p(w) e^{-\beta\lambda w} = \sum_{n=1}^N \log\int_{-\infty}^{\infty} \de w_n\, p(w_n) e^{-\beta\lambda w_n}.
 \end{align}
 Plugging in the definition of the probability of the work given in Eq.~\eqref{eq:TPMA1mt}, we then get:
 \begin{align}
 K^{-\beta w}(\lambda)&= \sum_{n=1}^{N-1}  \log\int_{-\infty}^{\infty} \de w_n\, p(w_n) e^{-\beta\lambda w_n} =\nonumber\\
 &=  \sum_{n=1}^{N-1}  \log\int_{-\infty}^{\infty} \de w_n\, \sum_{E^{(i)}_{n+1} -E^{(j)}_{n}  = w_n} \Tr\sqrbra{e^{-\beta\lambda E^{(i)}_{n+1} }\ket{E^{(i)}_{n+1} }\bra{E^{(i)}_{n+1}}  \ket{E^{(j)}_{n} }\bra{E^{(j)}_{n}} e^{\beta\lambda E^{(j)}_{n}} \varrho^{(j)}_n} = \nonumber\\
 &=   \sum_{n=1}^{N-1}  \log \bigg(\Tr\sqrbra{e^{-\beta\lambda H_{n+1}}e^{\beta\lambda H_{n}}\varrho_{n}}\bigg).
 \end{align}
 where we obtained the last equation by a resummation of the spectral decomposition of the exponential operators {\ms and we assumed $\rho_n$ to be diagonal in the eigenbasis of $H_n$}. This concludes the derivation of Eq.~\eqref{eq:CGF1II}.
 
 \section{Average and fluctuations of dissipated work in the slow driving limit }\label{app:FDR} 
 In this section we give the derivation of the slow driving approximation to average dissipated work and work fluctuations. The results of this appendix were already presented in~\cite{scandiThermodynamicLengthOpen2019, millerWorkFluctuationsSlow2019}, and are reproduced here due to their prototypical nature, which will be encountered in most of the derivations of the paper.
 
 We first derive Eq.~\eqref{eq:averageDiss} and Eq.~\eqref{eq:averageFluct} from the cumulant generating function, using the definition of cumulants Eq.~\eqref{eq:cumulants}. For what regards the average dissipated work we obtain:
 \begin{align}
 \beta\,\average{w_{\rm diss}}{}  & = - \frac{\de}{\de\lambda}\,K^{\text{diss}}(\lambda)\Big|_{\lambda=0} =- \sum_{i=1}^N\frac{\de}{\de\lambda}\log\Tr\sqrbra{\pi_{i+1}^\lambda \pi_i^{1-\lambda} } \Big|_{\lambda=0}= \nonumber\\
 &= - \sum_{i=1}^N \frac{\Tr\sqrbra{\pi_{i+1}^\lambda(\log \pi_{i+1}- \log \pi_i) \pi_i^{1-\lambda} }}{\Tr\sqrbra{\pi_{i+1}^\lambda \pi_i^{1-\lambda} }}\Big|_{\lambda=0} = \sum_{i=1}^N \Tr\sqrbra{\pi_{i}(\log \pi_{i}- \log \pi_{i+1})};\label{eq:A1}
 \end{align}
 in a similar fashion, the work fluctuations can be obtained as:
 \begin{align}
 \beta^2\, \sigma^2_{\rm diss} &=  \frac{\de^2}{\de\lambda^2}\,K^{\text{diss}}(\lambda)\Big|_{\lambda=0} = \sum_{i=1}^N\frac{\de}{\de\lambda} \frac{\Tr\sqrbra{\pi_{i+1}^\lambda(\log \pi_{i+1}- \log \pi_i) \pi_i^{1-\lambda} }}{\Tr\sqrbra{\pi_{i+1}^\lambda \pi_i^{1-\lambda} }}\Big|_{\lambda=0}=\nonumber\\
 &=\sum_{i=1}^N \Tr\sqrbra{\pi_{i}(\log \pi_{i}- \log \pi_{i+1})^2} - \Tr\sqrbra{\pi_{i}(\log \pi_{i}- \log \pi_{i+1})}^2.\label{eq:A2}
 \end{align}
 We can recognise in Eq.~\eqref{eq:A1} and Eq.~\eqref{eq:A2} the definition of relative entropy and relative entropy variance given in the main text. 
 
 Before passing to derive the slow driving limit of the above expression, it is useful to explain how to Taylor expand $\pi_{i+1}$ around $\pi_{i}$ for small variations of the Hamiltonian $\Delta H_i := (H_{i+1} - H_i)$. Using the Dyson series for the exponential operators we obtain~\cite{hiaiIntroductionMatrixAnalysis2014a, scandiQuantifyingDissipationThermodynamic2018}:
 \begin{align}
 e^{-\beta (H + \Delta H)} = e^{-\beta H} -\beta \int^1_0 \de x \, e^{-\beta (1-x) H} \Delta H e^{-\beta x H} +\bigo{\Delta H^2}.\label{eq:A3}
 \end{align}
 One can recognise in the expansion the operator $\J_{e^{-\beta H}}$ defined in the main text (Eq.~\eqref{eq:operatorJ}). Plugging Eq.~\eqref{eq:A3} in the definition of {\ms the} partition function, one can also prove the equality: $\mathcal{Z}(H+\Delta H) = \mathcal{Z}(H) -\beta\, \Tr\sqrbra{\Delta H e^{-\beta H}}+ \bigo{\Delta H^2}$, where we used the cyclicity of the trace. Then, it is straightforward to see that the Gibbs state $\pi_{i+1}$ can be expanded as:
 \begin{align}\label{eq:B4}
 \pi_{i+1}  = \pi_i -\beta\, \J_{\pi_i}[\Delta_{\pi_i} (\Delta H_i)] + \bigo{\Delta H_i^2},
 \end{align}
 where the average in the operator $\Delta_\varrho A := (A -\Tr\sqrbra{\varrho A})$ comes from the expansion of the partition function in the denominator of the Gibbs state. 
 
 Using matrix analysis one can also show that~\cite{hiaiIntroductionMatrixAnalysis2014a, scandiQuantifyingDissipationThermodynamic2018}:
 \begin{align}\label{eq:Brelativeentropy}
 S(\varrho || \varrho +\varepsilon \sigma) = \Tr\sqrbra{\varrho (\log \varrho - \log(\varrho +\varepsilon \sigma))} = \frac{\varepsilon^2}{2} \,\Tr\sqrbra{\sigma \,\J_\varrho^{-1}[\sigma]} + \bigo{\epsilon^3},
 \end{align}
 where $\varrho$ is definite positive, $\sigma$ is a traceless perturbation, and $\J_\varrho^{-1}$ is the inverse superoperator of the one appearing in the Dyson series, which arises from the Taylor expansion of the logarithm. Plugging $\pi_{i+1}$ in this expression we then obtain:
 \begin{align}
 \beta\,\average{w_{\rm diss}}{}  &=\sum_{i=1}^{N-1}  S(\pi_i || \pi_{i+1})=\sum_{i=1}^{N-1}  S(\pi_i || \pi_i -\beta\, \J_{\pi_i}[\Delta_{\pi_i} (\Delta H_i)])=
 \nonumber\\
 &= \sum_{i=1}^{N-1}  \Tr\sqrbra{\J_{\pi_i}[\Delta_{\pi_i} (\Delta H_i)]\,  \J_{\pi_i}^{-1}[\J_{\pi_i}[\Delta_{\pi_i} (\Delta H_i)] ]} = \frac{\beta^2}{2}  \sum_{i=1}^{N-1}   \Tr\sqrbra{\Delta H_i \J_{\pi_i}[\Delta_{\pi_i} (\Delta H_i)]},\label{eq:A6}
 \end{align}
 which holds up to corrections of order of $\bigo{\Delta H_i^2}=\bigo{N^{-2}} $. 
 In the limit $N\gg1$, if the change in $H_i$ is smooth enough, one can define an interpolating curve $H_t$ and convert all the sums in integrals. For this reason we use the substitution $\Delta H_i \rightarrow \frac{1}{N} \dot H_t$, so to rewrite Eq.~\eqref{eq:A6} as:
 \begin{align}
 \average{w_{\rm diss}}{} = \frac{\beta}{2N}  \sum_{i=1}^{N-1}  \frac{1}{N} \Tr\sqrbra{\dot H_i \J_{\pi_i}[\Delta_{\pi_i} (\dot H_i)]} \stackrel{N\gg 1}{\longrightarrow} \frac{\beta}{2 N}\int_\gamma \dt \Tr\sqrbra{\dot H_t\, \J_{\pi_t} [\Delta_{\pi_t}\dot H_t]},
 \end{align}
 where we used the definition of Riemann sum. This concludes the derivation of Eq.~\eqref{eq:avWork}.
 
 We can now pass to expand Eq.~\eqref{eq:A2}. Since we want only term up to order $\bigo{1/N^2}$ we can ignore the second part of the sum. Then the expansion takes the particularly simple form:
 \begin{align}
 \beta^2\, \sigma^2_{\rm diss} &= \sum_{i=1}^{N-1} \Tr\sqrbra{\pi_{i}(\log \pi_{i}- \log( \pi_i -\beta\, \J_{\pi_i}[\Delta_{\pi_i} (\Delta H_i)]))^2} + \bigo{\frac{1}{N^3}}= \nonumber\\&=\beta^2 \sum_{i=1}^{N-1} \Tr\sqrbra{\pi_{i}(\J_{\pi_i}^{-1}[\J_{\pi_i}[\Delta_{\pi_{i}} \Delta H_i] ])^2} = \frac{\beta^2}{N} \sum_{i=1}^{N-1}\frac{1}{N} \Tr\sqrbra{\pi_{i}(\Delta_{\pi_{i}} \dot H_i )^2},\label{eq:A8}
 \end{align}
 where to pass from the first to the second line, we used the Taylor expansion of the logarithm~\cite{hiaiIntroductionMatrixAnalysis2014a}. Taking the continuous limit completes the derivation of Eq.~\eqref{eq:fluctuations}, which also holds up to order $\bigo{1/N^2}$. 
 
 \section{Expansion of $\lambda$-R\'enyi divergence}\label{app:Renyi}
 
 In this section we prove the following expansion for the $\lambda$-R\'enyi divergence:
 \begin{align}\label{eq:appCRenyiexpansion}
 S_\lambda(\varrho+\varepsilon\sigma|| \varrho) = -\frac{\varepsilon^2}{2(\lambda -1)}\int_0^\lambda {\rm d}x \int_x^{1-x}{\rm d}y \, \text{cov}_\varrho^y(\J_\varrho^{-1}[\sigma], \J_\varrho^{-1}[\sigma]) +\bigo{\varepsilon^3},
 \end{align}
 where $\sigma$ is a traceless perturbation, $\J_{\varrho}^{-1}$ is the inverse of the operator in Eq.~\eqref{eq:operatorJ}, and we defined the $y$-covariance as:
 \begin{align}
 \text{cov}_\varrho^y(A, B) := \Tr\sqrbra{\varrho^{1-y}A \varrho^{y} B} - \Tr\sqrbra{A \varrho}\Tr\sqrbra{B \varrho}.
 \end{align}
 Due to the technical nature of the derivation, the beginning and the end of the proof are clearly marked.
 
 \begin{proof}
 	The two main ingredients to prove Eq.~\eqref{eq:appCRenyiexpansion} are the Dyson series of the exponential~\cite{hiaiIntroductionMatrixAnalysis2014a}:
 	\begin{align}\label{eq:C3}
 	e^{-A}e^{A + \varepsilon B} = \id + \varepsilon\int^1_0 \de x \, e^{-A x} \, B  \, \varrho^{A x} +\bigo{\varepsilon^2} = \id + \varepsilon\, e^{-A}\J_{e^{A}}[B]  +\bigo{\varepsilon^2} ,
 	\end{align}
 	and the Taylor expansion of the logarithm around the positive definite matrix $\varrho$~\cite{scandiQuantifyingDissipationThermodynamic2018}:
 	\begin{align}\label{eq:C4}
 	\log (\varrho + \varepsilon\sigma) = \log \varrho + \varepsilon\,\J_{\varrho}^{-1}[\sigma] - \varepsilon^2 \,\J_{\varrho}^{-1}\sqrbra{\int_0^1\de x_1 \int_0^{x_1}\de x_2 \, \varrho^{1-x_1}\J_{\varrho}^{-1}[\sigma]\varrho^{x_1-x_2}\J_{\varrho}^{-1}[\sigma]\varrho^{x_2}}.
 	\end{align}
 	Moreover, it is useful to rewrite the $\lambda$-R\'enyi divergence as:
 	\begin{align}\label{eq:C5R}
 	S_\lambda(\varrho || \sigma) =  \frac{1}{\lambda -1} \log \Tr\sqrbra{\varrho^\lambda \sigma^{1-\lambda}} =\frac{1}{\lambda-1} \log \sqrbra{1 + \int_0^\lambda \de x\, \Tr[\varrho^{x} (\log \varrho -\log\sigma) \sigma^{1-x}]},
 	\end{align}
 	where we simply differentiated the trace and integrated again, and we used the fact that $\Tr\sqrbra{\varrho^\lambda \sigma^{1-\lambda}}|_{\lambda=0} = 1$. This expression highlights the dependence of the $\lambda$-R\'enyi divergence  on the difference between $\varrho$ and $\sigma$.
 	
 	Passing to the expansion of $S_\lambda(\varrho+\varepsilon\sigma|| \varrho) $, we first focus on the trace inside of the logarithm in Eq.~\eqref{eq:C5R}:
 	\begin{align}
 	&\Tr[\varrho\, \varrho^{-x}(\varrho+\varepsilon\sigma)^{x} (\log (\varrho + \varepsilon \sigma)-\log \varrho)] =\nonumber\\
 	&= \Tr[\varrho\norbra{\id +\varepsilon\int^x_0 \de y \, \varrho^{-y} \, \J_{\varrho}^{-1}[\sigma]  \, \varrho^{y}} \norbra{ \varepsilon\,\J_{\varrho}^{-1}[\sigma] - \varepsilon^2 \,\J_{\varrho}^{-1}\sqrbra{\int_0^1\de x_1 \int_0^{x_1}\de x_2 \, \varrho^{1-x_1}\J_{\varrho}^{-1}[\sigma]\varrho^{x_1-x_2}\J_{\varrho}^{-1}[\sigma]\varrho^{x_2}}}].
 	\end{align}
 	The first order contribution is given by:
 	\begin{align}
 	\Tr[\varrho\,\J_{\varrho}^{-1}[\sigma]  ] = \Tr[\J_{\varrho}^{-1}[\varrho]\,\sigma]  = \Tr[\sigma] = 0,
 	\end{align}
 	where we used the hermiticity of $\J_{\varrho}^{-1}$ and the identity $\J_{\varrho}^{-1}[A] = \varrho^{-1}A $, whenever $[A, \varrho] = 0$. This last relation comes from the fact that the Fr\'echet derivative of the logarithm agrees with its usual derivative on the subspace of commutative operators.
 	
 	Passing to the study of the second order contribution, it should be noticed that one can use the change of variables $y \rightarrow1-y$ to obtain the relation:
 	\begin{align}
 	\int^x_0 \de y \, \Tr[\varrho^{1-y} \, \J_{\varrho}^{-1}[\sigma]  \, \varrho^{y}\J_{\varrho}^{-1}[\sigma]] = \frac{1}{2}\int^x_0 \de y \, \Tr[\varrho^{1-y} \, \J_{\varrho}^{-1}[\sigma]  \, \varrho^{y}\J_{\varrho}^{-1}[\sigma]] +\frac{1}{2} \int^1_{1-x} \de y \, \Tr[\varrho^{1-y} \, \J_{\varrho}^{-1}[\sigma]  \, \varrho^{y}\J_{\varrho}^{-1}[\sigma]].
 	\end{align}
 	Moreover, the presence of the trace allows us to simplify the double integration as:
 	\begin{align}
 	&\Tr[\varrho\J_{\varrho}^{-1}\sqrbra{\int_0^1\de x_1 \int_0^{x_1}\de x_2 \, \varrho^{1-x_1}\J_{\varrho}^{-1}[\sigma]\varrho^{x_1-x_2}\J_{\varrho}^{-1}[\sigma]\varrho^{x_2}}] =\int_0^1\de x_1\int_0^{x_1}\de x_2 \, \Tr[\varrho^{1-(x_1-x_2)}\J_{\varrho}^{-1}[\sigma]\varrho^{x_1-x_2}\J_{\varrho}^{-1}[\sigma]]=\nonumber\\
 	&=\int_0^1\de u\int_0^{u}\de v \, \Tr[\varrho^{1-v}\J_{\varrho}^{-1}[\sigma]\varrho^{v}\J_{\varrho}^{-1}[\sigma]]   = \int_0^1\de x\int_{x}^{1}\de y \, \Tr[\varrho^{1-y}\J_{\varrho}^{-1}[\sigma]\varrho^{y}\J_{\varrho}^{-1}[\sigma]],\label{eq:C8}
 	\end{align}
 	where passing from the first line to the second we used the substitution $u = x_1$ and $v = x_1-x_2$, and in the last equation the substitution $x = 1 - u$ and $y = 1 - v$. Adding together the last two equalities in Eq.~\eqref{eq:C8}, we obtain:
 	\begin{align}
 	\int_0^1\de x\int_0^{x}\de y \, \Tr[\varrho^{1-y}\J_{\varrho}^{-1}[\sigma]\varrho^{y}\J_{\varrho}^{-1}[\sigma]]  &= \frac{1}{2} \int_0^1\de x\int_0^{1}\de y \, \Tr[\varrho^{1-y}\J_{\varrho}^{-1}[\sigma]\varrho^{y}\J_{\varrho}^{-1}[\sigma]] =\nonumber\\
 	&=\frac{1}{2} \int_0^{1}\de y \, \Tr[\varrho^{1-y}\J_{\varrho}^{-1}[\sigma]\varrho^{y}\J_{\varrho}^{-1}[\sigma]] .
 	\end{align} 
 	Finally, we can pass to expand the logarithm which gives the result at second order:
 	\begin{align}
 	S_\lambda(\varrho+\varepsilon\sigma|| \varrho)  &= \frac{\varepsilon}{2(\lambda-1)} \int_0^\lambda {\rm d}x \norbra{\int_0^{x}{\rm d}y \, +\int_{1-x}^{1}{\rm d}y \, - \int_0^{1}{\rm d}y }\norbra{\Tr[\varrho^{1-y}\J_{\varrho}^{-1}[\sigma]\varrho^{y}\J_{\varrho}^{-1}[\sigma]] } =\nonumber\\
 	&=  -\frac{\varepsilon^2}{2(\lambda -1)}\int_0^\lambda {\rm d}x \int_x^{1-x}{\rm d}y \, \text{cov}_\varrho^y(\J_\varrho^{-1}[\sigma], \J_\varrho^{-1}[\sigma]),
 	\end{align}
 	which concludes the proof of  Eq.~\eqref{eq:appCRenyiexpansion}.
 \end{proof}
 
 \emph{Symmetry in the arguments}. The quadratic structure of Eq.~\eqref{eq:appCRenyiexpansion} hints at the approximate symmetry of the $\lambda$-R\'enyi divergence. In fact, using the identity $S_\lambda(\varrho|| \sigma) =\frac{-\lambda}{\lambda - 1} S_{1-\lambda}(\sigma || \varrho)$, we also obtain that:
 \begin{align}
 S_\lambda(\varrho|| \varrho +\varepsilon\sigma)\,&{\ms= \frac{-\lambda}{\lambda - 1} S_{1-\lambda}( \varrho +\varepsilon\sigma||\varrho) =}\nonumber\\ &=-\frac{\varepsilon^2}{2(\lambda -1)}\int_0^{1-\lambda} {\rm d}x \int_x^{1-x}{\rm d}y \, \text{cov}_\varrho^y(\J_\varrho^{-1}[\sigma], \J_\varrho^{-1}[\sigma]) =\nonumber\\ 
 &= -\frac{\varepsilon^2}{2(\lambda -1)}\norbra{\int_0^{1} {\rm d}x\int_x^{1-x}{\rm d}y \, \text{cov}_\varrho^y(\J_\varrho^{-1}[\sigma], \J_\varrho^{-1}[\sigma])+\int_1^{1-\lambda} {\rm d}x \int_x^{1-x}{\rm d}y \, \text{cov}_t^y(\J_\varrho^{-1}[\sigma], \J_\varrho^{-1}[\sigma])}=
 \nonumber\\ 
 &= -\frac{\varepsilon^2}{2(\lambda -1)}\int_0^{\lambda} {\rm d}x \int_x^{1-x}{\rm d}y \, \text{cov}_\varrho^y(\J_\varrho^{-1}[\sigma], \J_\varrho^{-1}[\sigma]) =S_\lambda(\varrho +\varepsilon\sigma|| \varrho ),
 \end{align}
 where the first integral in the second line goes to zero thanks to the symmetry of the $y $ integration bounds, and we performed the change of variables $x\rightarrow1-x$ in the second integral. This proves that the $\lambda$-R\'enyi divergence is symmetric at second order. 
 
 \emph{Relative entropy.} Moreover, it should be noticed that in the limit $\lambda\rightarrow1$ we regain the known expression of Eq.~\eqref{eq:Brelativeentropy} for the expansion of the relative entropy:
 \begin{align}
 S(\varrho|| \varrho +\varepsilon\sigma) &=\lim_{\lambda\rightarrow1}S_\lambda(\varrho|| \varrho +\varepsilon\sigma)  = \frac{\varepsilon^2}{2}\int_0^{1}{\rm d}y \, \text{cov}_\varrho^y(\J_\varrho^{-1}[\sigma], \J_\varrho^{-1}[\sigma])  =\nonumber\\
 &= \frac{\varepsilon^2}{2} \Tr\sqrbra{\J_\varrho^{-1}[\sigma] \, \J_\varrho[\J_\varrho^{-1}[\sigma]]} = \frac{\varepsilon^2}{2} \,\Tr\sqrbra{\sigma \,\J_\varrho^{-1}[\sigma]},
 \end{align}
 where in the first line we used L'Hospital's rule for limits of the form $0/0$, and we inserted in the last equation the definition of $\J_\varrho$.

 \section{Cumulant generating function in the slow driving limit}\label{app:CGF}
 
 In this section we will give the slow driving approximation to Eq.~\eqref{eq:CGFdissdef}, and we will further discuss the remarks made in sec.~\ref{sec:CGF}.
 
 \emph{Derivation and general results.} Starting from the expression of the dissipative CGF and applying the expansion in Eq.~\eqref{eq:appCRenyiexpansion} we obtain:
 \begin{align}
 K^{\text{diss}}(\lambda) =  \sum_{i=1}^{N-1} (\lambda-1)S_\lambda(\pi_{i+1} || \pi_{i}) = -\frac{1}{2} \sum_{i=1}^{N-1}  \int_0^\lambda {\rm d}x \int_x^{1-x}{\rm d}y \, \text{cov}_{\pi_i}^y(\J_{\pi_i}^{-1}[\bar{\Delta} H_i ], \J_{\pi_i}^{-1}[\bar{\Delta} H_i])  +\bigo{\bar{\Delta} H_i^3},
 \end{align}
 where we defined $\bar{\Delta} H_i := \pi_{i+1} -\pi_{i}$. As explained in Appendix~\ref{app:FDR}, in the limit in which the variation of $H_i$ is smooth, one can {\ms insert} the expansion of the Gibbs state  in Eq.~\eqref{eq:B4} to express $\bar{\Delta} H_i$, and consequently pass to the continuous limit using the definition of Riemann sum:
 \begin{align}
 K^{\text{diss}}(\lambda) &= -\frac{\beta^2}{2} \sum_{i=1}^{N-1}  \int_0^\lambda {\rm d}x \int_x^{1-x}{\rm d}y \, \text{cov}_{\pi_i}^y(\J_{\pi_i}^{-1}[\J_{\pi_i}[\Delta_{\pi_i} (\Delta H_i)]], \J_{\pi_i}^{-1}[\J_{\pi_i}[\Delta_{\pi_i} (\Delta H_i)]]) = \nonumber\\
 &= -\frac{\beta^2}{2 N} \int_\gamma  \int_0^\lambda {\rm d}x \int_x^{1-x}{\rm d}y \, \text{cov}_{t}^y(\dot H_t, \dot H_t) {\ms +\bigo{\frac{1}{N^2}}}.\label{eq:D2}
 \end{align}
 
 The dissipative CGF so obtained can be given in coordinates, in the instantaneous energy eigenbasis, as:
 \begin{align}
 K^{\text{diss}}(\lambda) = &\frac{\beta^2}{N}\norbra{\frac{\lambda^2-\lambda}{2}}\int_\gamma\sum_{i,j} \norbra{\pi_i |\dot H_{ii}|^2 -\pi_i\pi_j |\dot H_{ii}||\dot H_{jj}| }+\hspace{0.5 cm} \text{(commuting)}\\
 &+\frac{\beta^2}{2N}\int_\gamma\sum_{i>j} \frac{\pi_i^{\lambda}\pi_j^{1-\lambda}+\pi_j^{\lambda}\pi_i^{1-\lambda} - (\pi_i+ \pi_j)}{(\log \pi_j - \log \pi_i)^2} |\dot H_{ij}|^2,\hspace{0.5 cm} \text{(n.comm.)}\label{eq:D4}
 \end{align}
 where the subscript $t$ has been dropped for notation simplicity. This form makes particularly evident the fact that non Gaussian effects can arise solely from off diagonal terms in the driving. Indeed, defining the Wigner-Yanase-Dyson skew information:
 \begin{align}
 I^{y} (\varrho, L) :&= -\frac{1}{2}\Tr\sqrbra{[\varrho^{y}, L][\varrho^{1-y}, L]}=\nonumber\\
 &=\Tr\sqrbra{\varrho L^2}-\Tr\sqrbra{\varrho^y L \varrho^{1-y} L},\label{eq:D5}
 \end{align}
 (which provides a measure of the amount of information contained in $\varrho$ with respect to a non commuting observable $L$~\cite{wignerINFORMATIONCONTENTSDISTRIBUTIONS1963}), we can rewrite the cumulant generating function as:
 \begin{align}
 K^{\text{diss}}(\lambda) =&-\frac{\beta^2}{2N} \int_\gamma\int_0^\lambda {\rm d}x \int_x^{1-x}{\rm d}y \, \norbra{\Tr\sqrbra{\pi_t^{1-y}\dot H_t \pi_t^{y} \dot H_t} -\Tr\sqrbra{\dot H_t^2 \pi_t} + \Tr\sqrbra{\dot H_t^2 \pi_t} - \Tr\sqrbra{\dot H_t \pi_t}^2} =\nonumber\\
 =& -\frac{\beta^2}{2N} \int_\gamma\int_0^\lambda {\rm d}x \int_x^{1-x}{\rm d}y \norbra{\text{Var}_t[\dot H_t] - I^y(\pi_t, \dot H_t)} =\nonumber \\
 =&\,\frac{\beta^2(\lambda^2-\lambda)}{2N}\int_\gamma\text{Var}_t [\dot H_t] \;\;+ \;\; \frac{\beta^2}{2N} \int_\gamma\int_0^\lambda {\rm d}x \int_x^{1-x}{\rm d}y\,  I^y(\pi_t, \dot H_t),\label{eq:B20}
 \end{align}
 where in the last equation we explicitly carried out the $x$ and $y$ integration, {\ms thanks to the fact that} the variance does not depend on $y$. Since a Gaussian distribution has only a quadratic CGF, we see that the non Gaussian contribution can be linked in a precise manner with the lack of commutativity between $H_t$ and $\dot H_t$. This point will be further analysed in the next sections.
 
 We can now pass to verify some properties of the CGF so obtained. First, it is straightforward to check from Eq.~\eqref{eq:D2} that both normalisation ($K^{\text{diss}}(0) =\log \average{1}{} \equiv 0$) and Jarzynski equality ($K^{\text{diss}}(1) \equiv 0$)  are preserved by the approximation. The latter can be shown by explicitly writing the integral:
 \begin{align}\label{eq:D7}
 K^{\text{diss}}(1) &= \log \average{e^{-\beta (w-\Delta F)}}{} = -\frac{\beta^2}{2N} \int_\gamma\int_0^1 {\rm d}x \int_x^{1-x}{\rm d}y \, \text{cov}_t^y(\dot H_t,\dot H_t)\equiv 0 ,
 \end{align}
 and by noticing the fact that the $y$ integration bounds are antisymmetric around $x=1/2$,  while the $y$-covariance behaves as $\text{cov}^y_t(\dot H_t,\dot H_t) = \text{cov}_t^{1-y}(\dot H_t,\dot H_t)$. These results are trivial, since the original expression in Eq.~\eqref{eq:CGFdissdef} satisfies both conditions, but they work as a sanity check to guarantee that after the approximation we still have a probability distribution arising from a thermodynamic process.
 
 \emph{$y$-Covariance and KMS condition.} It is interesting to see how Eq.~\eqref{eq:D2} can be connected to linear response theory. Defining the Heisenberg picture for an operator $A$ by  $A^t(s) := e^{i H_t s} A e^{-i H_t s}$, and the thermal average as $\average{A}{\pi} := \Tr[\pi A]$, we can connect the  $y$-covariance with the autocorrelation function of the power operator $\dot H_t$ as:
 \begin{align}\label{eq:D82}
 \text{cov}_{t}^y(\dot H_t, \dot H_t) &= \Tr\sqrbra{\pi_t\,\pi_t^{-y} \dot H_t \pi_t^{y} \,\dot H_t} - \Tr\sqrbra{\pi_t \dot H_t} ^2= \Tr\sqrbra{\pi_t (\dot H_t -\langle \dot H_t\rangle_{\pi_{t}})\, e^{-\beta H_t y} \,(\dot H_t -\langle \dot H_t\rangle_{\pi_{t}})e^{\beta H_t y}} =\nonumber\\
 =&\average{\dot H_t(0) \dot H_t(i\beta y)}{\pi_t}-\average{\dot H_t(0)}{\pi_t}\average{ \dot H_t(i\beta y)}{\pi_t},
 \end{align}
 where for notational simplicity we suppressed the superscript in $\dot H_t^t$. This result is analogous in spirit to linear response theory~\cite{parisiStatisticalFieldTheory1988}. In fact, Eq.~\eqref{eq:D82} connects the work response arising from a linear perturbation with the connected two point correlation function of the power operator $\dot H_t$ {\ms evaluated in the equilibrium ensemble}. In this context, the KMS condition reads:
 \begin{align}
 \text{cov}_{t}^y(\dot H_t, \dot H_t) = \average{\Delta_{\pi_{t}}\dot H_t(0) \Delta_{\pi_{t}}\dot H_t(i\beta y)}{\pi_t} = \average{\Delta_{\pi_{t}}\dot H_t(0) \Delta_{\pi_{t}}\dot H_t(i\beta (1-y))}{\pi_t} =\text{cov}_{t}^{1-y}(\dot H_t, \dot H_t).
 \end{align}
 Since  we used this property to prove Eq.~\eqref{eq:D7}, this expression shows the close connection between the thermality of the state, encoded by the KMS condition, and Jarzynski equality.
 
 \emph{Entropy production rate and time reversal symmetry.}  The entropy production rate is connected with the breaking of time reversal symmetry between a trajectory and its inverse, as it is illustrated by fluctuation theorems~\cite{jarzynskiEqualitiesInequalitiesIrreversibility2011a}. We will show how this condition is encoded in the dissipative cumulant generating function. 
 
 Defining the time reversed protocol $\bar\pi_{i} := \pi_{N-i}$, we get the identity:
 \begin{align}\label{eq:D10}
 K^{\text{diss}}(\lambda) &=  \sum_{i=1}^{N-1} (\lambda-1)S_\lambda(\pi_{i+1} || \pi_{i})  = \sum_{i=1}^{N-1} {\ms \log}\,\Tr\sqrbra{\pi_{i+1}^{\lambda}  \pi_{i}^{1-\lambda}} = \nonumber\\ 
 &= \sum_{i=1}^{N-1} {\ms \log}\,\Tr\sqrbra{\bar\pi_{i}^{\lambda} \bar \pi_{i+1}^{1-\lambda}} =-\sum_{i=1}^{N-1} \lambda \,S_{1-\lambda}(\bar\pi_{i+1} || \bar\pi_{i}) = K^{\text{diss}}_{\text{rev}}(1-\lambda),
 \end{align}
 where we implicitly defined the time reversed CGF. The relation just obtained can be rewritten in terms of the probability distribution of the dissipated work (which, with an abuse of notation, we denote by $p(w)$):
 \begin{align}
 \int_{-\infty}^{\infty} \de w\, p(w) e^{-i \nu w} =\exp\norbra{K^{\text{diss}}\norbra{\frac{i\nu}{\beta}}} =  \exp\norbra{K^{\text{diss}}_{\text{rev}}\norbra{1-\frac{i\nu}{\beta}}}  = \int_{-\infty}^{\infty} \de w\, p_\text{rev}(w) e^{i \nu w}e^{-\beta w}.
 \end{align}
 Applying the inverse Fourier transform to the first and the last equation we obtain:
 \begin{align}
 p(w) &=\frac{1}{2\pi} \int {\rm d}\nu \int \de x\, e^{i \nu (w-x)} \, p(x) =\frac{1}{2\pi} \int {\rm d}\nu \int \de x\, e^{i \nu (w+x)} \,  p_\text{rev}(x) e^{-\beta x} =\nonumber \\
 &= \int \de x\, \delta(w+x) \,p_\text{rev}(x) e^{-\beta x} =p_\text{rev}(-w) e^{\beta w} .
 \end{align}
 In this way we see that Eq.~\eqref{eq:D10} is equivalent to the Crooks fluctuation relation~\cite{jarzynskiEqualitiesInequalitiesIrreversibility2011a}:
 \begin{align}\label{eq:D13}
 \frac{p(w)}{p_\text{rev}(-w)} = e^{\beta w},
 \end{align}
 which signals the fact that the ratio between the probability of dissipating an amount of work $w$ and the one of getting this work back by reversing the transformation is exponentially big in $w$. In this way, we can interpret Eq.~\eqref{eq:D13} as the underlying explanation for the emergence of the arrow of time.
 
 In the slow driving regime the CGF satisfies the condition:
 \begin{align}
 K^{\text{diss}}(\lambda)  &= -\frac{\beta^2}{2 N} \int_\gamma  \int_0^\lambda {\rm d}x \int_x^{1-x}{\rm d}y \, \text{cov}_{t}^y(\dot H_t, \dot H_t)  = -\frac{\beta^2}{2 N} \int_\gamma  \int_1^{1-\lambda} {\rm d}x \int_x^{1-x}{\rm d}y \, \text{cov}_{t}^y(\dot H_t, \dot H_t) =\nonumber\\
 &= -\frac{\beta^2}{2 N} \int_\gamma  \norbra{{\ms \int_0^{1-\lambda} {\rm d}x -\int_0^1 {\rm d}x} }\int_x^{1-x}{\rm d}y \, \text{cov}_{t}^y(\dot H_t, \dot H_t) =K^{\text{diss}}(1-\lambda) ,
 \end{align}
 where we first used the substitution $x\rightarrow1-x$, and then in the second line we applied Jarzynski equality to get rid of the {\ms second} integral in $\de x$. Comparing this relation with Eq.~\eqref{eq:D10}, we can deduce that the probability distribution for the entropy production during a protocol in the slow driving regime equals the one for its time reversed. This could also be inferred from the quadratic structure of the $y$-covariance, since it does not distinguish between $\dot H_t$ and $-\dot H_t$ . In this context, the Crooks relations becomes:
 \begin{align}
 \frac{p(w)}{p(-w)} = e^{\beta w},
 \end{align}
 also known as the Evans-Searles relations, which tell us that the probability of having a negative dissipation is exponentially suppressed.
 
 \section{Computation of the cumulants}\label{app:cumulants}
 In this section we explicitly derive a formula for all the cumulants of the distribution starting from Eq.~\eqref{eq:CGFcommsplit}:
 \begin{align}
 K^{\text{diss}}(\lambda) = \frac{\beta^2(\lambda^2-\lambda)}{2N}\int_\gamma\text{Var}_t [\dot H_t] + \frac{\beta^2}{2N} \int_\gamma\int_0^\lambda {\rm d}x \int_x^{1-x}{\rm d}y\,  I^y(\pi_t, \dot H_t).
 \end{align}
 First, we obtain again the expression for the average dissipated work :
 \begin{align}\label{eq:E2} 
 \average{w_{\rm diss}}{}  &= (-\beta)^{-1} \frac{\de}{\de\lambda}\,K^{\text{diss}}(\lambda)\Big|_{\lambda=0}=-\norbra{\frac{\beta(2\lambda -1)}{2N}\int_\gamma\text{Var}_t [\dot H_t] + \frac{\beta}{2N}\int_\gamma \int_\lambda^{1-\lambda} {\rm d}y \, I^y(\pi_t, \dot H_t)}\Big|_{\lambda=0}=\nonumber\\
 &=
 \frac{\beta}{2N}\int_\gamma\text{Var}_t [\dot H_t] - \frac{\beta}{2N}\int_\gamma \int_0^1 {\rm d}y \, I^y(\pi_t, \dot H_t),
 \end{align}
 where we can recognise the second term in the equality as the quantum correction $\mathcal{Q}$ defined in Eq.~\eqref{eq:Qdef}. The work fluctuations on the other hand are given by:
 \begin{align}\label{eq:E3} 
 \sigma^2_{\rm diss}  &=\norbra{ \frac{1}{N}\int_\gamma\text{Var}_t [\dot H_t] - \frac{1}{2N}\int_\gamma  (2 I^{\lambda}(\pi_t, \dot H_t) )}\Big|_{\lambda=0} = \frac{1}{N}\int_\gamma\text{Var}_t [\dot H_t],
 \end{align}
 where we used the fact that the skew information satisfies $I^{\lambda}(\pi_t, \dot H_t) = I^{1-\lambda}(\pi_t, \dot H_t)$, together with the identity $I^{0}(\pi_t, \dot H_t)=0$. 
 
 From the third cumulant onwards only the Wigner-Yanase-Dyson skew information contributes to the expression of the cumulants. In fact, if we further differentiate Eq.~\eqref{eq:E3} , we can see that:
 \begin{align}
 \kappa_{w}^{(n>3)} := (-\beta)^{-n} \frac{\de^n}{\de\lambda^n}\,K^{-\beta w}(\lambda)\Big|_{\lambda=0} =  \frac{\beta}{N(-\beta)^{n-1}} \int_\gamma \norbra{\frac{\de^{n-2}}{\de\lambda^{n-2}}  I^{\lambda}(\pi_t, \dot H_t)}.\label{eq:E4}
 \end{align}
 Using Eq.~\eqref{eq:D5} we can give a recursive formula of the equation just obtained. First, it is useful to give the expression for the derivative of the functional:
 \begin{align}
 \frac{\de}{\de\lambda}\Tr[\pi_t^{\lambda} A \pi_t^{1-\lambda} B] &= \Tr[\pi_t^{\lambda} (\log\pi_t A -A\log\pi_t)\pi_t^{1-\lambda} B]  = -\beta\, \Tr[\pi_t^{\lambda}\, [H_t,A] \pi_t^{1-\lambda} B] =\label{eq:E5}\\
 &= \Tr[\pi_t^{\lambda} A \pi_t^{1-\lambda} (B\log \pi_t - \log \pi_t B )]  = -\beta\, \Tr[\pi_t^{\lambda}\, A \pi_t^{1-\lambda} [B, H_t] ], \label{eq:E6}
 \end{align}
 where we used the fact that $\log\pi_t = -\beta H_t -\log \mathcal{Z}_t$. Then, by applying {\ms alternately} Eq.~\eqref{eq:E5} and Eq.~\eqref{eq:E6}, we can prove by induction the formula:
 \begin{align}
 \kappa_{w}^{(2n+1)} &= \frac{\beta^2 }{N (-\beta)^{2n+1} } \int_\gamma \norbra{\frac{\de^{2n-1}}{\de\lambda^{2n-1}} \Tr[\pi_t^{\lambda} \dot H_t \pi_t^{1-\lambda} \dot H_y] }\Big|_{\lambda=0} = \frac{1}{N } \int_\gamma \Tr[\pi_t\, C_{n-1}^\dagger C_n],\\
 \kappa_{w}^{(2n+2)} &= \frac{\beta^2 }{N (-\beta)^{2n+2} } \int_\gamma \norbra{\frac{\de^{2n}}{\de\lambda^{2n}} \Tr[\pi_t^{\lambda} \dot H_t \pi_t^{1-\lambda} \dot H_y] }\Big|_{\lambda=0} = \frac{1}{N } \int_\gamma \Tr[\pi_t\, C_{n}^\dagger C_n] ,\label{eq:E8}
 \end{align}
 where the index $n$ runs on integer values starting from $n=1$, and we recursively defined the operators $C_n$ by the relation: $C_0 = \dot H_t$ and $C_n = [H_t, C_{n-1}]$.
 
 From Eq.~\eqref{eq:E8} we can infer that all the even cumulants are positive. Moreover, since $I^{0}(\pi_t, \dot H_t)=0$ and $I^{\varepsilon}(\pi_t, \dot H_t)\geq0$ for any $\varepsilon\in(0,1)$, we can also deduce from Eq.~\eqref{eq:E4} that $\kappa_{w}^{(3)}\geq0$, with equality \emph{iff} $[H_t, \dot H_t]\equiv 0$ at all times.
 
 Finally, considering the coordinate expression of the CGF, we can also prove the positivity of the higher odd cumulants. First, it is useful to rewrite Eq.~\eqref{eq:D4} as:
 \begin{align}
 \eqref{eq:D4}=\frac{\beta^2}{N}\int_\gamma\sum_{i>j} \frac{(\pi_i+ \pi_j)(\cosh [(\log \pi_j - \log \pi_i)\lambda] -1)+(\pi_i- \pi_j)\sinh [(\log \pi_j - \log \pi_i)\lambda]}{(\log \pi_j - \log \pi_i)^2} |\dot H_{ij}|^2,
 \end{align}
 which can be verified by expanding the hyperbolic functions in terms of exponentials. In this way we can obtain even and odd cumulants from the expansion of the hyperbolic cosine and sine, respectively. Then, it is straightforward to give the explicit formula:
 \begin{align}
 \kappa_{w}^{(2n+1)} = (-\beta)^{-(2n+1)} \frac{\de^{(2n+1)}}{\de\lambda^{(2n+1)}}\,K^{\text{diss}}(\lambda)\Big|_{\lambda=0} = -\frac{1}{N \beta^{2n-1}}\int_\gamma\sum_{i>j} (\pi_i- \pi_j)(\log \pi_j - \log \pi_i)^{2n-1} |\dot H_{ij}|^2.
 \end{align}
 Since the logarithm preserves the order, the sum is negative. The additional minus {\ms sign} in front {\ms of the integral}, then, implies the positivity of $\kappa_{w}^{(2n+1)}$.
 
 \section{Cumulant generating function for a two level system }
 \label{app:qubit}
 In order to exploit the symmetries of the problem, we choose to parametrize the Hamiltonian of the two level system by spherical coordinates:
 \begin{align}
 H(r, \theta, \phi) = r \cos \phi \sin \theta\,\hat\sigma^x +  r \sin \phi \sin \theta\,\hat\sigma^y+ r \cos \theta\,\hat\sigma^z.
 \end{align}
 Our final goal is to write the $y$-covariance in matrix form. As one can straightforwardly verify, in fact, the covariance is a 2-form, hence it can be rewritten as:
 \begin{align}\label{eq:C1}
 \text{cov}_t^y(\dot H_t,\dot H_t) = \sum_{i,j} \dot x^i \dot x^j \text{cov}_t^y(\partial_{x_i},\partial_{x_j}),
 \end{align}
 where $x_i$ runs over the parameters $(r, \theta, \phi)$ and $\partial_{x_i}$ are defined by:
 \begin{align}\label{eq:C2}
 \left\{\begin{array}{l}
 \partial_r = \cos{\varphi}\sin{\theta} \, \hat\sigma_x+  \sin{\varphi}\sin{\theta} \, \hat\sigma_y +\cos{\theta} \, \hat\sigma_z \\
 \partial_\theta = r\cos{\varphi}\cos{\theta} \, \hat\sigma_x+  r\sin{\varphi}\cos{\theta} \, \hat\sigma_y - r\sin{ \theta} \, \hat\sigma_z \\
 \partial_\varphi =  - r\sin{\varphi}\sin{\theta} \, \hat\sigma_x+  r\cos{\varphi}\sin{\theta} \, \hat\sigma_y.
 \end{array}\right.
 \end{align}
 This form of the equation can be understood as a simple rewriting or, for the more mathematically inclined, can be read off as the fact that the $y$-covariance defines a metric on the space  of Hamiltonians parametrised by $x_i$, and with tangent space spanned by $\partial_{x_i}$. It is straightforward to verify that the form so obtained is also hermitian and positive definite, so its real part defines a Riemannian metric (notice that, since $\text{cov}_t^y(\dot H_t,\dot H_t) \equiv\overline{\text{cov}_t^y(\dot H_t,\dot H_t)}$, one can restrict to the real part of the covariance without affecting the physics). This interpretation allows one to exploit the geometrical picture arising to, e.g., devise optimal thermodynamic protocols~\cite{sivakThermodynamicMetricsOptimal2012,scandiThermodynamicLengthOpen2019}, but this direction will not be pursued further here. 
 
 From Eq.~(\ref{eq:C1}, \ref{eq:C2}) it is a problem of simple computation to obtain the $y$-covariance and the form of the CGF. The only two non zero components of the $y$-covariance are given by: 
 \begin{align}
 \left\{\begin{array}{l}
 \text{cov}^y(\partial_{r}, \partial_{r})=\text{sech}^2(\beta  r),\\
 \text{cov}^y(\partial_{\theta}, \partial_{\theta})= \text{sech}(\beta  r) \cosh (\beta  r (1-2 y))/2 \\
 \text{cov}^y(\partial_{\phi}, \partial_{\phi})= \text{sech}(\beta  r) \cosh (\beta  r (1-2 y))\sin ^2(\theta) /2.
 \end{array}\right.
 \end{align}
 
 Integrating these equations, we obtain $K^{\text{diss}}(\lambda)$ as:
 \begin{align}
 K^{\text{diss}}(\lambda) = \beta^2\int_\gamma (\dot r_t,  \dot \theta_t, \dot \phi_t) \left(
 \begin{array}{ccc}
 1 & 0 & 0 \\
 0 & r^2_t  & 0 \\
 0 & 0 & r^2_t\sin ^2(\theta_t ) \\
 \end{array}
 \right)\left(
 \begin{array}{ccc}
 e^{\rm c}_t(\lambda)& 0 & 0 \\
 0 & e^{\rm q}_t(\lambda) & 0 \\
 0 & 0 &e^{\rm q}_t(\lambda) \\
 \end{array}
 \right)\left(
 \begin{array}{ccc}
 \dot r_t \\
 \dot \theta_t\\
 \dot \phi_t \\
 \end{array}
 \right),\label{eq:20}
 \end{align}
 where the two eigenvalues are given by:
 \begin{align}
 e^{\rm c}_t(\lambda) &= \frac{1}{2} \left(\lambda ^2-\lambda \right) \text{sech}^2(\beta  r_t)\label{eq:21}\\
 e^{\rm q}_t(\lambda) &=\frac{\text{sech}(\beta  r_t) \cosh (\beta  r_t-2 \beta  \lambda  r_t)-1}{4 \beta ^2 r^2_t}.\label{eq:22}
 \end{align}
 
 This concludes the derivation of Eq.~\eqref{eq:CGFqubit}.
 
 \section{Cumulant generating function for an Ising chain in a transverse field }\label{app:Ising}
 We consider now a system described by an Ising chain in a transverse field, whose Hamiltonian reads:
 \begin{align}
 H(h)= -J \sum_{i=1}^L\hat \sigma_i^x\hat\sigma_{i+1}^x + h \hat\sigma_i^z,
 \end{align}
 where $J$ sets the energy scale, $h$ is the intensity of the transverse field and $L$ is the number of sites. We can apply a Jordan-Wigner transformation in order to map the problem to a free fermionic model:
 \begin{align}
 \hat\sigma^z_i &= 1 - 2c^\dagger_i c_i\\
 \hat\sigma^+_j &= \prod_{i<j}(1 - 2c^\dagger_i c_i)c_j\\
 \hat\sigma^-_j &=  \prod_{i<j}(1 - 2c^\dagger_i c_i)c_j^\dagger,
 \end{align}
 where $\curbra{c_i}$ are fermionic annihilation operators associated with each site. After the transformation we get the free Hamiltonian:
 \begin{align}
 H(h) =-J  \sum_{i=1}^L (c_i^\dagger - c_i)(c_{i+1}^\dagger + c_{i+1}) + h(1 - 2c^\dagger_i c_i).
 \end{align}
 Since this representation is quadratic in the creation and annihilation operators, the Hamiltonian is block diagonal in the momentum eigenbasis. For this reason, it is useful to decompose  $\curbra{c_j}$ in Fourier modes as:
 \begin{align}
 c_j= \frac{1}{\sqrt{L}}\sum_k c_k e^{i k j}.
 \end{align}
 Substituting in the Hamiltonian one gets:
 \begin{align}
 H(h) = \sum_{k>0} H_k(h) = \sum_{k>0} E_k(h) (c^\dagger_k c_k+c^\dagger_{-k} c_{-k}-1) -i \Delta_k (c^\dagger_{k} c^\dagger_{-k}+c_{k} c_{-k}),
 \end{align}
 where we used the shorthand notations $E_k(h) := J(2h -2 \cos(k))$ and $\Delta_k :=J(2\sin(k))$. Choosing the basis $\ket{k, -k}$, ordered as $\curbra{\ket{1, 1},\ket{0, 0}, \ket{1, 0},\ket{0, 1}}$, we can rewrite each $H_k$ as:
 \begin{align}\label{eq:D8}
 H_k(h) = \left(
 \begin{array}{cccc}
 E_k(h) & -i \Delta_k & 0 & 0 \\
 i \Delta_k & -E_k(h)  & 0 & 0\\
 0 & 0 &  0 & 0\\
 0 & 0 &  0 & 0
 \end{array}
 \right).
 \end{align}
 Then, thanks to the block diagonal form of the Hamiltonian, the exponential state factorises in the tensor product:
 \begin{align}
 e^{-\beta yH(h)} = \bigotimes_{k>0} \,e^{-\beta y H_k(h)} ,\label{eq:E9}
 \end{align}
 where the index $k$ runs over positive momenta only, so to account for the choice of the basis $\ket{k, -k}$. This rewriting also implies that it is sufficient to study the properties of the 4x4 matrix~\eqref{eq:D8} to understand the physics of the system. In particular, we can obtain the partition function as:
 \begin{align}
 \mathcal{Z}(h) = \Tr[e^{-\beta H(h)}] = \prod_{k>0} \,\Tr[e^{-\beta H_k(h)}].\label{eq:D12}
 \end{align}
 In the limit $L\gg 1$, the discrete set of momenta becomes approximately continuous in the Brillouin zone. In this regime, one can rewrite the logarithm of the partition function as:
 \begin{align}
 \log \mathcal{Z}(h) &=  \log\prod_{k>0} \,\Tr[e^{-\beta H_k(h)}] =  L \sum_{k>0} \frac{1}{L}\log \,\Tr[e^{-\beta H_k(h)}] =\nonumber\\
 &= L \int_0^\pi \de k\,\log\sqrbra{4 \cosh ^2\left(\beta \epsilon _k\right)} +\bigo{1},\label{eq:E11}
 \end{align}
 where in the last line we used the definition of Riemann sum and we defined the energy eigenvalue ${\epsilon _k = J \sqrt{h^2-2 h \cos (k)+1}}$. The exponential of Eq.~\eqref{eq:E11} gives the partition function. Moreover, the average power can be obtained as:
 \begin{align}
 \langle \dot H\rangle_{\pi(h)} = -\dot h \beta^{-1} \partial_h \log \mathcal{Z}(h)  = -2 \beta L  \dot h \int_0^\pi \de k\, \tanh (\beta   \epsilon _k)  \;(\partial_h \epsilon _k) +\bigo{1}.
 \end{align}
 
 We can now pass to evaluate the $y$-covariance. First, notice that the variation of the Hamiltonian also factorises as:
 \begin{align}
 \dot H = -J \sum_{i=1}^L  \dot h\,\hat \sigma_i^z = \dot h \sum_{k>0} \partial_h H_k  =  -J \dot h  \sum_{k>0}(c^\dagger_k c_k+c^\dagger_{-k} c_{-k}-1),
 \end{align}
 where $\partial_h H_k (h) $ can be written in matrix form, in analogy with Eq.~\eqref{eq:D8}, as:
 \begin{align}
 \partial_h H_k  =  \left(
 \begin{array}{cccc}
 2J & 0 & 0 & 0 \\
 0 & -2J & 0 & 0\\
 0 & 0 &  0 & 0\\
 0 & 0 &  0 & 0
 \end{array}
 \right).\label{eq:D11}
 \end{align}
 
 Thanks to the factorisation of the exponential in Eq.~\eqref{eq:E9} and the additive structure of $\dot H(h) $, we can rewrite:
 \begin{align}
 \text{cov}_h^y(\dot H,\dot H) &= 	\Tr\sqrbra{\pi(h)^{1-y}(\dot H-\langle \dot H\rangle_{\pi(h)} )\,\pi(h)^{y}\,(\dot H-\langle \dot H\rangle_{\pi(h)} )}=\nonumber\\
 &= \dot h^2 L \sum_{k>0} \frac{1}{L}\frac{ \Tr\sqrbra{e^{-\beta (1-y) H_k(h)}\partial_h H_k  e^{-\beta y H_k(h)} \partial_h H_k } }{\Tr[e^{-\beta H_k(h)}] } - \langle \partial_h H_k\rangle_{\pi(h)}^2= \nonumber\\
 &=\dot h^2 L \int_0^\pi \de k\, C(k, y, h) + \bigo{1},
 \end{align}
 where the function $C(k, y, h)$ is given by:
 \begin{align}
 C(k, y, h) &=\frac{2 \text{sech}^2\left(\beta  \epsilon _k\right)}{\epsilon _k^2} \left((h-\cos (k))^2+\sin ^2(k) \cosh \left(2 \beta  (1-2 y) \epsilon _k\right)\right).
 \end{align}

 The cumulant generating function of the dissipated work is then obtained by plugging this expression into the definition Eq.~\eqref{eq:CGFslowdriving}, which gives:
 \begin{align}\label{eq:G17}
 K^{\text{diss}}(\lambda) = -\frac{\beta^2 L}{2N} \int_\gamma \dot h^2\int_0^\lambda {\rm d}x \int_x^{1-x}{\rm d}y \, \int_0^\pi \de k\, C(k, y, h) . 
 \end{align}
 
 In Fig.~\ref{fig:fig5} we presented the dependence of ${\rm cov}^y_h$ on $y$ for different temperatures. As it was noticed from Eq.~\eqref{eq:CGFpolynomial}, if the $y$-covariance would be constant when varying $y$, we would regain a Gaussian distribution. This is the case at high temperature. On the other hand, one can see that for higher and  higher values of $\beta$ the non Gaussian effects become more evident, and the $y$-covariance starts being more and more sensitive to variations of $y$. Moreover, it should be noticed that this effect is more prominent for low values of $h$, a signal reminiscent of the zero temperature phase transition to a magnetic long range order. 
 
 In order to take the thermodynamic limit we consider the CGF of the variable $\frac{-\beta(w-\Delta F)}{L}$, which we will denote by $K^{\text{diss}}_{\text{resc}}(\lambda)$. This rescaling makes the average dissipation finite for $L\rightarrow\infty$. Moreover, from the definition of the CGF, it is straightforward to verify the general property: $K^{c X}(\lambda) = K^{X}(c \lambda)$. Then, we can rewrite Eq.~\eqref{eq:G17} as:
 \begin{align}
 K^{\text{diss}}_{\text{resc}}(\lambda) &= K^{\text{diss}}(\frac{\lambda}{L}) = -\frac{\beta^2 L}{2N} \int_\gamma \dot h^2\int_0^{\frac{\lambda}{L}} {\rm d}x \int_x^{1-x}{\rm d}y \, \int_0^\pi \de k\, C(k, y, h) =\nonumber\\
 &=-\frac{\beta^2 }{2N} \int_\gamma \dot h^2\int_0^{\lambda} {\rm d}x \int_{\frac{x}{L}}^{1-\frac{x}{L}}{\rm d}y \, \int_0^\pi \de k\, C(k, y, h) =\nonumber\\
 &=-\frac{\beta^2 \lambda}{2N} \int_\gamma \dot h^2 \int_{0}^{1}{\rm d}y \, \int_0^\pi \de k\, C(k, y, h) + \frac{\beta^2 \lambda^2}{2N L} \int_\gamma \dot h^2 \int_0^\pi \de k\, C(k, 0, h) + \bigo{\frac{1}{L^2}},\label{eq:G18}
 \end{align}
 where in the last line we expanded in powers of $x/L$ for $L\gg 1$. Finally, notice that Eq.~\eqref{eq:G18} can be cast in a more compact form as:
 \begin{align}
 K^{\text{diss}}_{\text{resc}}(\lambda) &= -\frac{\beta^2 }{2N}  \int_\gamma \dot h^2\norbra{ \lambda\Tr\sqrbra{\partial_h H\, \J_{\pi_t} [\Delta_{\pi_t}\partial_h H]} - \frac{\lambda^2}{L} \Tr\sqrbra{\partial_h H\, \operatorS_{\pi_t} [\Delta_{\pi_t}\partial_h H]}} +\bigo{\frac{1}{L^2}}.
 \end{align}
 
 \section{Asymmetry monotones and dissipation}\label{app:asymmetry}
 
 In this section we show how one can split the $y$-covariance in a coherent and a dephased contribution. Moreover, we will show how the two terms can be connected with a decomposition of the total protocol in two parallel processes, one in which the entropy production arises from the creation and dissipation of athermality resources, and one for the asymmetry resources. This will motivate the introduction of the two CGF $K^{\text{diss}}_{\text{deph}}(\lambda)$ and $K^{\text{diss}}_{\text{asym}}(\lambda)$.
 
 Using the definition of the coherent power operator $\dot H_t^{\text{c}} := \dot H_t-\mathcal{D}_t(\dot H_t)$, we can rewrite the  $y$-covariance as:
 \begin{align}\label{eq:H1}
 \text{cov}_{t}^y(\dot H_t, \dot H_t)  &= \text{cov}_{t}^y(\mathcal{D}_t(\dot H_t) + \dot H_t^{\text{c}}, \mathcal{D}_t(\dot H_t) + \dot H_t^{\text{c}})=\nonumber\\ 
 & =\text{cov}_{t}^y(\mathcal{D}_t(\dot H_t) , \mathcal{D}_t(\dot H_t) )+\text{cov}_{t}^y(\dot H_t^{\text{c}},  \dot H_t^{\text{c}})+2 \text{Re} \sqrbra{\text{cov}_{t}^y(\mathcal{D}_t(\dot H_t) , \dot H_t^{\text{c}})}.
 \end{align}
 We can now proceed to prove that the last term in the previous equation is zero. First, notice that the average of $\dot H_t^{\text{c}}$ is zero. Then, expressing the $y$-covariance in coordinates we have:
 \begin{align}
 \text{cov}_{t}^y(\mathcal{D}_t(\dot H_t) , \dot H_t^{\text{c}}) &=  \Tr\sqrbra{\pi_t^{1-y}\mathcal{D}_t(\dot H_t) \pi_t^{y} \dot H_t^{\text{c}}} = \Tr\sqrbra{\pi_t\mathcal{D}_t(\dot H_t) \dot H_t^{\text{c}}} =\nonumber\\
 &=\sum_i  (\pi_t)_i\mathcal{D}_t(\dot H_t)_i  \bra{i}  \dot H_t^{\text{c}}\ket{i} = 0,
 \end{align}
 where we denote by $\ket{i}$ the eigenbasis of the $H_t$, and we used the fact that $\dot H_t^{\text{c}}$ only has off-diagonal terms. This proves Eq.~\eqref{eq:covysplitting}. Plugging Eq.~\eqref{eq:H1} in the expression of the dissipative CGF in the quasistatic limit in Eq.~\eqref{eq:D2}  we also obtain:
 \begin{align}\label{eq:H3}
 K^{\text{diss}}(\lambda) = K^{\text{diss}}_{\text{deph}}(\lambda) +K^{\text{diss}}_{\text{asymm}}(\lambda),
 \end{align}
 which proves Eq.~\eqref{eq:splitCGF}. As explained in the main text, we can think of the two contribution in Eq.~\eqref{eq:H3} as coming from a decomposition of the main protocol in two steps: (a) a change of the Hamiltonian only along the diagonal $H_i \rightarrow H_i + \mathcal{D}_i(H_{i+1} - H_{i})$; (b) a {\ms rotation of} the energy basis $H_i + \mathcal{D}_i(H_{i+1} - H_{i}) \rightarrow H_{i+1}$. The dissipative CGF for the two protocols are respectively given by:
 \begin{align}\label{eq:H4}
 &K^{\text{diss}}_{\text{(a)}}(\lambda) =  \sum_{i=1}^{N-1} (\lambda-1)S_\lambda(\pi (H_i + \mathcal{D}_i(H_{i+1} - H_{i}) ) || \pi(H_i )),\\
 &K^{\text{diss}}_{\text{(b)}}(\lambda) =  \sum_{i=1}^{N-1} (\lambda-1)S_\lambda( \pi(H_{i+1}) || \pi (H_i + \mathcal{D}_i(H_{i+1} - H_{i}) )).\label{eq:H5}
 \end{align}
 To see how this two contribution build up the full CGF, we can now apply  Eq.~\eqref{eq:appCRenyiexpansion} to the two equations in the slow driving regime. For Eq.~\eqref{eq:H4} it is straightforward to obtain:
 \begin{align}
 K^{\text{diss}}_{\text{(a)}}(\lambda) =-\frac{\beta^2}{2 N} \int_\gamma  \int_0^\lambda {\rm d}x \int_x^{1-x}{\rm d}y \, \text{cov}_{t}^y(\mathcal{D}_t(\dot H_t),\mathcal{D}_t(\dot H_t)) = K^{\text{diss}}_{\text{deph}}(\lambda).
 \end{align}
 This term comes from the expansion of second laws of the form of~\cite{brandaoSecondLawsQuantum2015b}. In fact, from standard perturbation theory we can see that  $\pi (H_i + \mathcal{D}_i(H_{i+1} - H_{i}) )$ has the same spectrum as $\pi_{i+1}$. Also, Since $\mathcal{D}_t(\dot H_t)$ commutes with $\pi_t$ at all times, the $y$-covariance reduces to the usual variance: $\text{cov}_t^y(\mathcal{D}_t(\dot H_t),\mathcal{D}_t(\dot H_t)) = \text{Var}[\mathcal{D}_t(\dot H_t)]$. This means that in the quasistatic regime it is sufficient to constrain the work statistics of incoherent protocols with a single second law, arising from the expansion of the relative entropy which accounts for average quantities only.
 
 Applying Eq.~\eqref{eq:appCRenyiexpansion} to Eq.~\eqref{eq:H5} instead gives:
 \begin{align}
 K^{\text{diss}}_{\text{(b)}}(\lambda)=-\frac{1}{2} \sum_{i=1}^{N-1}  \int_0^\lambda {\rm d}x \int_x^{1-x}{\rm d}y \, \text{cov}_{\pi_{i+1}}^y(\J_{\pi_{i+1}}^{-1}[\tilde{\Delta} H_i ], \J_{\pi_{i+1}}^{-1}[\tilde{\Delta} H_i])  +\bigo{\tilde{\Delta} H_i^3},
 \end{align}
 where we defined $\tilde{\Delta} H_i:= (\pi (H_i + \mathcal{D}_i(H_{i+1} - H_{i}) ) - \pi_{i+1})$. It is a simple exercise in Taylor expansions to show that $\tilde{\Delta} H_i= \J_{\pi_{i+1}}[(H_{i+1} - H_{i})^{\text{c}} ]$ at first order. In the continuous limit we then get:
 \begin{align}
 K^{\text{diss}}_{\text{(b)}}(\lambda)=-\frac{\beta^2}{2 N} \int_\gamma  \int_0^\lambda {\rm d}x \int_x^{1-x}{\rm d}y \, \text{cov}_{t}^y(\dot H_t^{\text{c}},\dot H_t^{\text{c}}) = K^{\text{diss}}_{\text{asymm}}(\lambda).
 \end{align}
{\ms It is also interesting to notice that the same expression could have been derived for a process in which at each step the initial state of the system would be of the form $\mathcal{D}_i(\pi (H_{i+1}))$. In fact, $\pi (H_i + \mathcal{D}_i(H_{i+1} - H_{i})\equiv\mathcal{D}_i(\pi (H_{i+1}))$ at first order in perturbation theory. For this reason, we can substitute in Eq.~\eqref{eq:H5}  $S_\lambda( \pi(H_{i+1}) || \pi (H_i + \mathcal{D}_i(H_{i+1} - H_{i}) ))\rightarrow S_\lambda( \pi(H_{i+1}) || \mathcal{D}_i(\pi (H_{i+1})))$, giving a sum of terms akin to the one in the coherent second laws in Eq.~\eqref{eq:asymmetrycostraints}.}
 
 \section{Cumulant generating function for continuous evolution}
 \label{app:contEv}
 In this section we pass to the study of the cumulant generating function for a continuous process whose state can be approximated as $\varrho_t = \pi_t +\delta\varrho_t$, where $\delta\varrho_t$ is of order $\bigo{1/\tau}$, and we neglect higher order corrections.
 
 The cumulant generating function for a continuous process which initially starts in equilibrium is given by:
 \begin{align}
 K^{-\beta W}(\lambda) &=   \log\Tr\sqrbra{ e^{-\beta \lambda H_{\tau}} U_{\tau} e^{\beta \lambda H_{0}} \pi_0 U^{\dagger}_{\tau}} = \nonumber\\
 &=-\beta \lambda \Delta F + \log\Tr\sqrbra{ \pi_\tau^{\lambda} U_{\tau}  \pi_0^{1-\lambda}U^{\dagger}_{\tau}},\label{eq:F1}
 \end{align}
 where between the first and the second line we have multiplied and divided the trace by $(\mathcal{Z}_0/\mathcal{Z}_\tau)^\lambda$, and we used the definition of equilibrium free energy to isolate the first term in Eq.~\eqref{eq:F1}.
 
 By using the notation $\varrho_\tau :=U_{\tau}  \pi_0U^{\dagger}_{\tau}$, we can rewrite the dissipative CGF in the compact form:
 \begin{align}\label{eq:F2}
 K^{\text{diss}}(\lambda) =(\lambda - 1) S_\lambda(\pi_\tau || \varrho_\tau).
 \end{align}
 As explained in the main text, it is useful to make explicit the dependency of the CGF on the particular trajectory in the parameter space, and for this reason we rewrite Eq.~\eqref{eq:F2} as:
 \begin{align}\label{eq:68}
 K^{\text{diss}}(\lambda)  = \int_0^\tau \dt \norbra{\frac{\de}{\dt} \log\Tr\sqrbra{ \pi_t^{\lambda} \varrho_t^{1-\lambda}}}.
 \end{align}
 
 In order to compute~\eqref{eq:68}, it is useful to expand the trace in the form:
 \begin{align}
 \Tr\sqrbra{ \pi_t^{\lambda} \varrho_t^{1-\lambda}} = 1 + \int_0^\lambda \de x\, \Tr\sqrbra{ \pi_t^{x}(\log \pi_t - \log \varrho_t ) \varrho_t^{1-x}};
 \end{align}
 from this equation, it is easy to verify that the derivative in Eq.~\eqref{eq:68} can be expressed as:
 \begin{align}
 \frac{\de}{\dt} \log\Tr\sqrbra{ \pi_t^{\lambda} \varrho_t^{1-\lambda}} &= \lim_{\varepsilon\rightarrow 0} \frac{1}{\varepsilon}\log\frac{\Tr\sqrbra{ \pi_{t+\varepsilon}^{\lambda} \varrho_{t+\varepsilon}^{1-\lambda}}}{ \Tr\sqrbra{ \pi_{t}^{\lambda} \varrho_{t}^{1-\lambda}}}= \nonumber\\
 &=\lim_{\varepsilon\rightarrow 0} \frac{1}{\varepsilon} \log \norbra{1 + \frac{\varepsilon\int_0^\lambda \de x\,\partial_t \Tr\sqrbra{ \pi_t^{x}(\log \pi_t - \log \varrho_t ) \varrho_t^{1-x}}}{\Tr\sqrbra{ \pi_t^{\lambda} \varrho_t^{1-\lambda}}}+ \dots} =\nonumber\\
 &=  \frac{\int_0^\lambda \de x\,\partial_t \Tr\sqrbra{ \pi_t^{x}(\log \pi_t - \log \varrho_t ) \varrho_t^{1-x}}}{\Tr\sqrbra{ \pi_t^{\lambda} \varrho_t^{1-\lambda}}}.\label{eq:F5}
 \end{align}
 The computation of the derivative in Eq.~\eqref{eq:F5} is somehow more straightforward than the one in the original Eq.~\eqref{eq:68}. First, we split the derivative in three parts:
 \begin{align}
 \partial_t \Tr\sqrbra{ \pi_t^{x}(\log \pi_t - \log \varrho_t ) \varrho_t^{1-x}} &=  \Tr\sqrbra{ (\partial_t\pi_t^{x})(\log \pi_t - \log \varrho_t ) \varrho_t^{1-x}} + \Tr\sqrbra{ \pi_t^{x}(\partial_t \log \pi_t)  \varrho_t^{1-x}} +\label{eq:F6}\\
 &- \Tr\sqrbra{ \pi_t^{x}(\partial_t (\log \varrho_t\,  \varrho_t^{1-x}))}.\label{eq:F7}
 \end{align}
 It should be noticed that Eq.~\eqref{eq:F7} does not contribute to the final expression. In fact, using the definition of $\varrho_t = U_{t}  \pi_0U^{\dagger}_{t}$ and the cyclicity of the trace, we have:
 \begin{align}
 \Tr\sqrbra{ \pi_t^{x}(\partial_t (\log \varrho_t\,  \varrho_t^{1-x}))} = -i\, \Tr\sqrbra{ \pi_t^{x} H_t(\log \varrho_t\,  \varrho_t^{1-x})}+i \,\Tr\sqrbra{ \pi_t^{x} (\log \varrho_t\,  \varrho_t^{1-x})H_t} = 0,
 \end{align}
 since $[H_t, \pi_t] = 0$. Moreover, using the expansion of the thermal state {\ms provided} in Eq.~\eqref{eq:C3}, together with the identity $\partial_t \log \pi_t = \Delta_t \dot H_t $ we obtain the final result:
 \begin{align}
 \partial_t \Tr\sqrbra{ \pi_t^{x}(\log \pi_t - \log \varrho_t ) \varrho_t^{1-x}} &= -\beta \int_0^x\de y \,\Tr\sqrbra{ \pi_t^{y}\Delta_t \dot H_t \pi_t^{x-y}(\log \pi_t - \log \varrho_t ) \varrho_t^{1-x}}+\nonumber\\
 &-\beta	\, \Tr\sqrbra{ \pi_t^{x}\Delta_t \dot H _t \varrho_t^{1-x}}.
 \end{align}
 Plugging this expansion back into Eq.~\eqref{eq:68} we finally obtain the expression for the dissipative CGF presented in Eq.~\eqref{eq:CGFcontinuous2}:
 \begin{align}
 &K^{\text{diss}}(\lambda) = 
 -\beta\int_\gamma \int_0^\lambda {\rm d}x   \norbra{\frac{\Tr\sqrbra{ \pi_t^{x}\Delta_t \dot H_t  \varrho_t^{1-x}}}{\Tr\sqrbra{ \pi_t^{\lambda} \varrho_t^{1-\lambda}}} \; +\int_0^x\de y \frac{\Tr\sqrbra{ \pi_t^{y}\Delta_t \dot H_t \pi_t^{x-y}(\log \pi_t - \log \varrho_t ) \varrho_t^{1-x}}}{\Tr\sqrbra{ \pi_t^{\lambda} \varrho_t^{1-\lambda}}}}.\label{eq:F10}
 \end{align}
This equation is exact and it is the first result of the appendix. 
 
 We can now pass to the analysis of the CGF in the slow driving limit. As stated above we consider states that can be approximated as $\varrho_t \approx \pi_t +\delta\varrho_t$, in the trace sense:
 \begin{align}
 \Tr\sqrbra{A \varrho_t}= \Tr\sqrbra{A (\pi_t +\delta\varrho_t)}  +\bigo{1/\tau^{2}},
 \end{align}
 where $A$ is a generic observable. Then, using the two expansions~\cite{hiaiIntroductionMatrixAnalysis2014a, scandiQuantifyingDissipationThermodynamic2018}:
 \begin{align}
 &\log \varrho_t = \log \pi_t + \J_t^{-1}[\delta\varrho_t]+\bigo{\delta\varrho^2_t},\\
 &\varrho_t^{x} = e^{x(\log \pi_t +  \J^{-1}[\delta \varrho_t])} = \pi_t^x +  \int_0^x \de y \, \pi_t^{y} \J_t^{-1}[\delta \varrho_t] \pi_t^{-y} \pi_t^x +\bigo{\delta\varrho_t^2},
 \end{align}
 we can approximate all the terms in Eq.~\eqref{eq:F10}. For example, at first order the denominator is trivial:
 \begin{align}
 &\Tr\sqrbra{ \pi_t^{\lambda} \varrho_t^{1-\lambda}} = 1 +\cancel{ \Tr\sqrbra{\pi_t\J_t^{-1}[\delta\varrho_t]}} +\bigo{\delta\varrho^2_t},
 \end{align}
 where we used the cyclicity of the trace, the hermiticity of $\J_t^{-1}$, and the fact that $\delta\varrho_t$ is traceless.  For what regards the numerator, the two terms can be expanded as:
 \begin{align}
 &\Tr\sqrbra{  \varrho_t^{1-x}\pi_i^{x}\Delta_t \dot H_t} =\cancel{\Tr\sqrbra{  \pi_t\Delta_t \dot H_t}}  +   \int_0^{1-x} \de y \,\Tr\sqrbra{ \pi_t^{y} \J_t^{-1}[\delta \varrho_t] \pi_t^{1-y}\Delta_t \dot H_t}+\bigo{\delta\varrho_t^2},\label{eq:F15}\\
 &\Tr\sqrbra{ \pi_t^{-y}\Delta_t \dot H_t \pi_t^{x-y}(\log \pi_t - \log \varrho_t ) \varrho_t^{1-x}}=	-\Tr\sqrbra{\pi_t^{1-(x-y)} \Delta_t \dot H_t \pi_t^{x-y}\J_t^{-1}[\delta\varrho_t] } +\bigo{\delta\varrho_t^2},\label{eq:F16}
 \end{align}
 where in Eq.~\eqref{eq:F15} the first term cancels thanks to the definition of $\Delta_{\pi_t} \dot H_t$, and in Eq.~\eqref{eq:F16} the approximation $\varrho_t = \pi_t$ is sufficient, thanks to the presence of the difference of logarithms. Moreover, we can perform the change of variables $u = x$ and $v = x-y$ which gives:
 \begin{align}
 \int_0^\lambda {\rm d}x\int_0^x\de y \,\Tr\sqrbra{ \delta\varrho_t \J_t^{-1}[\pi_t^{1-(x-y)} \Delta_t \dot H_t \pi_t^{x-y}] } =  \int_0^\lambda {\rm d}u\int_0^u\de v \,\Tr\sqrbra{ \delta\varrho_t \J_t^{-1}[\pi_t^{1-v} \Delta_t \dot H _t\pi_t^{v}] }.
 \end{align}
 At this point we are ready to take the slow driving limit of Eq.~\eqref{eq:F10}, as:
 \begin{align}
 K^{\text{diss}}(\lambda) = &-\beta \int_\gamma  \int_0^\lambda {\rm d}x\int_0^{1-x} \de y \,\Tr\sqrbra{ \delta \varrho_t \J_t^{-1}[\pi_t^{1-y}\Delta_t \dot H_t \pi_t^{y}]}+\nonumber\\
 &+\beta \int_\gamma  \int_0^\lambda {\rm d}x\int_0^x\de y \,\Tr\sqrbra{ \delta\varrho_t \J_t^{-1}[\pi_t^{1-y} \Delta_t \dot H_t \pi_t^{y}] }=\label{eq:F17}\\
 &= -\beta \int_\gamma  \int_0^\lambda {\rm d}x\int_x^{1-x} \de y \,\Tr\sqrbra{ \delta \varrho_t \J_t^{-1}[\pi_t^{1-y}\Delta_t \dot H_t \pi_t^{y}]}.\label{eq:F19}
 \end{align}
 Since $\J^{-1}_t$ is hermitian, we can move it to $\delta \varrho_t$, obtaining:
 \begin{align}\label{eq:I20}
 K^{\text{diss}}(\lambda) =  -\beta \int_\gamma  \int_0^\lambda {\rm d}x\int_x^{1-x} \de y \,\Tr\sqrbra{ \J_t^{-1}[\delta \varrho_t] \pi_t^{1-y}\Delta_t \dot H_t \pi_t^{y}} =  -\beta \int_\gamma  \int_0^\lambda {\rm d}x\int_x^{1-x} \de y \,\text{cov}_t^y(\dot H_t, \J_t^{-1}[\delta \varrho_t ]).
 \end{align}
 This concludes the derivation of Eq.~\eqref{eq:CGFquasistaticcontinuous}.

 Comparing the derivation just presented with the one given in Appendix~\ref{app:Renyi}, we can also obtain the identity:
 \begin{align}
 \frac{\partial^2}{\partial t \partial s}\,  S_\lambda(\varrho+t \sigma_1|| \varrho + s \sigma_2) \bigg|_{t=s=0}= \frac{1}{2(\lambda -1)}\int_0^\lambda {\rm d}x \int_x^{1-x}{\rm d}y \, \text{cov}_t^y(\J_\varrho^{-1}[\sigma_1], \J_\varrho^{-1}[\sigma_2]) .
 \end{align}
 Then we can rewrite the CGF in Eq.~\eqref{eq:I20} as:
 \begin{align}
 K^{\text{diss}}(\lambda) =  (\lambda -1)\int_\gamma  \frac{\partial^2}{\partial \varepsilon_1 \partial \varepsilon_2}S_\lambda(\pi_{t+\varepsilon_1}|| \pi_{t}+\varepsilon_2\delta \rho_t),
 \end{align}
 so that we can divide a contribution coming from the driving (the derivative in $\varepsilon_1$) from the contribution arising from the non-equilibrium created during the protocol.
 
 As explained in the main text, we can consider the case where the reduced dynamics of the system takes a Lindblad form $\dot{\rho}_t=\lind_t(\rho_t)$. We suppose the Lindbladian is relaxing so that there exists a unique thermal fixed point at each instant of time:
 \begin{align}\label{eq:relax}
 \lim_{\nu\to\infty}e^{\nu \lind_t}(\rho)=\pi_t,
 \end{align}
 for any normalised state $\rho$. It can then be shown that the correction term in the slow driving regime is given by $\delta\varrho_t\equiv -\beta\lind_t^+[\J_t[\Delta_t \dot H]]$ \cite{cavinaSlowDynamicsThermodynamics2017}. Here $\lind_t^+$ is the Drazin inverse, which is formally given by \cite{scandiThermodynamicLengthOpen2019}
 \begin{align}\label{eq:drazin}
 \lind_t^+[.]:=\int^\infty_0 d\nu \ e^{\nu \lind_t}\big[\pi_t \Tr\sqrbra{(.)}-(.)\big].
 \end{align}
 Substituting $\delta\rho_t$ into~\eqref{eq:I20}, we find
 \begin{align}\label{eq:I21}
 K^{\text{diss}}(\lambda) =  \beta^2 \int_\gamma  \int_0^\lambda {\rm d}x\int_x^{1-x} \de y \,\text{cov}_t^y(\dot H_t, \J_t^{-1}[\lind_t^+[\J_t[\Delta_t \dot H]] ]).
 \end{align}
 We now introduce the following superoperator:
 \begin{align}
 \mathbb{M}^x_t(.):=\pi_t^x \  (.) \  \pi_t^{1-x},
 \end{align}
 which under integration yields $\J_t(.)=\int^1_0 dx \ \mathbb{M}^x_t(.)$. As a second assumption we suppose that the Lindbladian satisfies \textit{quantum detailed balance}, which implies the following \cite{Fagnola2010a}:
 \begin{align}\label{eq:detail}
 \lind_t \ \mathbb{M}^x_t(.)=\mathbb{M}^x_t \ \tilde{\lind}_t(.),
 \end{align}
 Here $\tilde{\lind}_t$ is the dual of the Lindbladian whose symmetric part coincides with that of $\lind_t$. Note that while we may assume~\eqref{eq:detail} \textit{a priori}, the condition naturally holds for weakly-coupled quantum systems connected to a single bath. Combining this with~\eqref{eq:I21} we find
 \begin{align}
 K^{\text{diss}}(\lambda) &=  \beta^2 \int_\gamma  \int_0^\lambda {\rm d}x\int_x^{1-x} \de y \,\text{cov}_t^y(\dot H_t, \J_t^{-1}\J_t\tilde{\lind}_t^+[\Delta_t \dot H]] )= \\
 &=  \beta^2 \int_\gamma  \int_0^\lambda {\rm d}x\int_x^{1-x} \de y \,\text{cov}_t^y(\dot H_t, \tilde{\lind}_t^+[\Delta_t \dot H]] )= \\
 &=  \beta^2 \int_\gamma  \int_0^\lambda {\rm d}x\int_x^{1-x} \de y \,\Tr\sqrbra{\dot H_t \mathbb{M}^y_t \tilde{\lind}_t^+[\Delta_t \dot H]] )}= \\
 &=  \beta^2 \int_\gamma  \int_0^\lambda {\rm d}x\int_x^{1-x} \de y \,\Tr\sqrbra{\dot H_t \lind_t^+ \mathbb{M}^y_t [\Delta_t \dot H]] )}= \\
 &=  -\beta^2 \int_\gamma  \int_0^\lambda {\rm d}x\int_x^{1-x} \de y \,\norbra{-\text{cov}_t^y([\lind_t^+]^\dagger(\dot H_t), \dot H_t)}.
 \end{align}
 Substituting in~\eqref{eq:drazin} into the above equation completes the derivation of~\eqref{eq:CGFquasistaticcontinuous2}. Finally, we note the positivity of the integrand:
 \begin{align}
 -\text{cov}_t^y([\lind_t^+]^\dagger(\dot H_t), \dot H_t)\geq 0.
 \end{align}
 This follows from the detailed balance relation~\eqref{eq:detail} and the fact that the non-zero eigenvalues of $\lind_t^+$ have a negative real part due to condition~\eqref{eq:relax} (see Appendix D in \cite{millerWorkFluctuationsSlow2019} for a detailed proof).

 \section{Notations used in the article}\label{app:notations}
 We list here the notations implicitly used throughout the paper:
 \begin{itemize}
 	\item we denote by $\mathcal{Z}(H)$ the partition function associated with the Hamiltonian $H$, which is defined by: $\mathcal{Z}(H) = \tr\sqrbra{e^{-\beta H}}$;
 	\item the Gibbs state associated with a Hamiltonian H is defined as $\pi(H) = \frac{e^{-\beta H}}{\mathcal{Z}(H)}$;
 	\item in most part of the paper the Hamiltonians will carry a discrete or continuous index; we will use the letter $i$ in the first case, and the letter $t$ for the latter one.  More precisely,  we will often use a continuous description by approximating the discrete path $H_0 \rightarrow H_1 \rightarrow ... \rightarrow H_N$ by a continuous one $\tilde{H}_t$ with  $t \in (0,1)$; so that $\tilde{H}_{i/N}=H_i$.  Then for example, we have that $\dot{\tilde{H}}_{i/N} = \lim_{N\rightarrow \infty} N(H_{i+1}-H_i)$. We will abuse notation and write $H_t$ instead of $\tilde{H}_t$ so that the index $t$ indicates a continuous description (whereas the subindex $i$ indicates a discrete one).
 	\item whenever an object is function of an indexed Hamiltonian alone, the index passes to the state. For example, $\mathcal{Z}_i$ stands for $\mathcal{Z}(H_i)$, or $\pi_i$ for $\pi(H_i)$;
 	\item we will denote by $\gamma$ the path in the parameters space defining the protocol{\ms, and we denote the integration over the protocol by $\int_\gamma\equiv\int^1_0 {\rm d}t$};
 	\item we will use the notation $\text{Var}_\varrho[A]$ to indicate the variance of the observable $A$: $\text{Var}_\varrho[A]:= \Tr\sqrbra{\varrho A^2} - \Tr\sqrbra{\varrho A}^2$.
 \end{itemize}

\end{document}